\documentclass[12pt,reqno]{article}
\usepackage{times}

\usepackage{amsmath}
\usepackage{amsthm}
\usepackage{amsfonts}
\usepackage{amssymb}
\usepackage{epic}
\usepackage{eepic}
\usepackage{epsfig}
\usepackage{times, macros}
\usepackage{graphics}
\usepackage{color}
\usepackage{framed}
\definecolor{shadecolor}{rgb}{1,0.9,0.7}

\def\bea{\begin{eqnarray}}
\def\eea{\end{eqnarray}}
\def\bal{\begin{align}}
\def\eal{\end{align}}

\newcommand{\SU}{{\rm SU(2)}}

\theoremstyle{plain}

\theoremstyle{remark}
\newtheorem{rem}{Remark}

\newcommand{\sst}{\scriptscriptstyle}

\newcommand{\ot}{\otimes}
\newcommand{\ra}{\to}

\newcommand{\fr}[2]{{\textstyle \frac{#1}{#2} }}

\newcommand{\al}{\alpha}
\newcommand{\be}{\beta}
\newcommand{\ga}{\gamma}
\newcommand{\Ga}{\Gamma}
\newcommand{\de}{\delta}
\newcommand{\De}{\Delta}
\newcommand{\ep}{\epsilon}
\newcommand{\la}{\lambda}

\newcommand{\si}{\sigma}

\newcommand{\vf}{\varphi}

\newcommand{\pa}{\partial}
\newcommand{\bv}{\mathbf{v}}
\newcommand{\bm}{\mathbf{m}}
\newcommand{\bw}{\bar{w}}

\newcommand{\bx}{\bar{x}}
\newcommand{\by}{\bar{y}}
\newcommand{\bz}{\bar{z}}
\newcommand{\CA}{{\mathcal A}}

\newcommand{\CC}{{\mathcal C}}
\newcommand{\CD}{{\mathcal D}}
\newcommand{\CE}{{\mathcal E}}
\newcommand{\CF}{{\mathcal F}}

\newcommand{\CH}{{\mathcal H}}
\newcommand{\CI}{{\mathcal I}}
\newcommand{\CL}{{\mathcal L}}
\newcommand{\CM}{{\mathcal M}}
\newcommand{\CN}{{\mathcal N}}
\newcommand{\CO}{{\mathcal O}}  

\newcommand{\CR}{{\mathcal R}}
\newcommand{\CS}{{\mathcal S}}
\newcommand{\CT}{{\mathcal T}}
\newcommand{\CU}{{\mathcal U}}
\newcommand{\CV}{{\mathcal V}}
\newcommand{\CW}{{\mathcal W}}

\newcommand{\CZ}{{\mathcal Z}}

\newcommand{\SH}{{\mathsf H}}

\newcommand{\SL}{{\mathsf L}}

\renewcommand{\SU}{{\mathsf U}}

\newcommand{\fg}{{\mathfrak g}}
\newcommand{\hfg}{\hat{{\mathfrak g}}}

\newcommand{\BR}{{\mathbb R}}

\newcommand{\BC}{{\mathbb C}}

\newcommand{\BP}{{\mathbb P}}
\newcommand{\BS}{{\mathbb S}}

\newcommand{\BZ}{{\mathbb Z}}

\newcommand{\tr}{\mathrm{tr}}

\newcommand{\rf}[1]{(\ref{#1})}

\begin{document}\thispagestyle{empty}
\title{A guide to two-dimensional conformal field theory}
\author{J\"org Teschner}
\address{
Department of Mathematics, \\
University of Hamburg, \\
Bundesstrasse 55,\\
20146 Hamburg, Germany,\\[1ex]
and:\\[1ex]
DESY theory, \\
Notkestrasse 85,\\
20607 Hamburg,
Germany}
\maketitle



Starting from the pioneering works on conformal field theory (CFT) from the 70's and early 80's of the
20th century, the subject has grown rather big. It has found many different applications to various 
parts of theoretical physics on the one hand, and it has inspired several new directions of mathematical
research on the other hand. However, as it often happens, the original unity of the subject got lost 
over time, it branched out in various directions to such an extent that communication may be hard 
between different branches of research on the subject of conformal field theory.

In such a situation it may be helpful to create a ``basic operating system'' for CFT by identifying an efficient 
formalism reaching as far as possible which can easily be extended by ``downloading suitable apps''
from the literature for more specialised or more advanced topics. This is what we are trying to do 
in these lecture notes.

More specific goals are:

i) Revisit 2d CFT from a more general perspective than often
done. Non-generic features are often assumed from the outset, 
we'll here emphasise whatever seems to hold more generically.

ii) Create bridges between the physicists point of 
view and  (some of) the mathematical approaches to CFT, 
wherever this seems beneficial for one or both disciplines.

iii) Discuss some connections between CFT and the theory of integrable models, 
a subject of growing 
importance which is rarely discussed in the existing literature.

Although the author was making efforts to reach a balanced presentation, a
certain bias caused by expertise and interests of the author is probably unavoidable.
There are at least two important areas about which almost nothing will be found in 
these notes. One is the subject of boundary CFT, for 
which \cite{ReS} offers a starting point, the other is the operator-algebraic approach to 
CFT, reviewed e.g. in \cite{Ka}.

Limitations of space and time forced me to give a fairly concise presentation, in which
many important aspects could not be covered, and many details had to be omitted. 

{\it A guide to the reader:}

As indicated above, the idea behind these notes is to 
present a ``skeleton'' of conformal field theory, consisting of some aspects most essential for 
many  applications, and most basic for the mathematical theory. Starting from 
this skeleton one can hopefully access more easily 
whatever more specialised topic one is interested in. The author was looking for a reasonable 
compromise between being mathematically precise, easily accessible for readers with a physics background,
and easy to adapt to more general situations.

These notes may  serve two tasks.  The main part may 
be used as a largely self-contained introductory course if the 
reader is studying these notes with some care, 
reading concentratedly, filling in the gaps and  solving the exercises.
One may, on the other hand, use them for getting an overview, 
a first idea about more advanced topics, 
and pointers towards further literature 
on the subject.
This is reflected in the structure of these lecture notes.
Sections having a title marked with an plus sign contain supplementary or more advanced material that 
could be omitted in a very first reading. Sections marked with a double plus are offering a guide to 
the literature on some further directions.

These notes have three parts. The first part discusses the notion of conformal symmetry and how the 
corresponding constraints on correlation functions get encoded into the representation-theoretic
definition of the 
conformal blocks as solutions to the conformal Ward identities. Extensions of conformal 
symmetry are briefly discussed at the end of this part. 

The second part discusses the conformal bootstrap. Starting from the gluing construction of 
conformal blocks we 
identify consistency conditions for the construction of physical correlation functions from 
conformal blocks. Minimal models and Liouville theory are discussed as examples where 
this program has been fully realised. Some approaches to the construction of more general
CFTs are indicated.

The third part outlines some of the known relations between CFT and integrable models,
using the connections to the isomonodromic deformation problem as main example. It is shown
that CFT can be understood as a quantisation of the isomonodromic deformation problem. 
A few relations to quantum integrable models are very briefly pointed out at the end.

At some occasions we are offering shortcuts to some key results which seem hard to 
extract from the existing literature. This concerns especially the material in 
Sections \ref{highergenus} and \ref{sec:FF}.

\newpage

\section{Part 1: Conformal symmetry}

A two-dimensional (2d) QFT will be called a (2d) conformal field theory if it has 
a symmetry group with Lie algebra containing the Virasoro 
algebra. 

\subsection{States}

The Virasoro algebra $\mathsf{Vir}_c$
is the infinite-dimensional Lie algebra with generators $L_n$, $n\in\BZ$ 
and relations
\begin{equation}
[\,L_n\,,\, L_m\,]\,=\,(n-m)L_{n+m}+\frac{c}{12}n(n^2-1)\de_{n+m,0}\,.
\end{equation}
A 2d QFT has Virasoro symmetry if 
\begin{itemize}
\item[(i)] its Hilbert space $\CH$ is a unitary representation
of  the product 
$\mathsf{Vir}_c\times\overline{\mathsf{Vir}}_c$ with generators\footnote{We use the same notation 
for generators of the Lie algebra $\mathsf{Vir}_c$ and for the operators 
representing them.} denoted by
$L_n$, $\bar{L}_n$, $n\in\BZ$, and
\item[(ii)] the Hamiltonian $\SH$ is given in terms of the operators  $L_0$ and $\bar{L}_0$ as
\begin{equation}
\SH\,=\,L_0+\bar{L}_0+\mathrm{(const.)}.
\end{equation}
\end{itemize}
It follows from (i), (ii) that $\CH$ has the form
\begin{equation}
\CH\,=\,\bigoplus_{R',R''}M_{R',R''}\otimes R'\otimes R''\,,
\end{equation}
where $R'$ and $R''$ are unitary irreducible  representations of the Virasoro 
algebras generated by $L_n$ and $\bar{L}_n$, respectively,
and $M_{R'R''}$ is a multiplicity space transforming trivially under $\mathsf{Vir}_c\times\overline{\mathsf{Vir}}_c$.
In order to have stability ($\SH>0$) we need to have representations  $R'$ containing vectors 
of lowest energy usually called highest weight representations containing a 
subspace $R_{0}'\subset R'$ such that
\begin{equation}
L_n\,e_{0}\,=\,\De_{R'}\de_{n,0}e_{0}\,,\qquad \forall v_0\in R_{0}'\,,\;\;n\geq 0\,.
\end{equation}
Vectors of the form $L_{-n_1}\cdots L_{-n_k}e_0$ with $k\geq 0$ and $n_l>0$ for $l=1,\dots,k$ span 
$R'$, but can be linearly dependent for some values of $c$ and $\De_0$. The situation is analogous  for $R''$.
The representation theory of the Virasoro algebra is discussed in many references including \cite{KR}.

We will furthermore mostly assume\footnote{This assumption may be weakened by allowing 
vacuum vectors $|0\rangle$ in the distributional sense, as the example of Liouville theory shows \cite{T01}.}
that the Hilbert space $\CH$ contains a distinguished vector $|0\rangle$ of lowest energy
called vacuum. The vacuum vector $|0\rangle$ satisfies $L_k|0\rangle =0$, $k=0,\pm1$.

\subsection{Fields and correlation functions}

Conformal field theories form a special class of quantum field theories (QFTs). We will mostly work 
in the framework of Euclidean QFT on a two-dimensional cylinder 
with coordinates $w=\tau+\mathrm{i}\si$, 
$w^*=\tau-\mathrm{i}\si$, $\si\sim \si+2\pi$. 
A Euclidean QFT on the cylinder is characterised by its set of fields 
$\{\Phi_v^{^{\rm\sst cyl}}(w,\bw);v\in\CF\}$, with $v\in\CF$ being a label for elements of a basis for the set 
of fields, together with  the collection of its Schwinger functions
$\CZ_{\mathbf{v}}^{^{\rm cyl}}(\mathbf{w},{\mathbf{w}^*})=\big\langle\prod_{r=1}^n\Phi_{v_r}^{^{\rm cyl }}(w_r^{},w_r^*)\big\rangle$, using tuple notations $\mathbf{w}=(w_1,\dots,w_n)$, $\mathbf{v}=(v_1,\dots,v_n)$ etc..
We will furthermore mostly restrict attention to QFTs having only bosonic fields. Key properties 
of the Schwinger functions are their single-valuedness 
and permutation symmetry, the invariance under exchange
of all variables with indices $r$ and $s$, $r,s=1,\dots,n$.

Under certain conditions on the Schwinger functions, the most important being called
reflection positivity, one may reconstruct\footnote{This follows from a variant of the Osterwalder-Schrader 
reconstruction theorem. See \cite{FFK1} for an account taylored to two-dimensional CFT.} from the collection of all
Schwinger functions $\CZ_{\mathbf{v}}^{^{\rm cyl}}(\mathbf{w},{\mathbf{w}^*})$
a Hilbert space $\CH$ with vacuum vector $|0\rangle$ acted upon by a family of operators 
$\Phi_v^{^{\rm\sst cyl}}(w,\bw)$ such that the Schwinger 
functions $\CZ_{\mathbf{v}}^{^{\rm cyl}}(\mathbf{w},{\mathbf{w}^*})$ with time-ordered arguments
$\tau_n>\tau_{n-1}>\dots>\tau_1$ can be represented as 
vacuum expectation values $\langle 0|\Phi_{v_n}^{^{\rm\sst cyl }}(w_n,w_n^*)\dots
\Phi_{v_1}^{^{\rm\sst cyl} }(w_1,w_1^*)|0\rangle$.
We are using the notation
$\langle v_1|v_2\rangle$ for the scalar product of two vectors $v_1,v_2\in\CH$.
%

Conformal symmetry will allow us to relate  fields 
$\Phi^{^{\rm\sst cyl }}_v(w,w^*)$ on the Euclidean cylinder 
to fields $\Phi^{}_v(z,\bz)$ on the complex plane $\BC$ having coordinates $z=e^w$.
In any ``reasonable'' CFT there should exist an analytic continuation of 
$\CZ_{\mathbf{v}}^{}(\mathbf{z},{\mathbf{z}^*})=\big\langle\prod_{r=1}^n\Phi_{v_r}^{}(z_r^{},z_r^*)\big\rangle$
to a multivalued analytic function 
$\CZ_{\mathbf{v}}^{}(\mathbf{z},\bar{\mathbf{z}})$ 
of independent variables $z_r$, $\bar{z}_r$, $r=1,\dots,n$, having an
expansion in 
$z_r/z_{r+1}$ and 
$\bz_r/\bz_{r+1}$ which is convergent for 
$|z_r/z_{r+1}|<1$ and
$|\bz_r/\bz_{r+1}|<1$, $r=1,\dots n$. The functions $\CZ_{\mathbf{v}}^{}(\mathbf{z},\bar{\mathbf{z}})$ 
defined in this way should have an analytic continuation to multivalued analytic 
functions on $\CC_n\times\CC_n$, 
$\CC_n=\BC^n\setminus\{z_r=z_s\,;\,r\neq s\,;\,r,s=1,\dots,n\}$ which are
single-valued when restricted to the so-called Euclidean domain $\bar{z}_r=z_r^*$,
$r=1,\dots,n$, with $z_r^*$ being the complex conjugate of $z_r$.

A further basic feature of conformal field theories is the state-operator
correspondence, an isomorphism: $v\mapsto \Phi^{}_v$ between $\CH$ and the space of fields,
which may be formulated in terms of the fields $\Phi^{}_v(z,\bz)$ on the complex plane 
conveniently as 
\begin{equation}
\lim_{z,\bz\ra 0}\Phi^{}_v(z,\bz)\,|0\rangle\,=\,v\in\CH\,.
\end{equation}
The special fields
\begin{equation}
T(z)\equiv\Phi_{L_{-2}|0\rangle}(z,\bz)=\sum_{k\in\BZ}L_n \,z^{-k-2}\,,\quad
\bar{T}(z)\equiv\Phi_{\bar{L}_{-2}|0\rangle}(z,\bz)=\sum_{k\in\BZ}\bar{L}_n \,z^{-k-2}\,,
\end{equation}
represent the only non-vanishing components of the energy-momentum tensor.
The translations $L_{-1}$ of the complex plane are realised on all fields $\Phi^{}_v(z,\bz)$
in a particularly simple way
\begin{equation}
[\,L_{-1}\,,\,\Phi^{}_v(z,\bz)\,]\,=\,\pa_z\Phi^{}_v(z,\bz)\,.
\end{equation}
The conditions on the Schwinger functions formulated above are necessary for getting a physically reasonable 
CFT. The bootstrap program discussed below is designed to construct examples satisfying these requirements.
Our next step will be to formulate the conformal invariance conditions in terms
of the Schwinger functions.

\subsection{Conformal Ward identities}

On the level of correlation functions one may express the conformal symmetry 
of a CFT using the conformal Ward identities \cite{BPZ}
\begin{align}\label{CWI}
0&=\bigg\langle\; \int_{\CC_t}dy\;\eta(y)T(y)\; \prod_{r=1}^n\Phi_{v_r}(z_r,\bz_r)\;\bigg\rangle\\
&=\;\sum_{r=1}^n\Bigg\langle\;\Phi_{v_n}(z_n,\bz_n)\cdots\bigg(
\int_{\CC_{z_r}}dy\;\eta(y)T(y)\Phi_{v_r}(z_r,\bz_r)\bigg)
\cdots \Phi_{v_r}(z_1,\bz_1)\Bigg\rangle\,,
\notag\end{align}
where  $\CC_t$ is a small circle on $\BP^1$ not encircling any element of $\{z_1,\dots,z_r\}$ 
which can be deformed  into 
the sum of circles  $\CC_r$ encircling only $z_r$ for $r=1,\dots,n$.
The identity \rf{CWI} is required to hold
for all meromorphic functions $\eta(y)$ on $\BC$ that are allowed to have poles of arbitrary order
only at $z_r$, $r=1,\dots,n$, and behave for $y\ra\infty$ as $\eta(y)=\CO(y^2)$.
Such functions $\eta(y)$ are in one-to-one correspondence with vector fields 
$\eta(y)\pa_y$ on $\BP^1\setminus\{z_1,\dots,z_n\}$.
\begin{quote}
{\small
{\bf Exercise 1.}
Derive \rf{CWI} using the analytic 
properties of the Schwinger functions mentioned above.
}\end{quote}

In order to discuss the consequences of the conformal Ward identities \rf{CWI}, 
let us note the following simple operator product expansion
\begin{equation}
T(y)\Phi_v(z,\bz)\,=\,\sum_{k\in\BZ}(y-z)^{k-2}\Phi_{L_{-k}v}(z,\bz)\,,
\end{equation}
following from the state-operator correspondence.
By considering functions $\eta(y)$ in \rf{CWI} having poles only at one point $z_r$ it 
is easy to see that \rf{CWI} allows us to rewrite the action of $L_{-k}$ on $v_r$ 
in terms of the action of $L_{m}$, $m\geq -1$ on $v_s$, $s\neq r$.

Equations \rf{CWI} are a linear system of equations relating the 
different correlation functions of a CFT. This suggests to look for the 
most general solution to these equations, and to construct the correlation functions 
as an expansion over a basis for the space of solutions to \rf{CWI}.
Let us note that \rf{CWI} only involves the copy of  the Virasoro algebra having 
generators $L_n$. There
is of course a similar identity for the other copy with generators $\bar{L}_n$.
This suggests that $\CZ_{\mathbf{v}}(\mathbf{z},\bar{\mathbf{z}})$ can be expanded as
\begin{equation}\label{holofact}
\CZ_{\mathbf{v}}(\mathbf{z},\bar{\mathbf{z}})\,=\,\sum_{\be',{\be}''}C_{\be'\be''\bm}^{}\CF_{\be',\bv'}^{}(\mathbf{z})
\CF_{\be'',\bv''}^{}(\bar{\mathbf{z}})\,,
\end{equation}
assuming that $\bv=(v_1,\dots,v_n)$ with $v_r=m_r\ot v'_r\ot v''_r\in\CH$, and that
$\{\CF_{\be,\bv}^{}(z);\be\in\CI\}$ is a basis for the space of solutions to the 
conformal Ward identities \rf{CWI}. In order to realise these
ideas more precisely let us clarify the mathematical 
meaning of the conformal Ward identities.

\subsection{A mathematical reformulation}

Let us consider the Riemann surface $C\equiv C_{0,n}=\BP^1\setminus\{z_1,\dots,z_n\}$, and
let $\CR=\bigotimes_{r=1}^nR_r$ be the tensor product of representations $R_r$
associated to the points $z_r$, respectively. 
We may define an action of the Lie-algebra $\mathsf{Vect}(C)$ of vector fields on $C$
by setting 
\begin{equation}
T_\xi:=\sum_{r=1}^n\sum_{k\in\BZ}\xi_k^{(r)}L_k^{(r)}\,,
\end{equation}
where $\xi_k^{(r)}$ are the Laurent expansion coefficients 
of the vector field $\xi\in \mathsf{Vect}(C)$, $\xi=\xi(y)\pa_y$ around $z_r$, defined by 
\begin{equation}
\xi(y)=\sum_{k\in\BZ}\xi_k^{(r)} (y-z_r)^{k+1}\,,
\end{equation}
and $L_k^{(r)}$ acts only on the r-th tensor factor of $\CR$, 
\begin{equation}
L_k^{(r)}\,=\,{\rm id}\otimes\dots\otimes{\rm id}\otimes\underset{\text{r-th}}{L_k}\otimes{\rm id}
\otimes\dots\otimes{\rm id}.
\end{equation}
Using these notations we will define conformal blocks as linear maps $f_C:\CR\ra \BC$ 
satisfying
\begin{equation}\label{CWIm}
f_C(T_\xi v)\,=\,0\,,\qquad\forall\;\,\xi\in\mathsf{Vect}(C)\,,\;\;
\forall\;\,v\in \CR\,.
\end{equation}
The set of linear equations \rf{CWIm} defines a subspace $\mathrm{CB}(C,\CR)$
in the dual  $\CR'$ of the vector space
$\CR$. The vector space $\mathrm{CB}(C,\CR)$ is  called
the space of conformal blocks. 

 It is not hard to verify that the Ward identities \rf{CWI} will hold provided that
the functions $\CF_{\be,\bv}^{}(z)$ appearing in \rf{holofact} are related to  conformal blocks 
$f^\be_C$ satisfying \rf{CWIm} as
\begin{equation}\label{blocksM/P}
f^\be_C(v):=\CF_{\be,\bv}^{}(z),\;\,\text{where}\;\, v=v_1\ot\dots\ot v_n\in\CR \;\,\text{if} \;\,\bv=(v_1,\dots,v_n).
\end{equation}
The definition of conformal blocks via \rf{CWIm}
is the formulation of the conformal Ward identities that has become
customary in the mathematical literature, see e.g. \cite{FB} and
references therein.

\begin{quote}
{\small
{\bf Exercise 2.}
\begin{itemize}
\item[a)] Verify in detail that \rf{holofact},  \rf{CWIm} and \rf {blocksM/P} imply \rf{CWI}.
\item[b)] Given a conformal block $\hat{f}_{C_{0,n+1}}\in \mathrm{CB}(C_{0,n+1},V_0\otimes \CR)$
show that the definition 
\begin{equation}
f_{C_{0,n}}(v_n\otimes\dots\otimes v_1):=
\hat{f}_{C_{0,n+1}}(e_0\otimes v_n\otimes\dots\otimes v_1)\,.
\end{equation}
yields a conformal block 
$f_{C_{0,n}}\in \mathrm{CB}(C_{0,n},\otimes \CR)$. Use this to conclude
that there exists an isomorphism 
$\mathrm{CB}(C_{0,n+1},V_0\otimes \CR)\simeq\mathrm{CB}(C_{0,n},\otimes \CR)$.
This isomorphism is often referred to as {\it propagation of vacua}.
\end{itemize}
}\end{quote}

Let us try to get a somewhat more concrete idea how the spaces of conformal blocks look like.
First note that by using the vector fields $\xi=(y-z_r)^{1-k}\pa_y$, $k=2,3,\dots,$
one may express the values of $f_C$ on arbitrary vectors $v$ in terms of 
its values on vectors $w$ of the form $w=\bigotimes_{r=1}^n (L_{-1})^{l_r} e_r$,
where the vectors $e_r$ satisfy the highest weight 
property $L_ne_r=\de_{n,0}\De_r e_r$, $n\geq 0$. One may next use the vector fields 
$\xi=y^k\pa_y$, $k=0,1,2$, to  express the values of $f_C$ on 
arbitrary vectors $v$ in terms of 
its values on vectors $w_\mathbf{n}$ of the form $w_\mathbf{n}=e_1\otimes e_2
\otimes e_3\otimes\bigotimes_{r=4}^n (L_{-1})^{\nu_r} e_r$,
$\mathbf{n}=(\nu_4,\dots,\nu_n)\in\BZ_{\geq 0}^{n-3}$. This means that a
conformal block $f_C$ is completely  characterised by the infinite collection 
of complex numbers $f_C(w_{\mathbf{n}})$, $\mathbf{n}\in\BZ_{\geq 0}^{n-3}$, 
which may be reformulated
as the statement that $\mathrm{CB}(C,\CR)\simeq (\CR_{-1})'$, where $(\CR_{-1})'$ is the dual of 
$\CR_{-1}=\bigotimes_{r=4}^n R_{-1,r}$, 
with $R_{-1,r}=\mathrm{Span}\{L_{-1}^\nu e_r;
\nu\in\BZ_{\geq 0}\}$. Note, in particular, that for $n=3$ it follows 
that the space of conformal blocks is at most one-dimensional. For $n\geq 4$ one finds an 
infinite-dimensional space of conformal blocks, in general.

It should be noted however, that the dimension of the space of conformal blocks may depend
sensitively on the choices of representations $R_r$ in $\CR=\bigotimes_{r=1}^N R_r$. 
For special choices of   the representations $R_r$ one may get a finite-dimensional space 
$\mathrm{CB}(C,\CR)$ of conformal blocks.

\begin{quote}
{\small
{\bf Example 1:}\\[1ex]
As an example let us consider the 
case  $n=4$, $R_r=V_{\al_r}$, where 
$V_{\al}$ is the irreducible highest weight representation of $\mathsf{Vir}_c$
generated from a highest weight vector $e_{\al}$ satisfying 
$L_ne_{\al}=\de_{n,0}\al(Q-\al)e_\al$, $n\geq 0$ if $c=1+6Q^2$. 
If $Q$ is parameterised as $Q=b+b^{-1}$, and 
$\al_2$ is chosen to be equal to $-b/2$,
one finds that there is the relation $(b^2 L_{-2}^{}+L_{-1}^2)e_{\al_2}=0$ in 
$V_{\al_2}$. \\[1ex]
{\bf Exercise 3:} 
Demonstrate that this relation, combined with the conformal Ward
identities \rf{CWIm} implies that the space of conformal blocks is two-dimensional in this
case.
}\end{quote}

Conformal field theories where all spaces of conformal blocks are finite-dimensional 
are called rational conformal field theories. Such CFT's are technically easier to study, but 
represent only a small subclass of the CFT's of interest for theoretical physics and
mathematics.

\subsection{Variations of insertion points}\label{FS:sec0}

The  definition of the conformal blocks is not quite complete yet. 

\subsubsection{Completing the definition of conformal blocks}

We shall complete the definition of the conformal blocks by adding the requirement that
\begin{equation}\label{FSconn}
f_C^{}(L_{-1}^{(r)}v)\,=\,\pa_{z_r}f_C^{}(v)\,.
\end{equation}
This is necessary to get
\begin{equation}\label{L-1}
\pa_z\Phi_v(z,\bz)\,=\,[\,L_{-1}\,,\Phi_v(z,\bz)\,]\,=\,\Phi_{L_{-1}v}(z,\bz)\,.
\end{equation}
The second equality in \rf{L-1} is a consequence of the state-operator correspondence 
together with $L_{-1}|0\rangle=0$.
Consistency with \rf{holofact} requires that we adopt \rf{FSconn}.

Note that the operation mapping a conformal block
$f_C^{}$ to the conformal block taking values $f_C^{}(L_{-1}^{(r)}v)$ 
on vectors $v\in \CR$ is a linear operator on $\mathrm{CB}(C,\CR)$. One may therefore
read \rf{FSconn} as the definition of a flat connection on the bundle of conformal blocks
over $\CM_{0,n}$.

\subsubsection{Consequences}

Let us 
introduce the {\it chiral partition function} $\CZ_f(C)$ as the value $f_C(e)$ of 
$f_C$ on $e=\bigotimes_{r=1}^n e_r$, the product of highest weight vectors.
Keeping in mind
that the value of the conformal blocks $f_C$ can be expressed in terms of the 
values $f_C$ has on vectors $w$ of the form $w=\bigotimes_{r=1}^n (L_{-1})^{l_r} e_r$ one sees that
the conformal Ward identities supplemented by
the definition \rf{FSconn} allow us to express the values $f_C(v)$ on arbitrary $v\in \CR$ 
as multiple derivatives of the chiral partition 
functions $\CZ_f(C)$.

\begin{quote}
{\small
{\bf Exercise 4:}\\[1ex] Show that for 
$C_{0,n}=\BP^1\setminus\{z_n,\dots,z_1\}$, 
$C_{0,n+1}=C_{0,n}\setminus\{y\}$, we have
\begin{equation}\label{CWI-T}
f_{C_{0,n+1}}^{}(L_{-2}e_0\otimes e_n\otimes\dots\otimes e_1)
=\sum_{r=1}^n \left(\frac{\De_r}{(y-z_r)^2}+\frac{1}{y-z_r}\pa_r\right)
\CZ^\be(z)\,.
\end{equation}
The resulting Ward identities for non-chiral correlation functions can be
written in the form
\begin{equation}
\bigg\langle\;T(y)\prod_{r=1}^n\Phi_{v_r}(z_r,\bz_r)\;\bigg\rangle
=\sum_{r=1}^n \left(\frac{\De_r}{(y-z_r)^2}+\frac{1}{y-z_r}\pa_r\right)
\bigg\langle\;\prod_{r=1}^n\Phi_{v_r}(z_r,\bz_r)\;\bigg\rangle\,,
\end{equation}
that one often finds in the literature.
}\end{quote}

Whenever the space of conformal blocks is finite-dimensional due to special properties of the
representations involved, one gets systems of
differential equations, as the following example illustrates.
\begin{quote}
{\small
{\bf Example 1 (ctd.):}\\[1ex]
As an example let us return to the 
case  $n=4$, $R_r=V_{\al_r}$, where 
$V_{\al}$ is the irreducible highest weight representation of $\mathsf{Vir}_c$
generated from a highest weight vector $e_{\al}$ satisfying 
$L_ne_{\al}=\de_{n,0}\al(Q-\al)e_\al$, $n\geq 0$ if $c=1+6Q^2$. 
For  $Q=b+b^{-1}$, and 
$\al_2=-b/2$
we had previously noted the relation $(b^2 L_{-2}^{}+L_{-1}^2)e_{\al_2}=0$ in 
$V_{\al_2}$. We may, without loss of generality, assume that 
$(z_1,z_2,z_3,z_4)=(0,z,1,\infty)$ and denote $C_z=\mathbb{P}^1\setminus\{0,z,1,\infty\}$. 
Use the 
conformal Ward identities to show that the chiral partition functions 
$\CZ_f(z)\equiv \CZ_f(C_z)$ satisfy the differential equation $\CD_{\rm\sst BPZ}\CZ_f(z)=0$, where
\begin{equation}
\CD_{\rm\sst BPZ}\,=\,\frac{1}{b^2}\frac{d^2}{dz^2}+\frac{2z-1}{z(1-z)}\frac{d}{dz}
+\frac{\De_1}{z^2}+\frac{\De_3}{(1-z)^2}+\frac{\kappa}{z(1-z)}\,,
\end{equation}
using the notations $\kappa=\De_1+\De_2+\De_3-\De_4$ and 
$\De_r=\al_r(Q-\al_r)$ for $r=1,\dots,4$.  Show that the equation $\CD_{\rm\sst BPZ}\CZ_f(z)=0$
can be reduced  to the hypergeometric differential equation $z(1-z)F''+[c-(a+b+1)z]F'-ab F=0$,  and
that the  two-dimensional space of solutions can be identified with the 
space of conformal blocks in this case.
}\end{quote}
We shall later use this example further.

\subsubsection{Geometric meaning}\label{geommean} 

To reformulate this observation in a more geometric way, let us introduce 
the space
\begin{equation}
\CM_{0,n}=\{(z_1,\dots,z_n)\in (\BP^1)^n;z_r\neq z_s,\;\,r,s=1,\dots,n\}/\mathrm{PSL}(2,\BR)\,,
\end{equation}
where elements of the group $\mathrm{PSL}(2,\BR)$ act via M\"obius transformations
$z_r\ra \frac{az_r+b}{cz_r+d}$.
The space $\CM_{0,n}$ can be identified with the moduli space of Riemann surfaces 
$C_{0,n}$
of genus $0$ and $n$ marked points.
 The universal cover $\widetilde{\CM}_{0,n}$ of the space $\CM_{0,n}$ 
can be identified with the Teichm\"uller space 
$\CT_{0,n}$, the space of deformations of the complex structures
on $C_{0,n}$.

One may observe that the values $f_C(w_\mathbf{n})$
characterising a conformal block $f_C$ can be identified with the 
Taylor expansion coefficients of $\CZ_f(z)$ around the values $z=(z_1,\dots,z_n)$
defining the given Riemann surface $C$. Considering the case $n=4$ as an example,
let us recall that the conformal blocks $f$ are in this case  fully characterised by
the infinite collection of complex numbers $\CZ_k=f(e_4\otimes e_3\otimes L_{-1}^k e_2\otimes e_1)$.
We may represent $C_{0,4}$ as $C_z\simeq\BP^1\setminus\{\infty, 1, z,0\}$, allowing us to
identify the parameter $z$ in this description as a coordinate for $M_{0,4}$.
The definition \rf{FSconn} relates the numbers $\CZ_k$ to the 
derivatives $\pa_z^k \CZ_f(C_z)$.  {\it Any} locally defined function 
$\CZ(z)$ on $\CT_{0,4}$ defines a conformal block in this way. This observation is easily generalised to $n>4$.

\begin{rem}
The converse to this statement is not true, in general. Given the collection of complex numbers $f_k$
characterising a conformal block on $C_{0,4}$ one gets a function $Z_f$ defined in a neighbourhood 
of the point with coordinate $z$ in $M_{0,4}$ only if the Taylor series 
$\sum_{k=0}^{\infty}t^k\CZ_k/k!$ converges, which will not be the case for arbitrary solutions of the 
conformal Ward identities.  However, for physical applications one will {\it not} 
be interested in the most general solution of the conformal Ward identities in the purely
algebraic sense, but rather in solutions for which the  corresponding chiral partition 
functions can be analytically continued over $\CT_{0,n}$.
The space of all such ``well-behaved'' conformal blocks generates a subspace
of the space of all algebraic solutions to the conformal Ward identities \rf{CWIm}.
\end{rem}

\subsection{Conformal blocks versus vertex operators}\label{CVO:sec}

Considering the special case $(z_3,z_2,z_1)=(\infty,z,0)$, $\CR= R_3\ot R_2\ot R_1$, one may use
the conformal blocks $f_{C_{0,3}}$ to define families of operators
$V_\rho(v_2,z):R_1\ra R_3$ labelled by vectors $v_2\in R_2$ and triples 
$\rho=\big[{}_{R_3}{}^{R_2}{}_{R_1}\big]$ of representations
such that 
\begin{equation}\label{CVOdef}
f_{C_{0,3}}(v_3\ot v_2 \ot v_1)\,=\,\langle\,v_3\,,\,V_\rho(v_2,z)\,v_1\,\rangle_{R_3}^{}\,.
\end{equation}
The 
operators
$V_\rho(v_2,z)$ defined in this way are called {\it chiral vertex operators}. When this becomes 
relevant we will make the dependence on the  triple  of representations $\rho$ involved 
more explicit, writing $V\big[{}_{R_3}{}^{R_2}{}_{R_1}\big](v_2,z)$ instead of 
$V_\rho(v_2,z)$.

\begin{quote}
{\small
{\bf Exercise 5:} 
\begin{itemize}
\item[a)] Demonstrate that the definition \rf{CVOdef} together with  the conformal Ward identities
imply the following commutation relations
\begin{equation}\label{CVOcomm}
L_n V_{\rho}(v_2|z)-V_{\rho}(v_2|z)L_n\,=\,
\sum_{k=-1}^{\infty}\bigg(\begin{matrix} n+1 \\ k+1 \end{matrix}\bigg) z^{n-k}\,V_{\rho}(L_{k}v_2|z)\,.
\end{equation}
\item[b)] Demonstrate that the vertex operators $V_{\rho}(v|z)$ are uniquely determined by the relations
\rf{CVOcomm} up to multiplication by a constant that may depend on $\rho$ and $v\in R_2$.
\item[c)] Use the conformal Ward identities to demonstrate that the vertex operators $V_{\rho}(v|z)$ 
furthermore satisfy the relations 
\begin{subequations}\label{descendants}
\begin{align}
&V_{\rho}(L_{-2}v|z)=T_< (z)V_{\rho}(v|z)+V_{\rho}(v|z)T_>(z)\,,\\
& V_{\rho}(L_{-1}v|z)=\pa_zV_{\rho}(v|z)\,,
\end{align}
\end{subequations}
where $T_<(z)$ and $T_>(z)$ are defined as
\begin{align}
T_<(z)=\sum_{n\leq -2}L_nz^{-n-2}\,,\qquad T_>(z)=\sum_{n> -2}L_nz^{-n-2}\,.
\end{align}
Relations \rf{descendants} allow
us to express $V_{\rho}(v|z)$, $v\in R_2$ in terms of $V_{\rho}(z)\equiv V_{\rho}(e_2|z)$,
where $e_2$ is the highest weight vector of the representation $R_2$, 
It follows that the chiral vertex operators $V_{\rho}(e|z)$ are uniquely defined by 
\rf{CVOcomm} and \rf{descendants} up to a constant that may depend on $\rho$. 
\item[d)]  Verify that
any vertex operator satisfying \rf{CVOcomm} and \rf{descendants} 
defines a conformal block via 
\rf{CVOdef}.
\end{itemize}
}
\end{quote}

In the special case where $v=e_2$, the highest weight vector of the representation $R_2$, 
one calls the vertex operator $V_{\rho}(z)\equiv V_{\rho}(e_2|z)$ a chiral {\it primary field}. The 
relations \rf{CVOcomm} simplify to 
\begin{equation}\label{CVOprim}
L_n V_{\rho}(z)-V_{\rho}(z)L_n\,=\,
z^n( z\pa_z+\De_{R_2}(n+1))V_{\rho}(z)\,,
\end{equation}
using \rf{FSconn} and $L_0e_2=\De_{R_2}e_2$.

The three-point functions of the physical vertex operators $\Phi_v(z,\bz)$ satisfy the conformal Ward identities.
It follows that $\Phi_v(z,\bz)$ can be decomposed as a sum over products of chiral vertex operators
for the representations of the Virasoro algebras generated by $L_n$ and $\bar{L}_n$, respectively.
\begin{quote}
{\small
{\bf Exercise 6:} 
Assume that the Hilbert space $\CH$ has the form 
\begin{equation}
\CH=\bigoplus_{R\in\CO}\CH_R\,,\qquad \CH_R=M_R\otimes R\,,
\end{equation}
where $\CO$ is a set of unitary highest weight representations of the Virasoro algebra
containing each representation $R$ only once, and $M_R$ is a multiplicity space
on which the Virasoro  generators $L_n$, $n\in\BZ$ act trivially (the $\bar{L}_n$ may 
act nontrivially on $M_R$). Demonstrate the 
``conformal Wigner-Eckart theorem'': Let $\rho=\big[{}_{R_3}{}^{R_2}{}_{R_1}\big]$
and $v=\mu_2\otimes v_2\in\CH_{R_2}$.
The operator $\Phi_v(z,\bz)$ then admits a decomposition
into CVO's of the following form:
\begin{equation}
\Pi_{R_3}\cdot\Phi_{v}(z,\bz)\cdot\Pi_{R_1} = \Xi_{R_3,R_1}^{\mu_2}\otimes V_{\rho}(v_2,z)\,,
\end{equation}
where $\Pi_R$ is the orthogonal projection 
onto $\CH_R$ and 
the operator $ \Xi_{R_3,R_1}^{\mu_2}:M_{R_1}\ra M_{R_3}$ is $z$-independent (but will depend on $\bar{z}$, in 
general).
}
\end{quote}

Another point of view is sometimes helpful. The commutation relations \rf{CVOcomm}  suggest to 
define an action of Virasoro algebra on a tensor product $R_2\otimes R_1$ of representations
by setting
\begin{equation}\label{co-product}
L_n (v_2\otimes v_1 )= v_2\otimes (L_nv_1)+\sum_{k=-1}^{\infty}\bigg(\begin{matrix} n+1 \\ k+1 \end{matrix}\bigg) z^{n-k}(L_k v_2)\otimes v_1
\,.
\end{equation}
We will denote the Virasoro-module defined using \rf{co-product} as $R_2\boxtimes R_1$. The  
chiral vertex operators $V_{\rho}(v_2|z)$ are thereby identified as close
relatives of the Clebsch-Gordan maps 
$C_z[{R_3}| R_2,R_1]$
intertwining the Virasoro module $R_2\boxtimes R_1$ defined using \rf{co-product} with
the standard action on $R_3$.
Note that we have chosen
not to include the dependence on $z$, manifest in \rf{co-product}, in the notation $R_2\boxtimes R_1$,
as this dependence will not be of interest whenever we will use this point of view, and since it
could be restored quite easily when needed.

This point of view puts conformal field theory into a useful analogy to group representation theory, as first 
emphasised in \cite{MS,FFK2}. Similar constructions can be introduced for extensions of the conformal
symmetry like current algebras. They have been used to exhibit profound relations between 
conformal field theory and quantum group representation theory \cite{KL}.

\subsection{Localising conformal blocks$^+$}\label{OneptLoc}

It is interesting and sometimes useful to use an alternative representation in which
conformal blocks in $\mathrm{CB}(C,\CR)$ get represented by elements of the dual $V_0'$ to the vacuum 
representation $V_0$ characterised by a modified invariance condition. 
This means that all information on the conformal blocks can be encoded in a vector in $V_0'$. 

To see this, let us first recall from Exercise 2b) that
the conformal Ward identities allow us to represent conformal blocks with insertions 
of the vacuum representation in terms of the conformal blocks without such insertions. 
Let us consider $\mathrm{CB}(C_{0,n+1},\CR\ot V_0)$, 
with $C_{0,n+1}=C_{0,n}\setminus\{z_0\}$ and 
vacuum representation $V_0$ associated to $z_0$. We had previously seen that the 
values of $f_{C_{0,n+1}}$ on arbitrary vectors can be expressed in terms of the
values $f_{C_{0,n+1}}(L_{-1}^{k_n}e_n\ot\dots \ot L_{-1}^{k_1}e_1\ot v)$, with $v\in V_0$ and 
$k_r\in\BZ_{\geq 0}$, $r=1,\dots,n$.
Using the {vector fields} 
\begin{equation}\label{vrdef}
\xi_r(y)\frac{\pa}{\pa y}=\left(\frac{y-z_0}{z_r-z_0}\right)^{3-n}
\prod_{\substack{s=1 \\ s\neq r}}^n\frac{y-z_s}{z_r-z_s}\frac{\pa}{\pa y}\,,
\end{equation}
one may compute $f_{C_{0,n+1}}(L_{-1}^{k_n}e_n\ot\dots \ot L_{-1}^{k_1}e_1\ot v)$ in terms of 
$f_{C_{0,n+1}}(e_n\ot\dots \ot e_1\ot v')$ for some $v'\in V_0$.  The functional 
$g:V_0\ra \BC$ defined by 
\begin{equation}\label{onept}
g(v):=f_{C_{0,n+1}}(e_n\ot\dots \ot e_1\ot v)\qquad\forall v\in V_0,
\end{equation}
satisfies a variant of the conformal Ward identities of the form 
\begin{equation}\label{modCWI}
g(T_\xi v)+\sum_{r=1}^n \xi_0^{(r)}\De_r g(v)\,=\,0\,,
\end{equation}
for all vector fields $\xi\in\mathsf{Vect}(\BP^1\!\setminus\!\{z_0\})$ which vanish at $z_r$ for all $r=1,\dots,n$. 
Using the observations above it is not hard to show that equations \rf{modCWI} 
define a subspace of the dual $V_0'$ of $V_0$ isomorphic to $\mathrm{CB}(C_{0,n+1},\CR\ot V_0)$,
which is furthermore isomorphic to $\mathrm{CB}(C_{0,n},\CR)$ by the propagation of vacua.

One may furthermore notice that the conformal Ward identities \rf{CWIm} specialised to the vector fields $\xi_r$
defined in \rf{vrdef} 
combined with \rf{FSconn}  imply that $g$ satisfies  identities of the form
\begin{equation}\label{modFSconn}
\pa_{z_r}g(v)=g(\mathsf{L}_{-1}^{(r)}v)\,,
\end{equation}
where $\mathsf{L}_{-1}^{(r)}$ are operators on $V_0$ satisfying $[\mathsf{L}_{-1}^{(r)},\mathsf{L}_{-1}^{(s)}]=0$ for all
$r,s=1,\dots,n$.

\begin{quote}
{\small
{\bf Exercise 7.}\\[-4.5ex]
\begin{itemize}
\item[a)] Let $t_g(x)\equiv g(T(x-z_0)e_0)$ 
be the expectation value of the energy-momentum tensor.
Given that $g$ satisfies \rf{modCWI} and \rf{modFSconn},
demonstrate that $t_g(x)$ extends to a function that is meromorphic on $C_{0,n}$ satisfying 
\begin{equation}
t_g(x)=\sum_{r=1}^n \left(\frac{\De_r}{(x-z_r)^2}+\frac{1}{x-z_r}\pa_r\right)g(e_0)\,.
\end{equation}
\item[b)] Let $t_f(x)=f(e_n\otimes\dots \otimes e_{2}\otimes T(x-z_1)e_1)$. Show that 
$t_f(x)=t_g(x)$ if $g$
is defined from a conformal block $f_{C_{0,n}}$ as above.
\end{itemize}
}\end{quote}

We conclude that there is a one-to-one correspondence between linear 
functionals $g:V_0\ra \BC$ satisfying \rf{modCWI} and conformal blocks 
$f_{C_{0,n}}\in\mathrm{CB}(C_{0,n},\CR)$. All information on the conformal block $f_{C_{0,n}}$ on 
the $n$-punctured sphere $C_{0,n}$ can be ``localised'' within the element $g\in V_0'$ 
assigned to 
an arbitrary point $z_0$ on $C_{0,n}$.  This type of representation for 
the conformal blocks will be referred to as one-point localisation. The operators $\mathsf{L}_{-1}^{(r)}$ 
appearing in \rf{modFSconn} realize an infinitesimal motion of the puncture at 
$z_r$ within the one-point localisation.

\subsection{Higher genus conformal blocks$^+$}\label{highergenus}

We will now briefly discuss how the conformal Ward identities can be generalised 
when $C_{0,n}$ is replaced by a Riemann surface $C\equiv C_{g,n}$ of higher 
genus $g$. For simplicity we will  here restrict  attention to the case where $n=1$ with 
vacuum representation $V_0$ inserted at a point $P\in C$.

\subsubsection{Conformal Ward identities}

Let $(C,P,x)$ 
be a Riemann surface $C$ with a marked point $P$, and a coordinate $x$ on a disc $D$ around $P$
such that $x(P)=0$. 
We may in this case define the conformal blocks as elements of the dual $V_0'$ of $V_0$ satisfying
\begin{equation}\label{CWIg>0}
f(T_\xi{v})\,=\,0\,,\quad \forall \xi\in\mathsf{Vect}(C\setminus P)\,,\quad
\forall \,{v}\in V_0\,,
\end{equation}
where $T_\xi$ is defined as 
\begin{equation}\label{Txidef}
T_\xi=\frac{1}{2\pi \mathrm{i}}\int _Cdx\;T(x)\xi(x)=
\sum_{n\in\BZ}\xi_nL_n, 
\end{equation}
if $\xi=\xi(x)\frac{\pa}{\pa x}\in\mathsf{Vect}(C\setminus P)$, and $\xi(x)=\sum_{n\in\BZ}\xi_n x^{n+1}$
is the Laurent expansion  around $x=0$. The space of conformal blocks,
defined as the solutions to \rf{CWIg>0}, may be finite-dimensional in some cases 
(minimal models, see below), but will be infinite-dimensional in general.

In order to understand the consequences of \rf{CWIg>0} more concretely, let us use the following 
consequence of the Riemann-Roch theorem, in this form proven in \cite[Appendix B]{AGMV}.
For generic\footnote{If $P$ is not a Weierstrass point, see \cite{AGMV}.} points $P$ on $C$ 
  there exist bases for $H^0(C\setminus P,K^2)$ and 
$H^0(C\setminus P,K^{-1})\equiv \mathsf{Vect}(C\setminus P)$, respectively,
generated by  elements having Laurent expansions around $x=0$ of the form
\begin{align}\label{qn-exp}
q_n(x)(dx)^2&=\Big(x^{n-2}+\sum_{m\geq h+2}Q_{nm}x^{m-2}\Big)(dx)^2\,,\qquad n\leq h+1\\
\xi_n(x)\frac{\pa}{\pa x}&= \Big(x^{n+1}+\sum_{m\geq -h-1}V_{nm}x^{m+1}\Big)\frac{\pa}{\pa x}\,,\qquad n\leq -h-2\,,
\end{align}
where $h=3g-3$.
The elements satisfy 
\begin{equation}\label{q-v-orth}
\int_{\CC}q_n \xi_m=0\,,\qquad
\forall\;m\leq -h-2, \quad \forall\;n\leq h+1,
\end{equation}
where $\CC$ is a small circle around $P$.

It then follows from \rf{CWIg>0} that we can express the values of $f(v)$ on arbitrary $v\in V_0$ in terms of the special 
values
$f_{\mathbf{n}}:=f(L_{-h-1}^{n_h}L_{-h}^{n_{h-1}}\cdots L_{-2}^{n_1}\,{e}_0)$,
where $\mathbf{n}=(n_1,\dots,n_h)$.

\subsubsection{Variations of the conformal blocks}

We are now going to show that variations of the defining data $(C,P,x)$ can be represented
using the Virasoro action on $V_0$. Let us consider general infinitesimal variations of 
a conformal block $f$ defined as 
\begin{equation}\label{canconn}
(\de_\zeta f)({v})\,=\,f(T_\zeta {v})\,,\qquad \zeta\in\mathsf{Vect}(A)\,,
\end{equation}
where $A$ is the annulus $D\setminus P$, and $T_\zeta$ is defined by 
replacing $\xi$ by $\zeta$ in \rf{Txidef}.  Our goal will be to identify conditions allowing 
us to interpret $g:=(1+\de_\zeta)f$ as a conformal block associated to an infinitesimal 
variation of the data $(C,P,x)$ defining $f$. To this aim let us note that 
\begin{equation}
[T_{\zeta},T_\xi]=-T_{[\zeta,\xi]}+\frac{c}{12}(\zeta,\xi)\,,\qquad
(\zeta,\xi):=\frac{1}{2\pi\mathrm{i}}\int_{\CC_P} dx \;\zeta'''(x)\xi(x)\,,
\end{equation}
leading to 
\begin{align}\label{varcalc}
&g(T_\xi{v})=f(T_\zeta T_\xi{v})=-f(T_{[\zeta,\xi]}{v})+\frac{c}{12}(\zeta,\xi)\,f({v})\,.
\end{align}
If we choose $\zeta$ such that
the bilinear form $(\xi,\zeta)$ vanishes for all $\xi\in\mathsf{Vect}(C\setminus P)$, 
we find that $g$ satisfies
\begin{equation}\label{CWIdef}
g(T_{\xi+[\zeta,\xi]}{v})\,=\,0+\CO(\zeta^2)\,,
\end{equation}
the conformal Ward identity for an infinitesimal variation of the vector fields $\xi$ generated by the
adjoint action of a vector field $\zeta$.

We may next notice that variations of the data $(C,P,x)$ will induce 
variations of the $\xi\in H^0(C\!\setminus\! P,K^{-1})$ 
appearing in \rf{CWIg>0} which can be represented by the adjoint action of
suitable vector fields $\zeta\in \mathsf{Vect}(A)$. 
Infinitesimal variations of $(P,x)$ can be represented by a vector field $\zeta$ that is 
holomorphic on $D$. In order to see that all variations of the complex structure of $C$
can be represented in this way one may use
the Virasoro uniformisation theorem 
describing the Teichm\"uller spaces  $\CT(C)$ in terms of 
vector fields on an annulus. This theorem states that
the Teichm\"uller space $\CT(C)$ can be represented as double
quotient
\begin{equation}\label{Viruni}
\mathcal{T}(C)\,=\, \mathsf{Vect}(C\!\setminus\! P)\setminus\mathsf{Vect}(A)\; /\; \mathsf{Vect}(D)\,.
\end{equation}
A proof can be found e.g. in \cite[Section 17]{FB} or \cite[Section 2.4]{Du}.

It remains to investigate the conditions for having $(\zeta,\xi)^{}=0$
for all $\xi\in\mathsf{Vect}(C\setminus P)$ in \rf{varcalc}. According to 
\rf{q-v-orth} this will be the case if  $\zeta'''(x)(dx)^2\in H^0(C\!\setminus\! P,K^2)$.
This condition is satisfied if $\zeta$ is an element of the subspace $\mathsf{V}_\CT$ of  $\mathsf{Vect}(A)$
spanned by the vector fields 
\begin{equation}
\zeta_n(x)\frac{\pa}{\pa x}= \bigg(\frac{x^{n+1}}{\nu_n}+\sum_{m\geq h+2}Q_{nm} \frac{x^{m+1}}{\nu_m}\bigg)
\frac{\pa}{\pa x}\,,\qquad  2\leq |n|\leq h+1\,.
\end{equation}
where $\nu_n=n(n^2-1)$ and the coefficients $Q_{nm}$ have been introduced in \rf{qn-exp}.
The subspace $\mathsf{V}_\CT$ is $6g-6$-dimensional and
carries
a non-degenerate symplectic form given by the restriction of the form $(.,.)$. The Virasoro uniformisation 
theorem identifies $\CT(C)$ as a quotient of $\mathsf{V}_\CT$ by the subspace 
generated by the $\zeta_n$ with $2\leq n\leq h+1$. This subspace is isomorphic to 
the space $H^0(C,K^2)$ of quadratic differentials on $C$, which is canonically
isomorphic to the cotangent fiber $T^*\CT(C)$ of $\CT(C)$. We thereby identify
$\mathsf{V}_\CT$ with the total space of the cotangent bundle $T^*\CT(C)$.

There is an interesting reformulation of the definitions above in terms of the chiral partition 
function $\CZ_f:=f(e_0)$, and the expectation value of the energy-momentum tensor
$\langle T(x)\rangle_f=f(T(x)e_0)/f(e_0)$. It can be shown \cite[Chapter 9.2]{FB} that the defining Ward
identity \rf{CWIg>0} is {\it equivalent}
to the condition that $t_f(x)\equiv \langle T(x)\rangle_f$ defines a holomorphic projective $c$-connection
on $C$, which means that the transformation of $t_f(x)$ from one patch on $C$ with coordinate $x$ to 
another patch with coordinate $y$ is represented as
\begin{equation}
{t}_f(x)=(y'(x))^2\,\tilde{t}_f(y(x))+\frac{c}{12}\{y,x\}\,,\quad \{y,x\}=
\frac{y'''}{y'}-\frac{3}{2}\left(\frac{y''}{y'}\right)^2\!.
\end{equation}
One may furthermore note that the definition of the canonical connection \rf{canconn} relates multiple
derivatives of the chiral partition function $\CZ_f$ with respect to the complex structure moduli of $C$
to the defining data of the conformal blocks, the values of $f$ on a sufficiently large collection
of vectors in $V_0$. First order derivatives of $\CZ_f$, in particular, are given by the expectation values
$\langle T_\zeta\rangle_f=\frac{1}{2\pi \mathrm{i}}\int _Cdx\,t_f(x)\zeta(x)$.


\subsubsection{Projective flatness}

We have seen that we may describe variations of the complex structure of $C$ in the form \rf{canconn}.
This defines a connection on the bundle of conformal blocks over $\CM(C)$, the 
moduli space of complex structures on $C$. The connection is not flat, but only projectively
flat, as the form $(\zeta_1,\zeta_2)$ will be non-vanishing for general elements $\zeta_1$, $\zeta_2$ of
$\mathsf{V}_\CT$. We are now going to discuss the implications of the projective flatness 
in a little more detail.

The variations $\de_\zeta$ may\footnote{Another condition for being
integrable is well-behavedness of the conformal blocks in the sense explained in Remark 1
at the end of Section \ref{geommean}, which will be assumed in the following.} be integrable 
if one restricts the choice of the vector fields 
$\zeta$ to Lagrangian subspaces $\mathsf{L}_\CT$ of $\mathsf{V}_\CT$ spanned by 
elements satisfying $(\zeta_1,\zeta_2)=0$ for all $\zeta_1,\zeta_2\in\mathsf{L}_\CT$.
It follows from the above that 
such Lagrangian subspaces are isomorphic to $\CT(C)$, but not canonically so. 
The definition of chiral partition functions $\CZ_f$ by means of integration of the 
canonical connection therefore
depends on the choice 
of a Lagrangian subspace in $\mathsf{V}_\CT$. One should note that the 
resulting ambiguity affects the chiral partition functions of all conformal blocks
in the same way. Modifying the choice of a Lagrangian subspace will modify the 
chiral partition functions of all conformal blocks 
by multiplication with the same locally defined function.

Having chosen families of Lagrangians in $\mathsf{V}_\CT$ varying holomorphically
over open subsets $M\subset\CM(C)$  may allow us to 
define chiral partition functions $\CZ_{f,M}$  on $M$ by
integrating the parallel transport defined using \rf{canconn}. 
The Lagrangians used to define the parallel transport in two neighbourhoods
$M$ and $N$ within $\CM(C)$ may differ on the overlap $M\cap N$. Keeping in 
mind that we have $\de_\zeta\log \CZ_f=\langle T_\zeta\rangle_f$, and noting that a change of 
the choice of Lagrangians in $\mathsf{V}_\CT$ changes the expectation values $t_f(x)$ for all 
conformal blocks $f$ by 
addition of the same quadratic differential, 
it is easy to see that the partition functions $\CZ_{f,M}$ and $\CZ_{f,N}$ defined  in this way 
in two neighbourhoods $M$ and $N$ differ by
an overall factor, $\CZ_{f,M}^{}=\chi_{MN}^{}\CZ_{f,N}^{}$, 
with $\chi_{MN}^{}$ being a function on $M\cap N\subset \CM(C)$
independent of the choice of $f$. The function $\chi_{MN}$ is defined by the choices used to define 
$\CZ_{f,M}$ and $\CZ_{f,N}$ up to an overall multiplicative constant. 
On triple overlaps we can therefore only require a weakened form of the usual consistency 
condition $\chi_{MN}^{}\chi_{NO}^{}\chi_{OM}^{}=\eta_{MNO}^{}$, with $\eta_{MNO}^{}$ being 
a constant which is easily seen to be representable in the form 
$\eta_{MNO}^{}=e^{\pi \mathrm{i} c\, \nu_{MNO}^{}}$.
The collection of functions $\chi_{MN}^{}$  associated to a cover of $\CM(C)$ defines what is called a projective 
line bundle $\CE_c$ over $\CM(C)$ in \cite{FS}.

By integrating the canonical connection, one can locally define bases for the 
bundle of conformal blocks spanned by horizontal sections.
One may thereby define a holomorphic vector bundle $\CV_c$ over $\CM(C)$.
More canonically defined is the projective vector bundle $\CW_c$ over $\CM(C)$ obtained
from $\CV_c$ by taking the tensor product with the projective line bundle $\CE_c^{-1}$. 
This removes the ambiguities from the choices in the local integration of the canonical 
connection, the price to pay is that the  consistency conditions for transition functions
of $\CW_c$ on triple overlaps are satisfied 
only up to a constant multiple of the identity.

The discussion above, together with its continuation in 
Section \ref{chirmod} below, reproduce the key ingredients of the perspective on conformal field theory
proposed in \cite{FS}.
The ambiguity observed above in the definition of the horizontal sections can be
physically interpreted as a generically unavoidable dependence on the choice of a renormalisation 
scheme in the definition of the energy-momentum tensor on higher genus surfaces \cite{FS}.

\subsection{Extensions of conformal symmetry$^{++}$}\label{ExtVOA}

We will see that conformal symmetry alone provides too little information to 
solve conformal field theories completely, in general.  There are, however, 
several cases where the conformal symmetry is extended to a larger algebraic
structure called a vertex operator algebra (VOA), allowing us to obtain further information
or even a complete solution of more complicated CFTs. We shall not attempt to give
a complete treatment, but  mention a few examples, and try to  indicate the main idea
behind the  definition of VOAs.

Extensions of the conformal symmetry can be generated by the Laurent expansion coefficients
of a set of fields $W^{(i)}(z)$, 
\begin{equation}\label{Wmodes}
W^{(i)}(z)\,=\,\sum_{m\in\BZ}W_m^{(i)}z^{-m-\De^{(i)}}\,,
\end{equation}
with $i$ being an index labelling the different fields,
and the parameter ${\De}^{(i)}$ is called the
spin of ${W}^{(i)}(z)$. The set of fields 
contains the energy-momentum tensor $T(z)$, and it is assumed that
\begin{equation}
[\,L_n\,,\, W_m^{(i)}\,]\,=\,(n(\De^{(i)}-1)-m)W_{n+m}^{(i)}\,,
\end{equation}
for the modes of all fields $W^{(i)}(z)$ except $T(z)$.

\subsubsection{Examples:} 
\begin{itemize}
\item[0)] {\it Free boson algebra}: Generated by modes $a_n$, $n\in\BZ$ satisfying $[a_n,a_m]=\frac{n}{2}\de_{n+m,0}$.
Out of a representation of the free boson algebra one may construct a one-parameter family of representations
of the Virasoro algebra using
\begin{equation}\label{FeiFu}
\begin{aligned}
L_n&=\mathrm{i}(n+1)Qa_n+\sum_{k\in\mathbb{Z}}a_ka_{n-k},\qquad n\neq 0,\\
L_0&=a_0^2+\mathrm{i}Qa_0+2\sum_{k>0}a_{-k}a_k.
\end{aligned}
\end{equation}
The representations of the Virasoro algebra defined in this way will have central charge 
$c=1+6Q^2$. If $\CF_\al$ is a representation in which 
the central element $a_0$ is represented as multiplication by $-i\al$ times the
identity operator one gets a representation of the Virasoro algebra with highest weight $\De_\al=\al(Q-\al)$.
\item[1)] {\it Affine Lie-algebra}: Generated by fields $J^a(z)$ with spin $1$, having Laurent modes
with relations
\begin{equation}
[J_n^a,J_m^b]\,=\,\mathrm{i}\,f^{abc}J_{n+m}^c+\mathsf{k}\eta^{ab}n\de_{n,-m}\,,
\end{equation}
where $f^{abc}$ are the structure constants of the semi-simple Lie-algebra $\mathfrak{g}$
with generators $T^a$, relations $[T^a,T^b]=\mathrm{i}f^{abc}T^c$ and invariant
bilinear form $(T^a,T^b)=\eta^{ab}$. $\mathsf{k}$ is the central element. Fixing $\mathsf{k}$ to 
a value $k$ defines the affine Lie algebra $\hfg_k$. 
The Virasoro algebra
gets embedded into the universal enveloping algebra $\CU(\hfg_k)$ of the affine Lie-algebra by means 
of the Sugawara construction.
\item[2)]{\it W-algebras}: The algebras $\CW_N$, $N\geq 3$ are generated by the Laurent
modes of fields $W^{(i)}(z)$, $i=2,\dots,N$ with $W^{(2)}=T(z)$, having complicated commutation 
relations. In the case $N=3$ we have, for example \cite{Za}:
\begin{align}\label{W3rels}
[W_n,W_m]=&\frac{c}{3\cdot 5!}(n^2-1)(n^2-4)\de_{n,-m}+\frac{16}{22+5c}\Lambda_{n+m}\\
& (n-m)\bigg(\frac{1}{15}(n+m+2)(n+m+3)-\frac{1}{6}(n+2)(m+3)\bigg)L_{n+m}\,,
\notag\end{align}
where $\Lambda_n=\frac{1}{5}x_nL_n+\sum_{k\in\BZ}:L_{n-k}L_k:$, $x_{2l}=1-l^2$, $x_{2l+1}=(2+l)(1-l)$.
We see from the example \rf{W3rels} 
that the modes of the $\CW_3$-algebra do not generate a Lie algebra.
The algebras $\CW_N$ for $N>3$ are even more complicated, and therefore best defined using the quantum Drinfeld-Sokolov reduction,
see \cite{FB} and references therein.

\end{itemize}

\subsubsection{Vertex operator algebras$^{++}$}\label{sec:VOA}

A flexible framework for the description of extended chiral symmetries in CFT 
is provided the the notion of a vertex operator algebra (VOA). The 
concept of a VOA may be motivated by the state-operator correspondence:
Given an extended symmetry algebra, it is natural to consider the 
vector space $\CA$ generated its action on the vacuum vector $|0\rangle$. 
It is then natural to label the currents $V(a,z)$ of a VOA  by vectors $a\in\CA$.
In the axiomatic definition of VOA given in the mathematical literature 
(see e.g. \cite{Bo,FLM,FB} and references therein) one considers the currents $V(a,z)$
as formal power series\footnote{Note that the conventions for mode expansions in the VOA
 literature often differ from those used in \rf{Wmodes}.} 
\begin{equation}
V(a,z)\,=\,\sum_{n\in\BZ} \mathsf{a}_{n}z^{-n-1}\,,
\end{equation}
with coefficients $\mathsf{a}_{n}$ being linear operators on $\CA$. These data satisfy 
certain axioms, including the conditions $\lim_{z\ra 0}V(a,z)|0\rangle =a$,
$V(|0\rangle,z)=\mathrm{id}$ 
and $[L_{-1},V(a,z)]=\pa_zV(a,z)$. The most important axiom is the locality 
axiom stating that the two formal power series $V(a,z)V(b,w)$ and
$V(b,w)V(a,z)$ coincide after multiplying them with a large enough power 
of $z-w$. This may informally be thought of as the condition 
that the commutator $[V(a,z),V(b,w)]$ can be expanded into a finite sum of 
derivatives of the delta-distribution supported on $z=w$.

It is possible to show that the vector space spanned by the modes $\mathsf{a}_n$
has a natural Lie algebra structure \cite[Chapter 4]{FB}. There may, however, be
additional relations among the modes $\mathsf{a}_n$, expressing some of them 
as composites of others. An example are the relations
expressing $\Lambda_n$ in terms of the $L_k$ in the case of the $\CW_3$-algebra
defined  above.

A simple generalisation of the notion of a VOA is a super VOA. The simplest example of a
super VOA is generated from $N$ species of 
{\it free fermions}, generated by the modes of fields $\psi_s(z)$, $\bar{\psi}_s(z)$, $s=1,\dots,N$,
\begin{equation}
\psi_s(z)=\sum_{n\in\BZ}\psi_{s,n} z^{-n-1}\,,\quad
\bar\psi_s(z)=\sum_{n\in\BZ} \bar\psi_{s,n} z^{-n}\,,
\end{equation}
having modes satisfying the anti-commutation relations
\begin{equation}
\{\psi_{s,n},\bar\psi_{t,m}\}=\de_{s,t}\de_{n,=m}\,,\qquad
\{\psi_{s,n},\psi_{t,m}\}=0\,,\qquad\{\bar\psi_{s,n},\bar\psi_{t,m}\}=0\,.
\end{equation}

It is possible to generalise the notion of conformal blocks to more general VOA. This is 
rather straightforward in genus $0$. If a current $W^{(i)}(z)$ has dimension $\De^{(i)}$, one 
simply needs to replace the vector fields $\xi$ in \rf{CWIm} by holomorphic $(1-\De^{(i)})$-differentials
on $C=C_{0,n}$. The definition of conformal blocks general VOA in higher genus can be found
in \cite{FB}.

Connections between the theory of VOA and the operator algebraic approach to conformal field theory
have recently been described in \cite{CKLW,Ten}.

\section{Part 2: Bootstrap}
\setcounter{equation}{0}

We will now describe a construction of large classes of conformal blocks from simpler 
building blocks called gluing construction. The conformal blocks that can be constructed in 
this way will be argued to coincide with the conformal blocks of interest for physics:
These are the conformal blocks which can appear in  
factorised representations of the form \rf{holofact} for physical correlation functions. 
The problem to construct 
the correlation functions of a CFT is thereby disentangled into a kinematic part, the construction of 
the conformal blocks, solved completely by exploiting only the symmetry constraints, and the 
problem to assemble the conformal blocks into single-valued Euclidean correlation functions.
The coefficients $C_{\be'\be''\bm}^{}$ appearing in the expansion \rf{holofact} carry the main 
dynamical information on a CFT, and are therefore of central interest in physical applications.
We will identify general consistency constraints on these coefficients, and briefly discuss
a family of cases where explicit solutions to these constraints are known.

\subsection{Gluing construction}\label{gluesect}

Our next goal will be to describe a recursive construction of large families of conformal
blocks. For some cases it is known that the resulting families of conformal blocks 
generate bases for the relevant (sub-)spaces of the spaces of conformal blocks. 

The gluing construction we are going to present is based on a geometric operation 
producing an $n$-punctured sphere 
$C=\BP^1\setminus\{x_1,\dots,x_n\}$ by gluing two spheres $C_1$ and $C_2$ with smaller numbers
of punctures.
For having convenient notations let us split 
$\{x_1,\dots,x_n\}=I_1'\cup I_2\cup I_1''$, where $I_1'=\{x_1,\dots, x_l\}$, 
$I_2=\{x_{l+1},\dots,x_m\}$, and $I_1''=\{x_{m+1},\dots,x_n\}$. If $l=0$ we set $I_1'=\emptyset$,
and similarly $I_1''=\emptyset$ for $m=n$. We then consider the following two punctured spheres 
\begin{equation}
C_1\,=\,\BP^1\setminus(\{x_i,i\in I_1\}\cup\{x\})\,,\qquad
C_2\,=\,\BP^1\setminus(\{\infty\}\cup\{y_i, i\in I_2\})\,,
\end{equation}
with $I_1=I_1'\cup I_1''$, and the positions of the punctures $y_i$ for $i\in I_2$ are related to $x_i$ via 
\begin{equation}
y_i=q^{-1}(x_i-x)\,.
\end{equation}
We want to describe how the 
$n$-punctured sphere $C_{0,n}$ can be represented as the result of a gluing operation
applied to $C_1$ and $C_2$.

Let us cut (sufficiently small) discs out of $C_1$ and $C_2$ giving us the open surfaces
\begin{equation}
D_1^\rho\,=\,\{z\in C_1;|z-x|\geq\rho\}\,,
\qquad
D_2^\rho\,=\,\{z\in C_2;|z|\leq q^{-1}\rho\}\,.
\end{equation}
with $q$ being a parameter we'll play with below. 
We are assuming that $D_1^\rho$ contains $x_i$, $i\in I_1$, 
and that $D_2^\rho$ contains  $y_i$, $i\in I_2$.
After scaling the disc $D_2^{\rho}$ by a factor of $q$ one may
glue it into the hole we've  cut out of $C_1$ to get $D_1^\rho$.
The result of this operation is the Riemann surface 
\begin{equation}
C\,=\,
\BP^1\setminus\{x_n,\dots,x_{1}\}\,,\;\,\text{where}\;\,x_{i}=qy_i+x,\;\,\text{if}\;\,
i\in I_2.
\end{equation}

The following operation produces a conformal block associated to 
$C$ from any two given conformal blocks $f_{C_1}$ and  $f_{C_2}$ associated to 
$\big(C_1,\big(\bigotimes_{i\in I_1'}R_i\big)\otimes R\otimes \big(\bigotimes_{i\in I_1''}R_i\big)\big)$, 
where $I_1=I_1'\cup I_1''$, 
and $\big(C_2,R\otimes\bigotimes_{i\in I_2}R_i\big)$,  
respectively\footnote{We adopt the convention that for $I=\emptyset$ we 
let $R\otimes \big(\bigotimes_{i\in I}R_i\big)=R$
and $ \big(\bigotimes_{i\in I}R_i\big)\otimes R=R$.}.
Let us introduce a non-degenerate invariant bilinear form 
$\langle \,.\,,\,.\,\rangle_R$ on 
$R\otimes R$. Such
a bilinear form can be defined uniquely by the properties 
\begin{equation}
\langle \,L_{-n}v\,,\,w\,\rangle_R\,=\,\langle \,v\,,\,L_n w\,\rangle_R\,,\qquad
\langle \,e_R\,,\,e_R\,\rangle_R\,=\,1\,.
\end{equation}
We may then use $f_{C_2}$ to define a map $V_{C_2}:\bigotimes_{i\in I_2}R_i\ra R$
which satisfies
\begin{equation}
f_{C_2}(v\otimes w)\,=\,\langle\,v\,,\,V_{C_2}(w)\,\rangle_R^{}\,.
\end{equation}
The conformal block $f_C$ is now defined by setting
\begin{equation}\label{gluing}
f_{C}(w'\ot v\ot w''):=f_{C_1}\big(\, w'\ot q^{L_0}V_{C_2}(v)\ot w''\,\big)\,,
\end{equation}
for all $v\in \bigotimes_{i\in I_2}^{m}R_i$, 
$w'\in\bigotimes_{i\in I_1'} R_i$ and $w''\in\bigotimes_{i\in I_1''} R_i$
The right hand side of \rf{gluing} is a series in powers of $q$ which is believed\footnote{This is known 
to be true in several cases. A general proof has not been given yet.}  to be 
convergent.

By using the gluing construction recursively, one may build conformal blocks 
for any n-puntured Riemann sphere $C_{0,n}$ 
from the conformal blocks associated to $C_{0,3}$. We had noted already that 
we have $\mathrm{dim}(\mathrm{CB}(C_{0,3},\CR))\leq 1$ in the case of the Virasoro algebra.
The choices made in the gluing construction therefore consist of two types of data:
\begin{itemize}
\item The different ways of gluing  $C_{0,n}$ from three-punctured spheres are called {\it pants
decompositions}. In order to distinguish pants decompositions related by 
monodromies\footnote{Variation of the position $z_r$ along a path encircling other
points, returning to the same position.} let us also introduce a trivalent graph 
ending in the points $z_r$, $r=1,\dots,n$, having exactly one vertex $\nu$ 
in each $C_{0,3}^\nu$ obtained in the pants decomposition. These
data will be called {\it gluing patterns}, and denoted by the letter
$\Gamma$. 
One may naturally equip the internal edges of $\Gamma$
with an orientation represented by an arrow ending at the vertex $\nu$ in  the three-punctured sphere
$C_{0,3}^\nu$ taking the role of $C_1$  in the construction above.
Two examples for gluing patterns on $C_{0,4}$ are depicted in Figure \ref{fmove}.
\begin{figure}[t]
\epsfxsize4cm
\centerline{\epsfbox{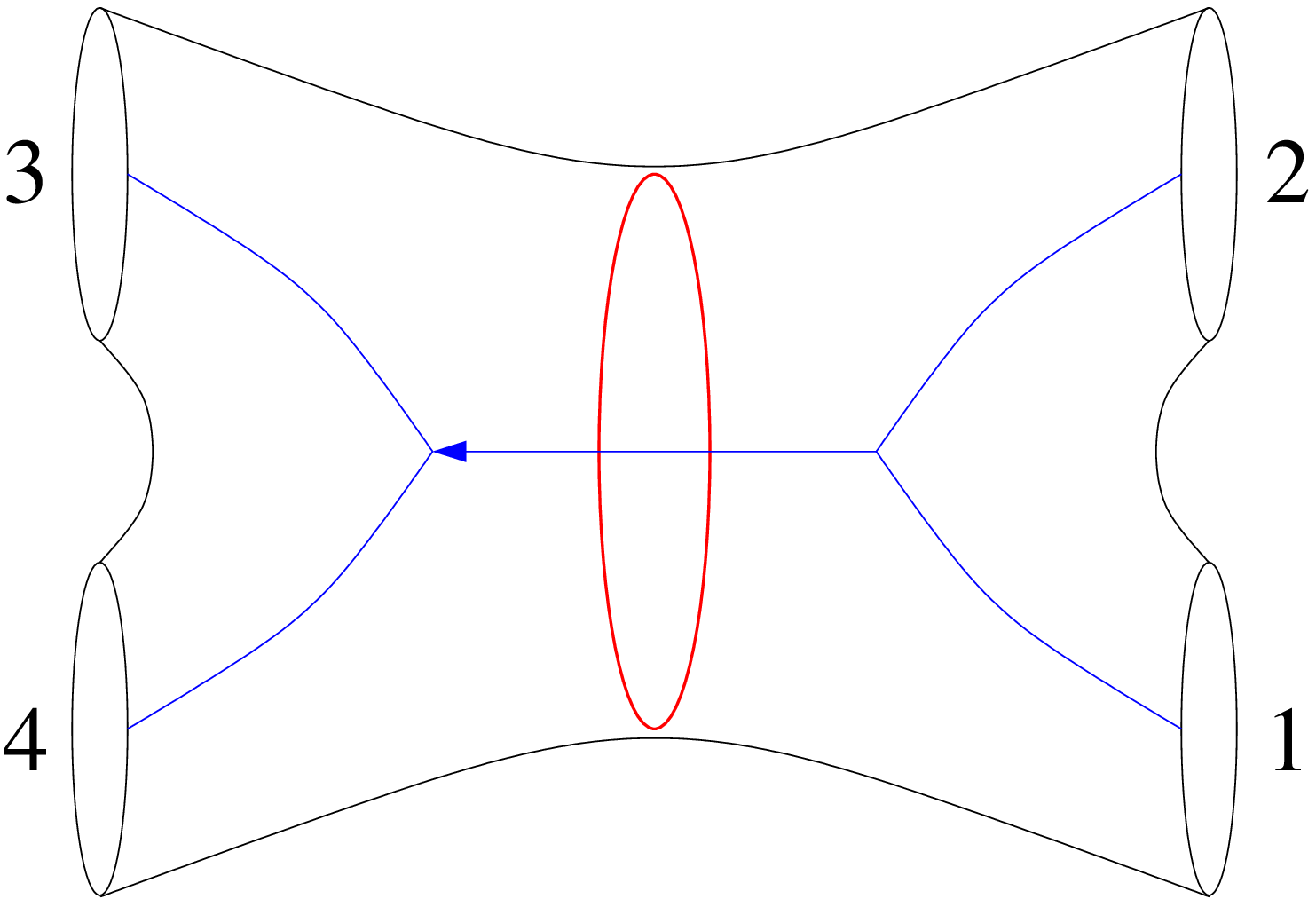}\hspace{1.5cm} 
\epsfxsize4cm\epsfbox{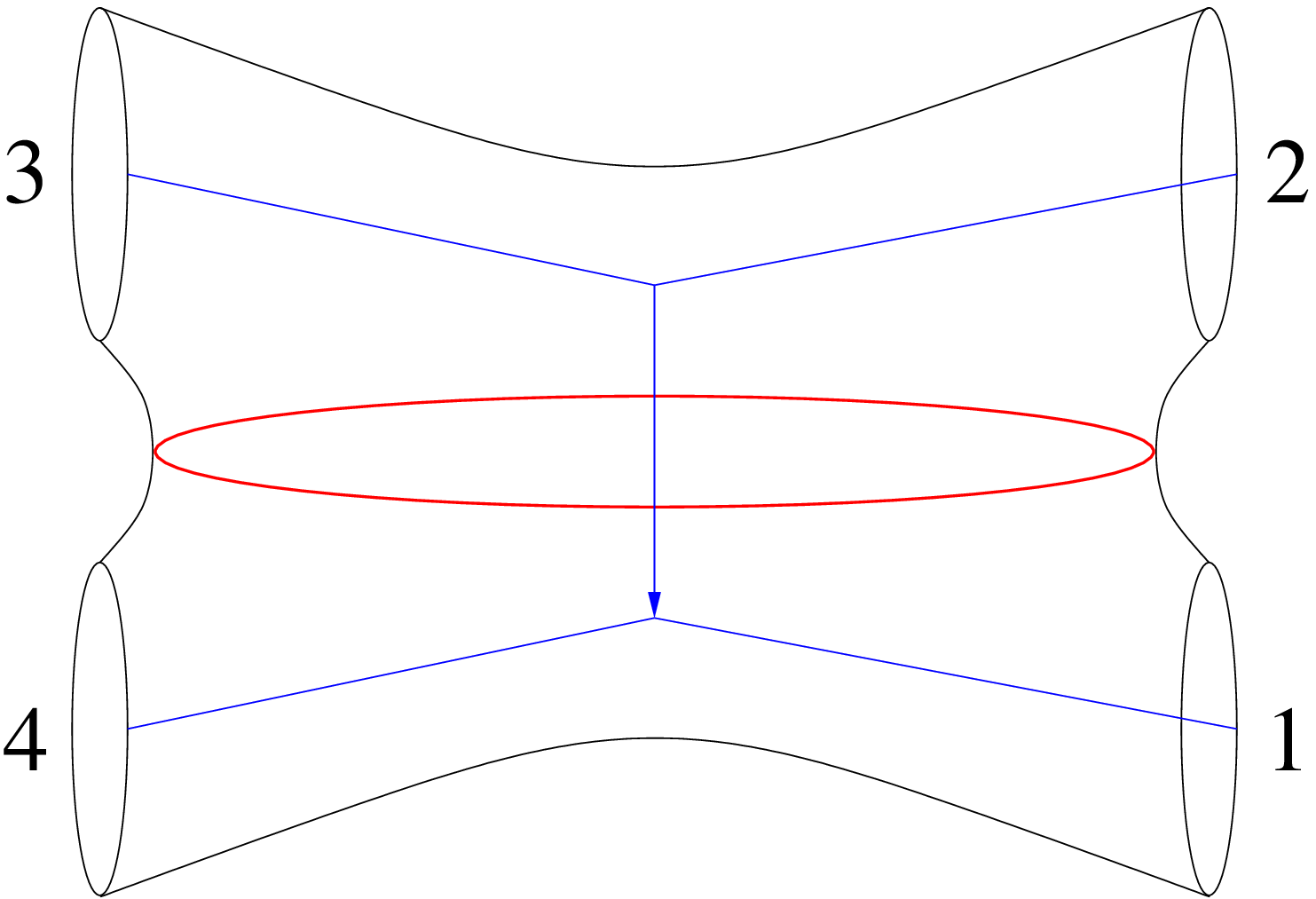}}
\caption{\it 
Two gluing patterns on the four-holed sphere obtained from $C_{0,4}$ by removing 
discs around the four punctures.
The gluing patterns on the 
left and on the right will be referred to 
as $\Gamma_s$ and $\Gamma_u$ (for s- and u-channel), 
respectively.}\label{fmove}\vspace{.3cm}
\end{figure}

\item For each edge $\epsilon$ of $\Gamma$ we need to specify the representation 
$R_\ep$ to be used in the gluing construction. 
These data will be collectively denoted by the letter $\be$. 
The set of all possible assignments of data $\be$ to graphs $\Gamma$ 
will be denoted by $\CC_{\Ga}$. \end{itemize}

We will use the notation $f^{\be}_{C,\Ga}(v)\equiv \CF^{\Ga}_{\be,\mathbf{v}}(\mathbf{z})$
for 
the conformal blocks constructed in this way
using the notation $v=\bigotimes_{r=1}^n v_i$ if $\mathbf{v}=(v_1,\dots,v_n)$ and $\mathbf{z}=(z_1,\dots,z_n)$,
and suppressing the choice of 
the representations $R_i$ attached to the punctures $z_i$ in the notations.
$\be$ only collects the labels of the representations used ``internally'' in the
gluing construction.

In Section \ref{CVO:sec} we had observed that conformal blocks define chiral vertex
operators and vice versa. It is not hard to see that the gluing patterns correspond
to the different ways to compose chiral vertex operators. Let us consider the case $n=4$ as an example.

\begin{quote} {\small {\bf Example:}
In the case $n=4$ one may consider two basic cases: 
In the first we take 
$C_2=\BP^1\setminus\{\infty,y_2,0\}$ and $C_1=\BP^1\setminus\{\infty,x_3,0\}$. 
It is easy to see that the gluing construction yields a result which may 
be represented in
vertex operator notation as 
\begin{equation}\label{s-channel}
f_C(v_4\otimes \dots\otimes v_1)=\big\langle \,v_4\,,\,V_{\rho'_s(R)}(v_3,x_3)q^{L_0} V_{\rho_s(R)}(v_2,y_2) \,v_1\,
\big\rangle\,,
\end{equation}
where $\rho'_s(R)=\big[{}_{R_4}{}^{R_3}{}_{R}\big]$, $\rho_s(R)=\big[{}_{R}\,{}^{R_2}{}_{R_1}\big]$.
We will associate the resulting conformal blocks with the gluing  pattern on the left
of Figure \ref{fmove}.

In the second case we shall take 
$C_2=\BP^1\setminus\{\infty,y_3,0\}$, $x_4=\infty$ and $x=x_2$. We then have
\begin{equation}
f_C(v_4\otimes \dots\otimes v_1)=\big\langle \,v_4\,,\,V_{\rho_u'(R)}\big(q^{L_0} V_{\rho_u(R)}(v_3,y_3) 
v_2,x_2\big)\,v_1\,
\big\rangle\,,
\end{equation}
where $\rho'_u(R)=\big[{}_{R_4}{}^{R}{}_{R_1}\big]$, $\rho_u(R)=\big[{}_{R}\,{}^{R_3}{}_{R_2}\big]$.
The conformal blocks defined in this way will be associated with the gluing  pattern on the right
of Figure \ref{fmove}.

We may, of course, further specialise to  $x_3=1$ in the construction above.
Choosing $y_2=1$ in the first case above we'll get
$q=x_2$, while in the second case $y_3=1$ gives $q=1-x_2$.}
\end{quote}

\subsection{Computing conformal blocks$^+$}

While it is straightforward to compute explicitly the very first few orders in the series expansions for chiral partition 
functions defined using the gluing construction, this will quickly become unmanageable for higher orders.
There are a few tools for more efficient calculations available. The free field representation briefly introduced in
the following section \ref{sec:free}  yields the most explicit formulae, but is in this form only 
applicable for special families of conformal blocks. We then quickly mention two further representations 
which are sometimes useful, with pointers to the relevant literature.

\subsubsection{Free field representation$^+$}\label{sec:free}

A subset of the conformal blocks can be elegantly represented using the free field representation 
for the Virasoro algebra introduced in Section \ref{ExtVOA}. The basic building blocks 
are the normal ordered exponential fields constructed from the generators $a_n$, $n\in\BZ$,
of the free boson algebra
\begin{equation}
e^{2\al\vf(z)}=
T_{\al}\exp\bigg( 2\mathrm{i}\al\sum_{k<0}\frac{a_k}{k}z^{-k}\bigg)e^{-2\mathrm{i}\al a_0\log(z)} 
\exp\bigg( 2\mathrm{i}\al\sum_{k<0}\frac{a_k}{k}z^{-k}\bigg),
\end{equation}
where
the operator $T_\al$ maps a Fock module $\CF_{\be}$ to $\CF_{\al+\be}$. The left hand side is
a short notation for the right hand side frequently used in the literature. It is straightforward to 
show that
\begin{equation}\label{interbos}
[L_n, e^{2\al\vf(z)}]=z^n(z\pa_z+(n+1)\De_\al)e^{2\al\vf(z)}\,,\qquad \De_\al=\al(Q-\al)\,,
\end{equation}
$L_n$ being the generators of the Virasoro algebra defined in \rf{FeiFu}.
Equation \rf{interbos} 
identifies $e^{2\al\vf(z)}$ as an intertwining operator between the Fock modules
$\CF_{\be}$ and $\CF_{\al+\be}$. We had seen that such intertwining operators 
allow us to define conformal blocks on the three-punctured sphere with
representations  assigned
to the three punctures 
having highest weights $\De_{\al}$, $\De_\be$ and $\De_{\al+\be}$.  

It is straightforward to calculate the expectation values of products of normal ordered exponentials, 
defined as
\begin{equation}\label{ExpVEV}
\big\langle \,e_0^*\,,\, e^{2\al_n(z_n)}\dots e^{2\al_1(z_1)}e_0\,\big\rangle\,=\,
\prod_{s>r}(z_s-z_r)^{-2\al_r\al_s},
\end{equation}
where $\sum_{r=1}^n\al_r=0$, and $e_0^*$ is the highest weight vector in the dual of the Fock space representation $\CF_0$. In \rf{ExpVEV}
we are assuming that $|z_s|>|z_r|$ for $s>r$, and the right hand side 
of \rf{ExpVEV} is defined by the principal value of the
logarithm when $z_r$ is real and positive for $r=1,\dots,n$.

A  generalisation of these vertex operators with one discrete parameter can be defined 
using the so-called screening operators, defined as
\begin{equation}
Q_\ga:=\int_\ga dz\; S(z),\qquad S(z):=e^{2b\vf(z)}.
\end{equation}
Note that $\De_b=1$, which implies that the commutator $[L_n,e^{2b\vf(z)}]$ is a total 
derivative. This implies that
products of normal ordered exponential with 
screening operators like 
\begin{equation}\label{freeCVO}
\mathsf{h}^\al_s(z)=
e^{2\al\vf(z)}\int_{\Ga_s}du_1\dots du_s\;S(u_1)\dots S(u_s)
\end{equation} 
will define more general 
intertwining operators mapping $\CF_{\be}$ to $\CF_{\be+\al+bs}$, {\it provided}
that the contour $\Ga_s$ of integration over $u_1,\dots,u_s$ is closed. 
The number $s$ of integration variables is also called the screening number.
When $b$ is real, and the real part of the parameter $\al$
is sufficiently negative one can replace the contour $\Ga_s$ by a product of
open contours $\ga_1\times\dots\times \ga_s$ starting 
and ending at $z$ in order to ensure absence of boundary terms. 

In this way one can 
obtain integral representations for the chiral partition functions associated to 
the special class of 
conformal blocks defined by the gluing construction satisfying the condition that
the triple $(\De_1,\De_2,\De_3)$ of highest weights associated to each pair of pants
is of the form $(\De_{\al},\De_{\be},\De_{\al+\be+bs})$. Different choices of contours $\Ga_s$
in \rf{freeCVO}
will define different bases for the subspace of the space of conformal blocks that can be
represented in this way.

\subsubsection{Recursion relations and AGT representation$^{++}$}

As noted above, one can not use the free field representation introduced above for general conformal
blocks. They define new types of special functions which will not have an integral representation with 
explicit integrand in general.\footnote{The functions defined using the free field representation as described
above have a finite-dimensional monodromy representation. This is not the case for generic conformal blocks, 
as will be seen below.}

For getting more explicit information on general
conformal blocks one may use two alternative types of representations. First, note that the developments
initiated by the discovery of relations between Virasoro conformal blocks and instanton partition functions in 
four-dimensional gauge theories \cite{AGT} led to explicit formulae for the expansion coefficients in the series 
expansions for conformal blocks, see \cite{AFLT} for a proof of the relations conjectured in AGT, and
the resulting formulae for the expansion coefficients.

Second, there exist recursion relations  
for the conformal blocks that are very useful for efficient numerical calculation of conformal blocks.
Such recursion relations were first obtained in 
\cite{Z87} for the conformal blocks on the four-punctured sphere, and recently 
generalised in \cite{CCY} to conformal blocks  associated to  more general Riemann surfaces.

\subsection{Crossing symmetry}\label{Crossing}


We are now going to explain why the conformal blocks constructed 
by the gluing construction are the ones of interest for applications in
theoretical physics. 

To see this we are first going to derive a more precise version of the holomorphically 
factorised form \rf{holofact} for the correlation functions of a CFT. Assuming, as before, that 
the CFT is unitary
one may decompose the Hilbert space $\CH$ in the following form
\begin{equation}
\CH\,=\,\bigoplus_{R',R''}M_{R',R''}\otimes R'\otimes R''\,,
\end{equation}
where $R'$ and $R''$ are unitary highest weight representations of the Virasoro 
algebras with generators $L_n$ and $\bar{L}_n$, respectively, and 
$M_{R',R''}$ is a multiplicity space on which the Virasoro algebras act trivially. Let 
us then study a correlation function $\big\langle\prod_{r=1}^n\Phi_{v_r}(z_r,\bz_r)\big\rangle$
of $n$ vertex operators $\Phi_{v_r}(z_r,\bz_r)$ associated to states $v_r\in\CH$ of the form
$v_r=m_r\ot v'_r\ot v''_r$, $v_r'\in R_r'$, $v_r''\in R_r''$, $m_r\in M_{r}$,
$r=1,\dots,n$.  We will see that such correlation functions can be represented 
in the following form
\begin{equation}\label{holofact2}
\CZ_{\mathbf{v}}(\mathbf{z},\bar{\mathbf{z}})=\sum_{\be',\be''}C^{\Ga}_{\be'\be''\mathbf{m}} \,
\CF_{\be',\bv'}^{\Ga}(\mathbf{z})\,\CF_{\be'',\bv''}^{\Ga}(\bar{\mathbf{z}})\,,
\end{equation}
where $\CF_{\be,\bv}^{\Ga}(\mathbf{z})$ are the conformal blocks constructed by the gluing construction 
as described in Section \ref{gluesect}. The expansion \rf{holofact2} is a more precise version of the 
holomorphically factorised representation \rf{holofact} postulated before.

Indeed, let us observe that the operator-state correspondence implies existence of an
operator product expansion (OPE) of the following form:
\begin{align}\label{OPE}
\Phi_{w_2} &(x,\bx) \Phi_{w_1}(y,\by)=  \Phi_{\Phi_{w_2}(x-y,\bw-\bz)w_1}^{}(y,\by)  \\
&= \sum_{\imath\in\CI} C_{w_2,w_1}^{v_{\imath}}\,
(x-y)^{\De_{v_\imath}-\De_{w_2}-\De_{w_1}} 
(\bx-\by)^{\bar{\De}_{v_{{\imath}}}-\bar{\De}_{w_2}-\bar{\De}_{w_1}}
\,\Phi_{v_{\imath}}(y,\by)\,.
\notag\end{align}
The first line is a consequence of the operator-state correspondence,  
identifying the composite field on the left of \rf{OPE} as the field associated to the state
$\Phi_{w_2}(x-y,\bx-\by)w_1$. The second line is then obtained by  picking a basis 
$\{v_\imath;\imath\in\CI\}$ for $\CH$ consisting of eigenvectors $v_\imath$
of $L_0$ and $\bar{L}_0$ with eigenvalues $\De_{v_\imath}$ and $\bar{\De}_{v_{\imath}}$, 
respectively, and expanding $\Phi_{w_2}(x-y,\bx-\by)w_1$ with respect to this basis. 

\begin{quote}
{\small
{\bf Exercise 8:}
Consider the case of a four-point function $\langle \prod_{r=1}^4\Phi_{v_r}(z_r,\bz_r)\rangle$.
Demonstrate that applying the 
OPE \rf{OPE} to the pair $\Phi_{v_2}(z_2,\bz_2)\Phi_{v_1}(z_1,\bz_1)$ yields an expansion of the 
form \rf{holofact2}, with conformal blocks $\CF^{\be}_{\Ga,\mathbf{v}}(\mathbf{z})$ constructed using the 
gluing pattern on the left of Figure \ref{fmove}, while application of \rf{OPE} to the pair
$\Phi_{v_3}(z_3,\bz_3)\Phi_{v_2}(z_2,\bz_2)$ yields an expansion  
of the 
form \rf{holofact2}, with conformal blocks $\CF^{\be}_{\Ga,\mathbf{v}}(\mathbf{z})$ constructed using the 
gluing pattern on the right of Figure \ref{fmove}. 
Hint: Use the conformal Wigner-Eckart theorem from Section \ref{CVO:sec}.
}
\end{quote}

It is a basic physical requirement that the correlation functions 
$\CZ_{\mathbf{v}}(\mathbf{z},\bar{\mathbf{z}})$ are single-valued and real analytic in the variables $z_r$,
$r=1,\dots,n$ away from the diagonals $z_i=z_j$. This implies that 
$\CZ_{\mathbf{v}}(\mathbf{z},\bar{\mathbf{z}})$ defines a function on $\CM_{0,n}$. 
The expansions \rf{holofact2} yield (presumably convergent) series expansions
around the component of the boundary of $\CM_{0,n}$ specified by the 
gluing pattern $\Ga$, corresponding to an ordering presecription for 
successively  performing OPEs. In order for $\CZ_{\mathbf{v}}(\mathbf{z},\bar{\mathbf{z}})$ to be 
analytic away from the diagonals $z_i=z_j$, we need that the conformal blocks
$\CF_{\be,\bv}^{\Ga}(\mathbf{z})$ admit an analytic continuation to the universal 
cover $\widetilde{\CM}_{0,n}$ of the configuration space ${\CM}_{0,n}$. We will see that 
this is typically the case for the conformal blocks coming from the gluing construction.

The correlation function $\CZ_{\mathbf{v}}(\mathbf{z},\bar{\mathbf{z}})$ will admit many equivalent representations
of the form \rf{holofact2}, obtained by performing OPEs in different orders and represented
by different choices of gluing patterns. The equivalence of the different representations is
expressed in relations of the form
\begin{equation}\label{crossing}
\sum_{\be',\be''}C^{\Ga_1}_{\be'\be''\bm} \,
\CF_{\be',\bv'}^{\Ga_1}(\mathbf{z})\,\CF_{\be'',\bv''}^{\Ga_1}(\bar{\mathbf{z}})\,=\,
\sum_{\be',\be''}C^{\Ga_2}_{\be'\be''\bm} \,
\CF_{\be',\bv'}^{\Ga_2}(\mathbf{z})\,\CF_{\be'',\bv''}^{\Ga_2}(\bar{\mathbf{z}})\,;
\end{equation}
it may be necessary to analytically continue $\CF_{\be',\bv'}^{\Ga_1}(\mathbf{z})$ and 
$\CF_{\be'',\bv''}^{\Ga_1}(\bar{\mathbf{z}})$ w.r.t. $z_1,\dots,z_n$ in order to define the left
hand side in a domain of $\CM_{0,n}$ where the right hand side of \rf{crossing}
is defined through convergent power series expansions. These relations, often 
referred to as the crossing symmetry conditions, can be regarded as 
a system of equations constraining the remaining unknowns, the coefficients $C^{\Ga}_{\be'\be''\bm}$
in \rf{holofact2}, with coefficients $\CF_{\be',\bv'}^{\Ga}(\mathbf{z})\,\CF_{\be'',\bv''}^{\Ga}(\bar{\mathbf{z}})$ fully
determined by conformal symmetry. The next step in the bootstrap program
will be to observe that  there often exist linear relations between the conformal blocks 
$\CF_{\be,\bv}^{\Ga_1}(\mathbf{z})$ and $\CF_{\be,\bv}^{\Ga_2}(\mathbf{z})$ associated to different 
gluing patterns $\Ga_1$ and $\Ga_2$,
allowing us to exhibit the mathematical content of the 
crossing symmetry conditions more clearly.

\subsection{Fusion and braiding}\label{fusbr}

We are now going to observe that there often exist relations of the form
\begin{equation}\label{fusion}
\CF_{\be_1,\bv}^{\Ga_1}(\mathbf{z})
\,=\,\sum_{\be_2\in\CS}F^{\Ga_1\Ga_2}_{\be_1\be_2}\,
\CF_{\be_2,\bv}^{\Ga_2}(\mathbf{z})\,,
\end{equation}
allowing us to turn the relations \rf{crossing} into more tractable problems. 
The set $\CS$ is a subset of the set of all possible ways to color a gluing 
pattern $\Ga$ with choices of intermediate representations.

\begin{figure}[t]
\epsfxsize8cm
\centerline{\epsfxsize8cm\epsfbox{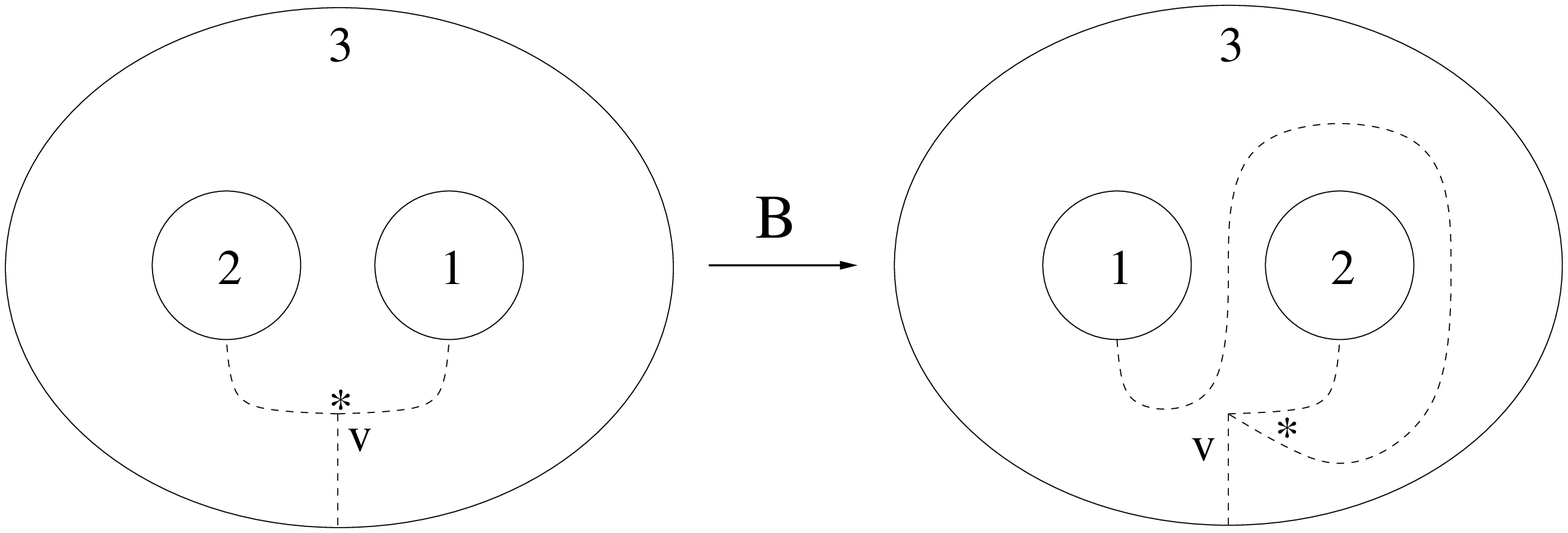}}
\caption{\it The braiding operation}\label{bmove}\vspace{.3cm}
\end{figure}

The derivation of relations of the form \rf{fusion} is a difficult problem in general, of 
central importance for the mathematics of conformal field theories.
In order to demonstrate validity of relations of the form \rf{fusion} it is first of all
useful to observe that it suffices to derive such relations in the cases $n=3$ and $n=4$.
This follows from the known fact \cite{MS,BK} that the transition between any two
gluing patterns $\Ga_1$ and $\Ga_2$ can be broken up into a sequence of elementary
operations localised in subsurfaces isomorphic to $C_{0,3}$ and $C_{0,4}$. 
These elementary operations are called braiding and fusion, braiding being depicted
in Figure \ref{bmove} while fusion is the passage from the gluing
pattern on the left to the one on the right of Figure \ref{fmove}. It therefore suffices to establish 
\rf{fusion} in these two cases.


We shall begin by discussing the braiding operation.
\begin{quote} {\bf Exercise 9}:
{\small Let $\Ga_1$ and $\Ga_2$ be the conformal blocks depicted in the 
left and right parts of Figure \ref{bmove}, respectively. Recall that there is no
internal label $\be$ needed in this case, but the conformal blocks depend on the 
choices of three representations $R_1,R_2, R_3$ associated to the three punctures.
Demonstrate that
\begin{equation}\label{braid}
\CF^{\Ga_2}_{\bv}
\,=\,e^{\pi \mathrm{i}(\De_{R_3}-\De_{R_2}-\De_{R_1})}\,
\CF^{\Ga_1}_{\bv}\,,
\end{equation}
in this case, where $\De_R$ is the $L_0$-eigenvalue of the highest weight vector 
of the Virasoro representation $R$.
}
\end{quote}

This main difficulty is to derive relations of the form \rf{fusion} for $n=4$. The following example illustrates
how such relations can be calculated in a simple case.

\begin{quote} {\bf Example 1} (ctd.):
{\small Let us return to the case 
$C=C_{0,4}$ with $\al_2=-b/2$ considered before. It is instructive 
to check that the conformal blocks defined using the gluing pattern
$\Ga_s$ introduced on the left of Figure \ref{fmove} have chiral partition 
functions $\CZ^s_{\beta}(z)\equiv \CZ^s_{\pm{1}/{2}}(z)$
given by the following linearly independent solutions of the 
differential equation $\CD_{\rm\sst BPZ}\CZ(z)=0$:
\begin{align}
&\CZ^s_{+\frac{1}{2}}(z)\,=\,z^{b\al_1}(1-z)^{b\al_3}F(A,B,C;z)\,,
\end{align}
where $F(A,B,C;z)$ is the hypergeometric function. with arguments given as
\begin{align}
\begin{aligned}
& \quad A=b(\al_1+\al_3+\al_4-3b/2)-1\,,\\
& \quad B=b(\al_1+\al_3-\al_4-b/2)\,,
\end{aligned}\qquad 
C=b(2\al_1-b)\,.
\end{align}
The  second linearly independent solution
$\CZ^s_{-{1}/{2}}(z)$ is given by the expressions obtained by replacing 
$\al_i\ra Q-\al_i$ for $i=1,3,4$ throughout, 
Similar formulae will of course represent the chiral partition 
functions $\CZ^u_{\pm{1}/{2}}(z)$ representing the conformal blocks 
defined from the gluing pattern
$\Ga_u$ introduced on the right of Figure \ref{fmove}. Note that the 
representations associated to the edges marked with an arrow in Figure
\ref{fmove} are given as,
\begin{equation}\label{furu}
\be\,=\,\al_1\mp \frac{b}{2} \;\;\text{for}\;\;\CZ^s_{\pm\frac{1}{2}}\,,\qquad
\be\,=\,\al_3\mp \frac{b}{2} \;\;\text{for}\;\;\CZ^u_{\pm\frac{1}{2}}\,,
\end{equation}
respectively.
The restrictions on the set of representations that may be used in the
gluing construction following from the presence of null vectors are
often called {\it fusion rules}.

Well-known relations satisfied by the hypergeometric functions
then give us the following relations 
\begin{equation}\label{F-deg}
\left(\begin{matrix} 
\CZ^s_{+\frac{1}{2}}(z)\\[1ex]
\CZ^s_{-\frac{1}{2}}(z)
\end{matrix}
\right)=
\left(\begin{matrix} 
\frac{\Ga(C)\Ga(C-A-B)}{\Ga(C-A)\Ga(C-B)} & \frac{\Ga(C)\Ga(A+B-C)}{\Ga(A)\Ga(B)}\\[1ex]
\frac{\Ga(2-C)\Ga(C-A-B)}{\Ga(1-A)\Ga(1-B)} & \frac{\Ga(2-C)\Ga(A+B-C)}{\Ga(!+A-C)
\Ga(1+B-C)}
\end{matrix}
\right)
\left(\begin{matrix} 
\CZ^u_{+\frac{1}{2}}(z)\\[1ex]
\CZ^u_{-\frac{1}{2}}(z)
\end{matrix}
\right),
\end{equation}
from which one may easily read off the explicit formulae for the 
coefficients $F^{\Ga_1\Ga_2}_{\be\be'}$ in this case.
}\end{quote}

A first family  of examples for which relations of the form \rf{fusion} are known to 
hold in general
is found by considering the cases where $c=1-6(\be-\be^{-1})^2$, $\be=\sqrt{p/p'}$ with $p$, $p'$
being positive coprime integers satisfying $p<p'$. The corresponding CFTs are called minimal models.
In order to parameterise the relevant set of representations of $\mathsf{Vir}_c$ let us introduce the set
\begin{equation}
\mathbb{A}_{\mathsf{KT}}=\Big\{a_{mn};\;a_{mn}=\frac{\be}{2}(1-m)-\frac{1}{2\be}(1-n),\;m=1,\dots,p',\; n=1,\dots,p\,\Big\}\,.
\end{equation}
Representations having highest weights $\De_a=a(a-q)$, $q=\be^{-1}-\be$,  with $a\in \mathbb{A}_{\mathsf{KT}}$ are
referred to as representations contained in the Kac-table.
For conformal blocks having external representations in the Kac-table  the existence
of the relations \rf{fusion} was established in \cite{Hu}.

There is another family of cases, for which relations of the form \rf{fusion} have been established. 
This is the case when $c\geq 25$,  
$C=C_{0,4}$ with $\al_r=Q/2+iP_r$, $P_r\in\BR$. For this case it was found in 
\cite{T01} that there exist relations of the form 
\begin{equation}\label{Vir-fusion}
\CF_{P,\bv}^{\Ga_s}(q)
\,=\,\int_{\BR_+}dP' \;F^{\Ga_s\Ga_u}_{}(P,P')\,
\CF_{P',\bv}^{\Ga_u}(q)\,.
\end{equation} 
The relations \rf{Vir-fusion} were derived in \cite{T01} using a generalisation 
of the free field representation described in Section \ref{sec:free} to the case of non-integer 
screening numbers $s$. A similar result for $c=1$ was obtained in \cite{ILTy}, see also Section \ref{Impl}.

Even more general cases  can be deduced from \rf{Vir-fusion}
using analytic continuation in  $\al_r$, $r=1,\dots,4$, or in the central charge $c$, which exist 
as long as $c>1$ \cite{TV13}. This gives the complete
answer for the fusion transformations of the conformal blocks of arbitrary irreducible 
highest weight representations
of the Virasoro algebra with $c>1$. 

The existence of fusion relations of the form \rf{Vir-fusion} for generic representations 
is a remarkable and highly non-trivial fact that deserves to be better understood. It implies that the conformal 
blocks obtained from the gluing construction generate a subspace that is closed under 
the representation of the braid group defined by composing braiding and fusion 
operations. These results offer a starting point for the development of a harmonic analysis on 
spaces of conformal blocks, related to the harmonic analysis
on the Teichm\"uller spaces themselves \cite{T01,TV13}.

Combining \rf{crossing} and \rf{fusion} yields a system of equations for the remaining 
undetermined quantities $C^{\Ga_1}_{\be'\be''}$ in \rf{holofact2},
\begin{equation}\label{crossing2}
\sum_{\be'_1,\be''_1}C^{\Ga_1}_{\be_1'\be_1''\bm} \,
F^{\Ga_1\Ga_2}_{\be_1'\be_2'}
F^{\Ga_1\Ga_2}_{\be_1''\be_2''}
\,=\,C^{\Ga_2}_{\be_2'\be_2''\bm}
\end{equation}
According to the discussion above one may consider the coefficients 
$F^{\Ga_1\Ga_2}_{\be_1\be_2}$ as known data, fully determined by conformal symmetry alone.
The system of equations \rf{crossing2} expresses the constraints on the as yet unknown 
data $C^{\Ga}_{\be'\be''\bm}$ following from crossing symmetry. 

Finding the most general solution of \rf{crossing2}  would be equivalent to a full
classification and solution of all CFTs, which seems pretty hopeless in general. 
For special CFTs one may exploit additional constraints on the  form of the coefficients 
$C^{\Ga}_{\be'\be''\bm}$,  allowing us to determine them completely. One class of theories
where this has been realised fairly completely will be described in Section \ref{sec:minmod}.

One may hope that the powerful numerical techniques that have recently been developed
for getting constraints on the coefficients 
$C^{\Ga}_{\be'\be''\bm}$ can lead to some progress in this direction. We feel unable to 
offer a good survey of the literature on this line of research here. As one possible starting 
point containing further references we would like to mention the talk \cite{Yi} and 
the paper \cite{CKLY} describing numerical evidence for the uniqueness of the Liouville CFT
discussed in Section \ref{sec:minmod} within a certain class of CFTs. It would be very 
interesting if such techniques can be used to explore what comes beyond this class.

\subsection{Rational and non-rational minimal models${}^+$} \label{sec:minmod}

We will now discuss a few examples where the bootstrap strategy has been fully realised. 
The examples are known under the names of (generalised) minimal models and 
Liouville theory, respectively. The main simplifying features of these CFTs are (i) the absence of 
multiplicities, and (ii) a diagonal pairing of the representations of the two copies of 
$\mathsf{Vir}_c$ generating the Hilbert space of the theory, leading to the form
\begin{equation}\label{spec}
\CH_{\rm MM}=\bigoplus_{R\in \mathsf{KT}}R\otimes R,\qquad
\CH_{\rm LT}=\int_{\mathbb{S}} d\al \;R_{\al}\otimes R_{\al},
\end{equation} 
for the Hilbert spaces $\CH_{\rm MM}$ of the minimal models, and $\CH_{\rm LT}$ of Liouville
theory, respectively. The sets of representations appearing in $\CH_{\rm MM}$ and in 
$\CH_{\rm LT}$ are denoted by $\mathsf{KT}$ for "Kac table" and $\mathbb{S}$, respectively.
The representations appearing in $\CH_{\rm MM}$ and $\CH_{\rm LT}$ will be unitary highest weight 
representations of the Virasoro algebra with $c<1$ and $c> 25$, respectively. 

\subsubsection{Factorisation}

Due to the absence of multiplicities we now have $C^{\Ga}_{\be'\be''\bm}=C^{\Ga}_{\be'\be''}$.
In order to deduce the resulting restrictions on the coefficients
$C^{\Ga}_{\be'\be''}$ in the holomorphic factorisation \rf{holofact2} let us 
recall the link between the OPE and an expansion over a basis for the Hilbert space $\CH$
noted after \rf{OPE}. We may assume that $w_i\in  R_{\al_i}\otimes R_{\al_i}$ for $i=1,2$, and
the summation over a basis for $\CH$ may be split into a summation or integration over a representation label
denoted $\al_3$, and a summation over vectors $w_3$ forming a basis for $R_{\al_3}\otimes R_{\al_3}$. 
The coefficients  $C_{w_2,w_1}^{w_3}$
in \rf{OPE} can be expressed in terms of the coefficient $C(\al_3|\al_2,\al_1)$ 
associated to $w_i=e_i$, the products of highest weight vectors in $R_{\al_i}\otimes R_{\al_i}$ for $i=1,2,3$,
respectively.

\begin{quote}
{\small
{\bf Exercise 10:} \\[1ex]
a) Use the conformal Ward identities to demonstrate that two-and three-point functions
of primary fields in Liouville field theory have the form 
\begin{align}\label{twopoint}
\langle\,V_{\al_2}(z_2,\bz_2)V_{\al_1}(z_1,\bz_1)\,\rangle =\;& \de_\BS(\al_2,\al_1)\, B(\al_1)\,
|z_2-z_1|^{-4\De_{\al_1}}\,,\\
 \langle V_{\al_3}(z_3,\bz_3)V_{\al_2}(z_2,\bz_2)V_{\al_1}(z_1,\bz_1)\rangle =\;
 &|z_2-z_1|^{2(\De_{\al_3}-\De_{\al_2}-\De_{\al_1})}
|z_3-z_2|^{2(\De_{\al_1}-\De_{\al_3}-\De_{\al_2})}\notag \\
 \times &|z_1-z_3|^{2(\De_{\al_2}-\De_{\al_3}-\De_{\al_1})} C(\al_3,\al_2,\al_1)\,,
\label{threepoint}\end{align}
where $\de_\BS(\al_2,\al_1)$ is the delta-distribution on $\BS$.\\[1ex]
b) Use the results from part a) together with the OPE \rf{OPE} to demonstrate that the as yet
undetermined coefficients $C(\al_3,\al_2,\al_1)$ in 
three-point functions \rf{threepoint}, $B(\al)$ in the two-point functions \rf{twopoint}
and the OPE coefficients $C(\al_3|\al_2,\al_1)$ are related
as
\begin{equation}
C(\al_3,\al_2,\al_1)\,=\,C(\al_3|\al_2,\al_1)B(\al_3)\,.
\end{equation}
}
\end{quote}

Using the observations above
it becomes easy to see that the coefficients 
$C^{\Ga}_{\be'\be''}$ in the holomorphic factorisation \rf{holofact2}
are supported on the diagonal $\be'=\be''$, and  factorise as
\begin{equation}\label{factor}
C^{\Ga}_{\be'\be''}= \de(\be',\be'')
\prod_{\nu\in\mathsf{vert}(\Ga)} C(\al_1^\nu, \al_2^\nu,\al_3^\nu)\prod_{e\in\mathsf{i}\text{-}\mathsf{edg}(\Ga)}B(\be_e)
\,,
\end{equation}
where $\mathsf{vert}(\Ga)$ and $\mathsf{i}\text{-}\mathsf{edg}(\Ga)$ are the sets of 
vertices and internal edges of $\Ga$, respectively, with triples of representation 
labels $(\al_1^\nu, \al_2^\nu,\al_3^\nu)$ assigned to the punctures of 
the three-punctured sphere $C_{0,3}^\nu$ appearing in the pants decomposition associated to $\Gamma$, and 
representation labels $\be_e$ associated to the internal  edges of $\Ga$.

\subsubsection{Functional equations}

It turns out that
the conditions following from crossing symmetry \rf{crossing} can be written more explicitly
in the special case where $n=4$ with $\al_2=-b/2$ considered above, not imposing 
further restrictions on $\al_1$, $\al_3$ and $\al_4$ for the time being. Using \rf{factor} one may write the expansion
\rf{holofact2} more explicitly as 
\begin{align}\label{simplecrossing-1}
\CZ(z,\bz)&=\sum_{m=\pm\frac{1}{2}}
C(\al_4,\al_3,\al_1-m b)C(\al_1-m b|\al_2,\al_1) \,
\CZ_m^s(z)\,\CZ_m^s(\bz)\\
&=\sum_{m=\pm\frac{1}{2}}C(\al_4,\al_3-m b,\al_1)C(\al_3-m b|\al_3,\al_2) \,
\CZ_m^u(z)\,\CZ_m^u(\bz)\,,
\label{simplecrossing-2}\end{align}
where $\CZ_m^s(z)$ and $\CZ_m^u(z)$, $m=\pm 1/2$, were introduced in Example 1c).
Inserting \rf{F-deg} into 
\rf{simplecrossing-1} one obtains an expression  apparently containing terms proportional 
to $\CZ^{\pm 1/2}_s(z)\CZ^{\mp 1/2}_s(\bz)$. No such terms occur in \rf{simplecrossing-2}. 
Consistency therefore requires that the coefficients in front of the terms $\CZ_{\pm 1/2}^s(z)\CZ_{\mp 1/2}^s(\bz)$ vanish.
This is easily seen to be the case
if the coefficients $C(\al_3,\al_2,\al_1)$ satisfy a 
functional relation of the following form:
\begin{equation}\label{shiftrel}
\frac{C(\al_1,\al_2,\al_3)}{C(\al_1,\al_2,\al_3+b)}=
d(\al_3)D(\al_1,\al_2,\al_3)\,,
\end{equation}
where $D(\al_1,\al_2,\al_3)$ can be assembled from the 
elements of the matrix \rf{F-deg} leading to the expression
\begin{equation}
D(\al_1,\al_2,\al_3)=\prod_{s_1,s_2=\pm}\ga\big(b\al_3+s_1b(\al_1-\fr{Q}{2})
+s_2b(\al_2-\fr{Q}{2})\big)\,, 
\end{equation}
written in terms of the function $\ga(x)=\Ga(x)/\Ga(1-x)$. The function 
$d(\al_3)$ in \rf{shiftrel} is the product of some known and some unknown functions depending on $\al_3$ only.
We will later fix the resulting indeterminacy.
A similar functional equation is obtained by noting that 
$0=(b^2L_{-1}^2+L_{-2})e_{-1/2b}$
holds in $V_{-b/2}$. It leads to another functional equation obtained
from \rf{shiftrel} by replacing $b\ra b^{-1}$ everywhere.

\subsubsection{Explicit solutions}

The functional relations \rf{shiftrel} can be solved for $\mathrm{Re}(b)>0$ corresponding to $c>1$
by an 
expression of the form 
\begin{equation}\label{DOZZ}
C(\al_1,\al_2,\al_3)\,=\, \frac{C_0\prod_{r=1}^3 N(\al_r)}{\prod_{s_1,s_2=\pm}
\Upsilon_b\big(\al_3+s_1\big(\al_1-\fr{Q}{2}\big)
+s_2\big(\al_2-\fr{Q}{2}\big)\big)}\,,
\end{equation}
using the fact that the special 
function 
\begin{equation}\label{Updef}
\log\Upsilon_b(x)\,=\,\int_{0}^{\infty}\frac{dt}{t}\bigg[\!\left(\frac{Q}{2}-x\right)^2
\!\!e^{-2t}
-\frac{\sinh\big(\frac{t}{2}({Q}-2x)\big)}{\sinh(bt)\sinh(b^{-1}t)}\,\bigg]\,,
\end{equation}
satisfies the functional equations
\begin{equation}
\Upsilon_b(x+b)\,=\,b^{1-2bx}\ga(bx)\Upsilon_b(x)\,,\qquad
\Upsilon_b(x)\,=\,\Upsilon_{1/b}(x)\,.
\end{equation}
As we did not determine the function $d(\al_3)$ in \rf{shiftrel} yet, we are not yet able
to fix the functions $N(\al)$ in \rf{DOZZ}. One may notice, however, that the function 
$N(\al)$ can be changed by changing the normalisation of the 
primary fields $\Phi_{e_{\al}}(z,\bz)$.

The bootstrap equations \rf{shiftrel} remain valid
for $c\leq 1$. However, it turns out that
the formula \rf{DOZZ} representing a solution to the 
equations \rf{shiftrel} for $c\geq 1$ can not be
used in this case. The  special function $\Upsilon_b$ 
can only be used for ${\rm Re}(b)>1$ covering the cases with $c\geq 1$.
In order to solve the functional equations  \rf{shiftrel} in the case
$c\leq 1$ it is convenient to use the parameters $a=-i\al$, $\be=ib$. 
Instead of \rf{DOZZ} one may now use an expression of the following form \cite{Z05}
\begin{equation}\label{DOZZm}
C_{\rm M}(a_1,a_2,a_3)\,=\, \frac{\prod_{s_1,s_2=\pm}
\Upsilon_\beta\big(\be+a_3+s_1\big(a_1-\fr{q}{2}\big)
+s_2\big(a_2-\fr{q}{2}\big)\big)}{C_0\prod_{r=1}^3 N_{\rm M}(\al_r)}\,,
\end{equation}
using the notation $q=\be^{-1}-\be$.
It can be shown that \rf{DOZZm} can not
be obtained from \rf{DOZZ} by analytic continuation \cite{Z05}.

\subsubsection{Solutions for $c\leq 1$: (Generalised) minimal models}

If all representations used in the definition of the conformal blocks 
appearing in \rf{holofact2} are highest weight 
representations with highest weights $\De_a=a(a-q)$, $a\in\mathbb{A}_{\mathsf{KT}}$
one may show that crossing symmetry is satisfied with $C(a_3,a_2,a_1)$
chosen as the 
restriction of $C_{\rm M}(a_1,a_2,a_3)$ to $a_i\in\mathsf{KT}$, $i=1,2,3$.
It can furthermore be verified that this restriction 
reproduces the expressions for the three point functions of the minimal models 
first calculated in \cite{DF} for a suitable choice of field normalisation
function $N_{\rm M}(\al_r)$ and normalisation factor $C_0$ in \rf{DOZZm}.

However, this turns out not to be the end of the story yet.
It was verified numerically in 
\cite{RS} for various cases that
there exists a {\it continuous} set of intermediate dimensions to be 
used to construct  the four-point functions in the form 
\rf{holofact2} such that choosing the 
coefficients to be given in terms of \rf{DOZZm}  indeed solves
the crossing symmetry conditions \rf{crossing}. Following 
\cite{Z05} we shall refer to the 
theory defined in terms of these three point functions as generalised 
minimal models. Even if the CFT defined thereby is non-unitary in general, it appears 
to define an interesting model of statistical mechanics \cite{IJS}.

For rational values of $c\leq 1$ there exist further solutions $\tilde{C}_{\rm M}(a_1,a_2,a_3)$
to the crossing symmetry conditions \cite{RS}. The functions $\tilde{C}_{\rm M}(a_1,a_2,a_3)$ 
differ from ${C}_{\rm M}(a_1,a_2,a_3)$ 
by a non-analytic factor containing step-functions. In the case $c=1$ one may identify 
$\tilde{C}_{\rm M}(a_1,a_2,a_3)$ as a limit of the expressions for the minimal model three point 
functions \cite{RW}, or as a limit of the Liouville three point functions \cite{S} to be discussed below.
This implies crossing symmetry of the correlation functions build using  $\tilde{C}_{\rm M}(a_1,a_2,a_3)$ for 
$c=1$. Crossing symmetry has been verified numerically for other rational values of $c$ in \cite{RS}. 

The non-uniqueness of the solutions to the crossing symmetry conditions for $c\leq 1$ 
following from the
results summarised above
is a remarkable phenomenon
which deserves to be better understood.

\subsubsection{Solution for $c>1$: Liouville theory}

In order to round off the bootstrap solution of first examples for interesting
CFTs let us observe,
on the one hand, that it has been verified analytically 
for the case $\al_r\in \frac{Q}{2}+i\BR$, $r=1,\dots,4$
in \cite{T01} that
the three point functions defined in \rf{DOZZ}, when used 
as coefficients  in  \rf{holofact2}, will satisfy the crossing conditions \rf{crossing}.
A key ingredient in this verification are the fusion relations \rf{Vir-fusion}
identifying
the set of intermediate representations that are summed (in fact integrated) over
in \rf{holofact2} to be $\{V_{\al};\al\in\frac{Q}{2}+i\BR_+\}$. By a suitable 
choice of normalisation of the conformal blocks one can identify the crossing conditions
as expression of the unitarity of the fusion transformations \rf{Vir-fusion} with respect to 
a natural scalar product.  

It is furthermore remarkable that the correlation functions $\CZ_{\mathbf{v}}(\mathbf{z},\bar{\mathbf{z}})$
turn out to be entire analytic in the parameters $\al_r$, $r=1,\dots,n$, allowing us to 
obtain representations of arbitrary 
correlation functions $\CZ_{\mathbf{v}}(\mathbf{z},\bar{\mathbf{z}})$ for $c>1$ 
in the form \rf{holofact2}
by analytic continuation \cite{T01,TV13}.

It remains to notice \cite{T01}, on the other hand, that
there exists a choice of the normalisation-dependent 
function $N(\al)$ in \rf{DOZZ} such that the fields $\phi(z,\bz)$
and $e^{2b\phi(z,\bz)}$ defined as
\begin{equation}\label{Lioudef}
\phi(z,\bz):=\frac{1}{2}\frac{\pa}{\pa \al} \Phi_{e_\al}(z,\bz)\Big|_{\al=0}\,,\qquad
e^{2b\phi(z,\bz)}:=\Phi_{e_\al}(z,\bz)\big|_{\al=b}\,,
\end{equation}
satisfy the equation 
\begin{equation}\label{EOM}
\pa_z\bar\pa_{\bz}\phi(z,\bz)\,=\,\pi\mu \,e^{2b\phi(z,\bz)}\,.
\end{equation}
The choices of $C_0$ and $N(\al)$ which will do the job are 
\begin{equation}\label{N-C}
N(\al)=\big(\pi \mu\ga(b^2)b^{2-2b^2}\big)^{-\frac{\al}{b}}\Upsilon(2\al)\,,\qquad
C_0=\big(\pi \mu\ga(b^2)b^{2-2b^2}\big)^{\frac{Q}{b}}\Upsilon(b)\,.
\end{equation}
This observation may be used to identify the CFT characterised by the 
three point function \rf{DOZZ} as Liouville theory. 
It follows from \rf{Lioudef} and \rf{EOM} that the classical limit 
of the field $\phi(z,\bz)$ will satisfy the classical  Liouville equation of motion.

\begin{rem} It had been conjectured in \cite{DO,ZZ} that the function defined
in \rf{DOZZ} with $N(\al)$, $C_0$ given in \rf{N-C} represents the 
three point function of Liouville theory. The paper \cite{ZZ} describes several
highly nontrivial checks of this proposal. 
The method described above for finding this result was introduced in \cite{T95}. 
Using the probabilistic framework for the construction of Liouville theory introduced in 
\cite{DKRV}, the method from \cite{T95} was recently used in the proof of 
\rf{DOZZ}, \rf{N-C}  proposed in \cite{KRV}.
A different method to derive \rf{DOZZ}, \rf{N-C}  
had previously  been outlined in \cite{T01,T03}. 
The conformal bootstrap program for Liouville field theory
was completed in \cite{T01} for genus zero, and in \cite{TV13} for 
Riemann surfaces $C$ of higher genus. 
\end{rem}

\subsection{Higher genus$^+$}

\subsubsection{Gluing construction}

In order to construct conformal blocks associated to Riemann surfaces of genus
larger than zero one needs a generalisation of the gluing construction. 
Geometrically one needs an operation that creates a surface $C$ with $n$
punctures and $g>0$ 
handles by identifying annular
neighborhoods of a pair of punctures $(P,P')$ on a single surface 
$\hat{C}$ of genus $g-1$ and $n+2$ punctures.
Let $t$ and $t'$ be coordinates in discs $D_P$ and $D_{P'}$ around 
$P$ and $P'$, respectively such that $t(P)=0=t'(P')$. By
identifying points $Q$ and $Q'$ in annular neighborhoods of $P$ and $P'$ 
which satisfy $t(Q)t'(Q')=q$ one may define a Riemann surface $C$ of
genus $g$ with $n$ punctures. The parameter $q$ represents one of the coordinates for the moduli 
space $\CM(C)$ of complex structures on the surface of interest.

This operation has a counterpart on the level of conformal blocks 
defined as follows. Let $\hat{f}$ be a conformal block associated 
to $\hat{C}$ with representations $R,R_1,\dots,R_n,R$ assigned
to the punctures $P,P_1,\dots,P_n,P'$, respectively. 
One may then define
\begin{equation}
f(v_1\otimes\dots\otimes v_n):=\sum_{\mu.\nu\in\CI_R}B^{\mu\nu}\,
\hat{f}( v_\mu\otimes v_1\otimes\dots\otimes v_n\otimes q^{L_0} v_\nu)\,
\end{equation}
where $\{v_{\mu};\mu\in \CI_R\}$ is a basis for the representation $R$, and
$B^{\mu\nu}$ are the matrix elements of the inverse of the matrix formed out of
$B_{\mu\nu}=\langle v_{\mu},v_{\nu}\rangle_R$, $\mu,\nu\in\CI_R$, 
with $\langle.,.\rangle_R$ being
the invariant bilinear form on $R$. It can be checked that
$f$ defines a conformal block on $C$.
This operation can be interpreted as taking the  trace 
$\tr_R^{}(q^{L_0}V({w}_{[n]}^{},\hat{C}))$ of the 
generalised chiral vertex operators $V({w}_{[n]}^{},\hat{C}):R\ra R$, 
${w}_{[n]}^{}\in R_1\otimes\dots \otimes R_n$,  
defined such that
\begin{equation}
\big\langle\, v\,,\,V({w}_{[n]}^{},\hat{C})\, v'\,\big\rangle_R^{}= 
\hat{f}( v \otimes {w}_{[n]}^{} \otimes v')\,,
\end{equation}
holds for all $v,v'\in R$, $
{w}_{[n]}^{}\in R_1\otimes\dots\otimes R_n$.

By using this generalised form of the gluing construction one can define, in much the same way as before,
families of conformal blocks associated to gluing patterns $\Ga$ coloured by the choices of
intermediate representations.

\subsubsection{Modular transformation}

Of particular importance is the case where $g=1$, where the definition above
reduces to taking traces of composites of chiral vertex operators. Writing
the parameter $q$ as $q=e^{2\pi i \tau}$, $\Im(\tau)>0$, 
one may represent $C=C_{1,1}^\tau$ as cylinder 
$\{w\in\mathbb{C};w\sim w+2\pi\}$ with additional identification 
$w\sim w+2\pi \tau$. We will use the notation
\begin{equation}
\chi_{R}^{R_1}(\tau):= \tr_R^{}\big(q^{L_0}V\big[{}_R{}^{R_1}{}_R\big](e_1,0)\big),
\end{equation} 
for the one-point torus conformal blocks with external representation $R_1$.

The conformal transformation $w\ra w'=-w/\tau$ maps $C_{1,1}^\tau$ to a torus $C_{1,1}^{-1/\tau}$ 
obtained 
from the cylinder by the additional identification 
$w\sim w-2\pi/\tau$. This transformation maps the $a$-cycle 
$\Im{w}=0$ and 
$b$-cycle $\{w\in\mathbb{C};w=2\pi\tau \vartheta;\vartheta\in[0,1]\}$
on  $C_{1,1}^\tau$ to 
$b$-cycle and $a$-cycle on $C_{1,1}^{-1/\tau}$, respectively.

It is quite remarkable that there exist linear relations 
between the conformal blocks $\chi_{R}^{R_1}(\tau)$ of the following general form
\begin{equation}\label{modtrsf}
\chi_{R}^{R_1}(\tau)= (-i\tau)^{-\De_{R_1}}
e^{\pi \mathrm{i} \frac{c}{12}(\tau+\frac{1}{\tau})}\sum_{R'}S_{RR'}^{R_1}  
\chi_{R'}^{R_1}(-1/\tau).
\end{equation}
These relations represent the modular transformation  $w\ra w'=-w/\tau$ 
on the level of the conformal blocks. The existence of such transformations
is known in the case of minimal models (see \cite{HL} for a guide to the relevant
mathematical literature), and in the case of Liouville theory \cite{HJS}.

\subsubsection{Modular invariance of physical correlation functions}

The physical correlation functions on higher genus surfaces may be represented in 
a form generalising the structure \rf{holofact2} in an obvious way. One thereby defines
(presumably convergent) series 
expansions for the generalisations
of the Schwinger functions associated to higher genus Riemann surfaces. Basic
physical consistency requirements are the existence of (real) analytic continuations over
the Teichm\"uller space of deformations of $C$ which is essentially\footnote{The relation between
partition functions defined from different gluing patterns may 
involve overall factors represented by the absolute value of the transition 
functions of the $c$-th power of the Hodge line bundle.} independent of the 
gluing pattern used in \rf{holofact2}.

It can be shown \cite{MS} that these requirements are satisfied if (i) crossing symmetry 
holds for genus zero correlation functions and (ii) if modular invariance holds for the 
partition functions $\CZ_v^{1,1}$ associated to the one-punctured torus,
\begin{equation}\label{modinv}
\CZ_v^{1,1}(\tau,\bar{\tau})= |\tau|^{-2\De_{R_1}}e^{- \frac{\pi c}{12}\Im(\tau+\frac{1}{\tau})} 
\CZ_v^{1,1}(-1/\tau,-1/\bar{\tau}),
\end{equation}
where here $v\in R_1$.
This condition is interesting already in the case where  $R_1$ is the vacuum representation,
corresponding to the torus $C_{1,0}$ without insertions. 
For the partition functions constructed from the representations 
in the Kac table there exists a complete classification of all modular invariant combinations of the 
form \rf{holofact2} with finite multiplicities \cite{CIZ}.

\subsubsection{Chiral modular bootstrap$^{++}$}\label{chirmod}

It can be shown that any two gluing patterns $\Ga_1$ and $\Ga_2$ can be related by sequences of 
fusion, braiding, and modular transformations \cite{MS,BK}. A transformation of the 
mapping class group, the fundamental group of the moduli space of Riemann surfaces 
$\CM(C)$, is obtained from these transformation by noting that the elements $\mu$ of the 
mapping class group can be realised as diffeomorphisms of the surface to itself, mapping
one gluing pattern $\Ga$ to another $\mu.\Ga$. Whenever one can 
realise the transition from $\Ga$ to $\mu.\Ga$
in terms of fusion, braiding, and modular transformations one may define a representation of the
mapping class group on the spaces of conformal blocks involved.
The matrices or integration
kernels representing these transformations therefore represent data of considerable interest
for CFT. 

Comparing with the discussion of higher genus conformal blocks in Section \ref{highergenus}
one may identify 
the conformal blocks defined by the gluing construction as sections of the bundle $\CV_c$ 
which are horizontal with respect to the canonical connection defined by the 
energy-momentum tensor. Fusion coefficients  $F^{\Ga_1\Ga_2}_{\be_1\be_2}$, the braiding factors appearing in 
\rf{braid}, and modular  transformation coefficients $S_{R_1R_2}^{R_0}$ 
represent the constant transition functions defining the projectively flat bundle $\CW_c$.
The 
prefactor $(-i\tau)^{-\De_{R_1}}e^{\pi \mathrm{i} \frac{c}{12}(\tau+\frac{1}{\tau})}$ in \rf{modtrsf}, on the other hand, 
represents a transition 
function of the particular projective line bundle $\CE_c$ relating 
$\CW_c$ to the vector bundle $\CV_c$ in the case of conformal blocks defined using the 
gluing construction, 
as was explicitly shown in \cite{TV13}. 

Having defined the flat vector bundle $\CV_c$ in terms of its transition functions allows us to characterise the conformal
blocks as horizontal sections of $\CV_c$. To find 
multivalued analytic functions on the moduli space $\CM(C)$ with given monodromies
is a mathematical problem of Riemann-Hilbert type. This Riemann-Hilbert problem 
provides an alternative characterisation of the 
relevant spaces of conformal blocks \cite{TV13}.

\subsection{Verlinde loop operators${}^{+}$}\label{verloop}

The spaces of conformal blocks on Riemann surfaces $C$ carry another important algebraic structure, the structure
as a module over a non-commutative algebra $\CA$ which can be identified as a quantum deformation
of the algebra of functions on the moduli space of flat $\mathrm{SL}(2,\BC)$-connections on $C$. 

Using the fact that fusion relations simplify when one of the representations
involved is a degenerate representation $V_{-b/2}$ or $V_{-1/2b}$
we will define certain operators acting on the spaces of conformal blocks defined 
by the gluing construction. The resulting additional algebraic structure on spaces of conformal blocks 
has turned out to be relevant in several applications, and it offers a useful 
alternative way to characterise the conformal blocks \cite{TV13}.

The definition of this structure is based on the comparison of the spaces of
conformal blocks on a Riemann surface $C$ to the spaces of conformal blocks
on a modified surface $\hat{C}=C\setminus\{y_0,y\}$ obtained by cutting out a disc $D$ from $C$ and re-gluing
a twice punctured disc $D\setminus\{y_0,y\}$ into $C\setminus D$. The representations associated to the 
punctures  at $y_0$ and $y$ will both be degenerate representations $V_{-b/2}$.
The gluing construction of conformal blocks on $C$ may easily be modified to provide a construction 
of conformal blocks on $\hat{C}$ such that one of the three-punctured spheres appearing in this 
construction contains $\hat{D}$. 
We had seen that in this case there exist only two
possibilities for the choice of representation on the boundary $\pa\hat{D}$ of $\hat{D}$, namely $V_0$
and $V_{-b}$. Fixing it to be the vacuum representation $V_0$, one obtains 
a space of conformal blocks from the 
gluing construction which is isomorphic to the space of conformal blocks on $C$ 
thanks to the propagation of vacua.  

It follows from the differential equations expressing the null vector decoupling that the 
conformal blocks are analytic in $y$ at least as long as $y\in\hat{D}$.
An analytic continuation of these conformal blocks to arbitrary 
$y\in C$ can be defined by a sequence of changes of pants decomposition, 
all described in terms of the
elementary fusion and braiding moves explicitly computed above. This allows us to 
consider the monodromies defined by analytic continuation of the conformal blocks
introduced  above along closed curves on $C$, starting and ending in $\hat{D}$.
It is not hard to see, and illustrated by the examples explicitly calculated in \cite{AGGTV,DGOT} 
that the conformal blocks obtained by this analytic continuation are linear combinations
of the conformal blocks associated to $\hat{C}$ one had started from. The
choices of intermediate representations may, however, get modified. If the representation
$V_{\be_e}$ was assigned to the  edge $e$ of the gluing pattern $\hat\Ga$ used to define 
conformal blocks on $\hat{C}$, one will get a linear combinations of the 
conformal blocks with representations $V_{\be_e+mb}$, $m\in\BZ$ assigned to $e$  as the result of
the analytic continuation. This means in particular that the result of the analytic
continuation will be a linear combination of conformal blocks defined by assigning  representations
$V_0$ and $V_{-b}$  to $\pa\hat{D}$, in general. 
An operator from the spaces of conformal blocks on $C$ to itself is obtained by projecting 
the result of the analytic continuation to the contribution having $V_0$ assigned to $\pa\hat{D}$.
The operator associated to a closed curve
$\ga$ on $C$ is called the Verlinde loop operator $\SL_{\ga}$.

The explicit computations of Verlinde loop operators
described in \cite{AGGTV,DGOT} identify the algebra $\CA$ generated by these operators
as a non-commutative deformation of the 
algebra of functions on the moduli space of flat $\mathrm{SL}(2,\BC)$-connections on $C$, with 
$\SL_\ga$ representing the quantised 
counterpart of the trace of the holonomy of a flat connection along $\ga$. 

It can be shown that this 
action characterises the conformal blocks of the Virasoro algebra essentially uniquely as solutions
to a natural Riemann-Hilbert problem defined in terms of the $\CA$-module structure \cite{TV13}.
Indeed, the construction described above defines Verlinde loop operators 
$\SL_{\Ga,\ga}$ for each pants decomposition $\Gamma$. Different choices of $\Gamma$ 
yield different representatives $\SL_{\Ga,\ga}$ for the generators 
$\CL_\ga$ of one and the same abstract algebra $\CA$. 
Fusion, braiding and the modular transformations define the intertwining 
operators $\SU_{\Gamma_2\Gamma_1}$ 
between the representations generated by the operators $\SL_{\Ga_1,\ga}$
and $\SL_{\Ga_2,\ga}$, satisfying 
$\SL_{\Ga_2,\ga}\SU_{\Gamma_2\Gamma_1}=\SU_{\Gamma_2\Gamma_1}
\SL_{\Gamma_1,\ga}$.
Under certain conditions one can argue that the 
fusion, braiding and the modular transformations are completely characterised by this
intertwining property. This is the case e.g. for Virasoro representation with $c=1+6Q^2\geq 25$, 
$\De=Q^2/4+P^2$, $P\in\BR$. In this case one can define a scalar product making the Verlinde 
operators self-adjoint. The fusion and the modular transformation may then be identified 
with the unitary operators mapping some of the operators $\SL_{\Ga,\ga}$ to diagonal 
form, explaining why they are essentially uniquely defined by this property.

Together with the discussion in Section \ref{chirmod} one may conclude that the spaces of physically relevant  conformal 
blocks are essentially determined by the representation theory of the algebra $\CA$ of quantised
functions on the moduli space of flat $\mathrm{SL}(2,\BC)$-connections on $C$. This can be
regarded as a dual topological characterisation of the spaces of conformal blocks 
that are relevant for physical applications. More details on this point of view can be found in \cite{TV13}.
A variant of this story will be discussed in Part \ref{Intpart} of these lectures in connection with integrability.

\subsection{Extended chiral symmetry$^{++}$}



As remarked above, there is little hope that we can solve CFTs having only Virasoro symmetry
exactly, in general. If, for example, one is considering a CFT with a Lagrangian formulation with 
many fundamental fields, the best one would generically hope for would be partial control over
certain composites of the fundamental fields thanks to Virasoro symmetry. If, however, Virasoro 
symmetry is extended to a symmetry under a larger chiral algebra, one regains hope that 
exact results may be within reach. The ``size'' of the symmetry must be in a reasonable proportion
to the ``complexity'' of the field content. Out goal in this subsection will be to indicate some directions
of research aiming at the solution of CFTs with higher chiral symmetries. 

It is fairly straightforward to generalise the definition and the gluing construction of conformal
blocks to cases with higher symmetries. This amounts to forming useful combinations of 
Virasoro conformal blocks having intermediate states related by the higher symmetries. 
It is furthermore not hard to formulate suitable generalisations of the crossing symmetry
and modular invariance conditions. An additional complication that will generically arise is 
due to the fact that the spaces of conformal blocks for higher chiral symmetries may have
dimensions higher than one already in the case $C=C_{0,3}$.
A first example where this happens is found in the 

\begin{quote}
{\small
{\bf Exercise 11:} Let $\CA=W_3$, the extension of $\mathsf{Vir}_c$ by generators $W_n$ with $\De=3$
introduced in Section \ref{ExtVOA}.                                                                                                                                                                                                                                                                                                                                                                                                                                                                                                                                                                                                                                                                                                                                                                                                                                                                                                                                                                                                                                                                                                                                                                                                      
Show that $\mathrm{dim}(\mathrm{CB}(C_{0,3},\CR))=\infty$ for generic representations
$\CR$.\\[1ex]
{\it Sketch of solution:} Use (generalised) 
Ward identities to compute the values of a conformal block
$f_{C_{0,3}}$ on arbitrary vectors $v\in\CR$ in terms of 
$f_{C_{0,3}}(e_1\ot e_2\ot (W_{-1})^k e_3)$. One may therefore define conformal blocks 
$f^l_{C_{0,3}}$ satisfying $f_{C_{0,3}}^l(e_1\ot e_2\ot (W_{-1})^k e_3)=\de_{k,l}$.
The conformal blocks obviously generate a (possibly overcomplete) basis 
for $\mathrm{CB}(C_{0,3},\CR)$. $\hfill\square$
}
\end{quote}
Additional features of the representations assigned to the punctures may 
make the dimension of the space of conformal blocks finite. Whenever 
$\mathrm{dim}(\mathrm{CB}(C_{0,3},\CR))$ is {\it not}
smaller or equal to one as is the case for the Virasoro algebra, one has to face additional 
complications in the bootstrap program.
For certain VOA one may nevertheless establish 
a picture reasonably close to the case of the Virasoro VOA. We shall now offer a few pointers
to the available literature.

\subsubsection{Relation to braided tensor categories}

As pointed out in Section \ref{CVO:sec} for the case of the 
Virasoro VOA, one may use the  Ward identities defining the 
conformal blocks to define a generalised notion of tensor product of representations
called fusion product. Generalisations of this notion can be defined for many  
VOAs.
Viewing the chiral vertex operators as generalisation of the Clebsch-Gordon maps intertwining 
the representation on tensor products with irreducible representations one is naturally
led to identify the coefficients of fusion transformations as analogs of the Racah-Wigner
$6j$-symbols, relating two natural ways of decomposing triple tensor products into
irreducible representations by first projecting tensor products of pairs of representations
onto irreducible ones.


A more abstract point of view has turned out to be useful. Given a VOA $\CA$ representing
an extension of conformal symmetry one may attempt to find categories of representations
on which the fusion product defined using the $\CA$-Ward identities is closed. 
One such category of representations of affine Lie algebras $\hat{\fg}_k$ at negative
levels $k$ was identified in the work of Kazhdan and Lusztig \cite{KL} to be 
the category denoted $\CO_k$ defined by certain finiteness conditions.

One needs to note that commutativity and associativity of the fusion product
are not manifest, in general. However, braiding and fusion transformation of 
conformal blocks can be used to define natural isomorphisms between 
(iterated) fusion products with different order of tensor factors, or different orders
in which tensor products are iterated. The isomorphisms defined in this way 
express the sense in which the fusion products are commutative and associative. 
The resulting structures can be formalised using the mathematical notion of a braided tensor 
category. This perspective has first been developed for the case of VOAs associated to 
affine algebras $\hat{\fg}_k$ at negative level in the work \cite{KL}. It was furthermore shown 
in \cite{KL} that the braided tensor category defined in this way is equivalent to a certain 
category of quantum group representations. As indicated above, this amounts to a complete 
characterisation of the fusion and braiding transformations for the conformal blocks defined from 
the category $\CO_k$ of representations of $\hat{\fg}_k$.

\subsubsection{Rationality and modular tensor categories}

Under certain finiteness conditions on the considered category of representations one can prove that there
exist modular transformations similar to \rf{modtrsf} among the conformal blocks of a VOA
associated to surfaces of genus one \cite{Zh}.
For such VOAs it may furthermore be shown that the 
spaces of conformal blocks associated to surfaces of {\it arbitrary} genus are finite-dimensional. 
Such CFTs are called rational CFTs (RCFTs).
Categories of representations of VOAs having this property are examples of 
what is called modular tensor
categories.
A guide to 
the literature on the mathematical foundations of this theory
and some non-rational generalisations 
can be found in \cite{HL}.
The book \cite{BK} develops 
a similar perspective, together with
mathematical background and relations to the theory of three-manifold invariants. 


This theory provides the basis for the powerful characterisation 
of the set of solutions to the conditions
of crossing symmetry and modular invariance characterising physical correlation functions 
in terms of the theory of modular tensor categories developed in \cite{FRS} 
and references therein.

\subsubsection{Seriously non-rational cases} 

From the point of view of theoretical physics, it seems unlikely that
the finiteness conditions characterising RCFTs are realised for generic CFTs.
Extending the existing theory for RCFTs to seriously non-rational CFTs like Liouville theory is an 
important mathematical problem. The results on
Liouville theory outlined above indicate that the development of such an extension
is not hopeless, after all. Indeed, one may view the mathematical 
structure defined by the known fusion and modular transformations for 
unitary representations of the  Virasoro algebra with $c>1$  as
an example for a natural non-rational 
generalisation of modular tensor categories \cite{TV13}. It seems very likely 
that there exists a vast class of conformal field theories of non-rational type which 
is insufficiently understood at the time of writing.





\section{Part 3: Relations to integrable models$^+$}\label{Intpart}
\setcounter{equation}{0}

There are various relations between CFT and the theory of integrable models, the main subject of this
Les Houches summer school. It won't be possible to discuss them all, so we will pick a particular
one, with a short guide to other relations between CFT and integrable models at
the end.  We have chosen to discuss the relation between CFT and the theory of isomonodromic  deformations
of ordinary differential equation on Riemann surfaces for the following reasons.
\begin{itemize}
\item The corresponding classically integrable
equations, especially the Painlev\'e VI equation, appear in many different problems of mathematical physics. 
The relation to CFT appears to 
be beneficial for the study of the isomonodromic deformation 
problem \cite{GIL}. 
\item There are 
interesting connections to $\CN=2$ supersymmetric
gauge theories in the context of the AGT-correspondence \cite{GIL,ILT}.
\item These connections feed back into the conformal bootstrap for $c=1$ \cite{ILTy}.
They may furthermore serve to illustrate many elements 
of the CFT formalism presented above in an interesting example.
\item This relation is part of a family of relations between CFT and the quantisation 
of moduli spaces of flat connections on Riemann surfaces, which reveal important aspects of the 
mathematical nature of CFT.
\end{itemize}
After a review of the relation between the isomonodromic deformation equations and the 
Riemann-Hilbert problem, we will in the following explain why solving the Riemann-Hilbert 
problem is equivalent to constructing  particular conformal blocks for the free fermion
super VOA. The isomonodromic 
tau-functions are thereby identified as chiral partition functions. This relation is 
a particular instance of the general relations between infinite Grassmannians, free fermions and integrable 
hierarchies reviewed and elaborated e.g. in \cite{SW}, and in the physics papers \cite{AGMV,Di}. 
It is then shown how the representation theory of the Virasoro algebra can be used to give a {\it constructive}
solution of the Riemann-Hilbert problem, basically by constructing 
the relevant conformal blocks for the free fermion VOA in terms of the conformal blocks 
for the Virasoro algebra. 
 Other relations between CFT and the theory
of integrable models are discussed at the end of this section.


\subsection{Isomonodromic deformations and tau-function}

The isomonodromic deformation problem is one of the 
most important classically integrable systems in mathematical physics.
It  arises naturally in the study
of flat connections on $C_{0,n}$. Any flat connection
on $C_{0,n}$ is gauge equivalent to a holomorphic connection of the
form $\pa_y-A(y)$,  with matrix-valued functions $A(y)$ of the form
\begin{equation}\label{Aform}
A(y)\,=\,\sum_{r=1}^n \frac{A_r}{y-z_r}\,.
\end{equation}
We will assume that  $A_1,\dots A_n$ satisfy $\sum_{k=1}^nA_k=0$.
Let us then consider the equation for parallel transport with respect to the connection $A(y)$,
\begin{equation}\label{horizontality}
\frac{\pa}{\pa y}\Psi(y)\,=\,A(y)\Psi(y)\,,
\end{equation}
In order to define a unique solution one may impose the initial condition
$\Psi(y_0)={\rm id}$ at a point $y_0\in C_{0,n}$.
Equation \rf{horizontality} determines the 
analytic continuation of a solution $\Psi(y)$ along closed paths $\ga_r$
on $C_{0,n}$, denoted as $\Psi(\ga_k^{}.y)$, to be of the form
\begin{equation}\label{monoact}
\Psi(\ga_r^{}.y)\,=\,\Psi(y)M_r\,, 
\end{equation}
with $y$-independent matrices  $M_r$ called monodromy matrices. 

The fundamental group $\pi_1$
of $C_{0,n}:=\BP^1\setminus\{z_1,\dots,z_n\}$
has $n$ generators $\ga_1^{},\dots,\ga_n^{}$ subject to one relation
$\ga_1^{}\circ\ga_2^{}\circ\dots\circ\ga_n^{}=1$. Having fixed an initial point $y_0$
allows us to represent the generators $\ga_r$ of $\pi_1$ by closed paths starting and
ending at $y_0$. The monodromies $M_r$ define
representations\footnote{Here understood as {\it anti}-homomorphisms
$\rho:\pi_1(C_{0,n})\ra G$} $\rho$ of
$\pi_1(C_{0,n})$ in $G={\rm GL}(N,\BC)$  if $A(y)$ is a family of $N\times N$-matrices. 
This means that
$M_k:=\rho(\ga_r^{})\in G$, $r=1,\dots,n$
satisfy the relation $M_n\cdot M_{n-1}\cdot \dots\cdot M_1=1$. 
A change of initial point $y_0$ results in an overall
conjugation with elements of $G$.

Variation of the positions
$z_r$ of the punctures on $C_{0,n}$, which appear as parameters in the 
differential equation \rf{horizontality}, will generically  modify the monodromies. It is, 
however, always possible to compensate the resulting changes by variations
of the matrix residues $A_r$ of $A(y)$.
We will see below that variations of the positions $z_r$ will not change the
monodromies of the connection $A(y)$ provided that the
matrix residues $A_k=A_k(z)$ satisfy the following equations,
\begin{equation}\label{preSchlesinger}
\begin{aligned}
&\pa_{z_k}A_k\,=\,-\sum_{l\neq k}
\frac{[A_k,A_l]}{z_k-z_l}\,,\\
&\pa_{z_l}A_k\,=\,\frac{y_0-z_k}{y_0-z_l}\frac{[A_k,A_l]}{z_k-z_l}\,,\quad
k\neq l\,,
\end{aligned}
\qquad \pa_{y_0}A_k\,=\,-\sum_{l\neq k}\frac{[A_l,A_k]}{y_0-z_l}\,.
\end{equation}
In the limit $y_0\ra\infty$ one finds the Schlesinger equations
\begin{equation}\label{Schlesinger}
\begin{aligned}
&\pa_{z_k}A_k\,=\,-\sum_{l\neq k}
\frac{[A_k,A_l]}{z_k-z_l}\,,\\
&\pa_{z_l}A_k\,=\,\frac{[A_k,A_l]}{z_k-z_l}\,,\quad
k\neq l\,.
\end{aligned}
\end{equation}
The Schlesinger equations are nonlinear partial differential equations for the 
matrices $A_r$. In special cases $n=4$ it is known that one may reduce 
these equations to the Painlev\'e VI-equation.

The
Schlesinger equations define Hamiltonian flows, generated by
the Hamiltonians
\begin{equation}
H_r:=\frac{1}{2}\;\underset{y=z_r}{\rm Res}\operatorname{tr}A^2(y)=\sum_{s\neq r}
\frac{{\rm tr}(A_rA_s)}{z_r-z_s}\,,
\end{equation}
using the Poisson structure
\begin{equation}
\big\{\,A\left(y\right)\,\substack{\otimes\vspace{-0.1cm} \\ ,}\,A\left(y'\right)\,\big\}\,=\,
 \left[\,\frac{\mathcal{P}}{y-y'}\,,\,A\left(y\right)\otimes 1+1\otimes A\left(y'\right)\,\right],
\end{equation}
where $\mathcal{P}$ denotes the permutation matrix.
The tau-function $\tau(\mathbf{z})$ is defined as the generating function for the
Hamiltonians $H_k$,
\begin{equation}\label{taudef}
H_k\,=\, \pa_{z_k}\log\tau(\mathbf{z})\,.
\end{equation}
Integrability of \rf{taudef} is ensured by the Schlesinger equations
\rf{Schlesinger}. The concept of a tau-function plays a key role in the
theory of many classically integrable models. 

\subsection{The Riemann-Hilbert problem}

Solving the Schlesinger equations is closely related to the Riemann-Hilbert
problem, as we shall now explain. 

We consider the cases where
the matrices $M_r$ are diagonalizable, $M_r=C_r^{-1}e^{2\pi \mathrm{i}D_r}C_r^{}$, for a fixed choice of
diagonal matrices $D_r$. 
The Riemann-Hilbert problem is to find a multivalued
analytic
matrix
function $\Psi(y)$ on $C_{0,n}$ such that the monodromy along
$\ga_r$ is represented by \rf{monoact}.
The solution to this problem is unique up to left multiplication with
single valued matrix functions. In order to fix this ambiguity we need to
specify the singular behaviour of $\Psi(y)$ at $y=z_r$, leading to the following
refined version of the Riemann-Hilbert problem: 
\begin{quote}
{\it Find a matrix
function $\Psi(y)$ such that the following conditions are satisfied.
\begin{itemize}
\item[i)] $\Psi(y)$ is a multivalued, analytic and invertible
on $C_{0,n}$, and sarisfies $\Psi(y_0)\,=\,1\,$.
\item[ii)] There exist neighborhoods of $z_k$, $k=1,\dots,n$ where
$\Psi(y)$ can be represented as
\begin{equation}\label{asym}
\Psi(y)\,=\,\hat{Y}^{(k)}(y)\cdot(y-z_k)^{D_k}\cdot C_k\,,
\end{equation}
with $\hat{Y}^{(k)}(y)$  holomorphic and invertible at $y=z_k$,
$C_k\in G$, and $D_k$ being diagonal  matrices for $k=1,\dots,n$.
\end{itemize}
If such a function $\Psi(y)$ exists, it is uniquely determined by the
monodromy data $\mathbf{C}=(C_1,\dots,C_n)$ and the diagonal matrices $\mathbf{D}=(D_1,\dots,D_n)$.
}\end{quote}
It is known that generic representations $\rho:\pi_1(C_{0,n})\ra
G$ can be realised as monodromy representation of such a
Fuchsian system, which means that a solution to the Riemann-Hilbert problem
formulated above will generically exist. We shall now briefly indicate how the Riemann-Hilbert
problem is related to the isomonodromic deformation problem.

Given a solution $\Psi(y)$
to the Riemann-Hilbert problem we may define a  connection $A_y(y)$
as
\begin{equation}
A_y(y):=\,(\pa_y \Psi(y))\cdot (\Psi(y))^{-1}\,,
\end{equation} 
It follows from ii) that $A_y(y)$ is a rational function of $y$ which has the form
\begin{equation}
A_y(y)\,=\,\sum_{r=1}^{n}\frac{A_r(z)}{y-z_r}\,.
\end{equation}
One may similarly deduce from ii) that the variations of $\Psi(y)$ with respect to $z_r$
can be represented in the form
\begin{equation}
A_r(y):=\,(\pa_{z_r} \Psi(y))\cdot (\Psi(y))^{-1}=-\frac{A_r(z)}{y-z_r}\,.
\end{equation}
Commutativity of the partial derivatives $\pa_y$ and $\pa_{z_r}$ acting on 
$\Psi(y)$ implies differential equations for the matrices $A_r$
which turn out to be the Schlesinger equations. This is how we get 
solutions to the isomonodromic deformation equations from solutions to the Riemann-Hilbert problem.

Given a solution to the  isomonodromic deformation  equations, one may, on the other hand, 
construct the connection $A(y)$ via \rf{Aform} and 
study the fundamental matrix solution
$\Psi(y)$ of the differential equation \rf{horizontality}
normalized
by $\Psi(y_0)=1$. If the eigenvalues  of
$A_k$ do not differ by integers, the resulting function $\Psi(y)$
will satisfy the conditions i) and ii) above for certain
matrices $C_1,\dots,C_n$ and diagonal matrices $D_1,\dots, D_n$. 

The monodromies $M_r$ of $\Psi(y)$
play the role of conserved quantities the existence of which is 
ensuring the integrability of the Schlesinger system 

\subsection{Relation to free fermion CFT and infinite Grassmannians}\label{sec:FF}

It is known for a while that there are deep relations between
classical  integrable models,  the geometry of infinite Grassmannians and
 the conformal field theory of free fermions. The relation to infinite Grassmannians is discussed in 
 the classic reference \cite{SW}, reviews of these relations together with the link to free fermions
 can be found e.g. in the physics papers \cite{AGMV,Di}.
We are now going to explain how  the CFT
formalism introduced above can be used to get a 
unified picture of these relations.

To this aim we are first going to show how to encode a solution $\Psi(y)$
to the Riemann Hilbert problem into the definition of conformal blocks for the free fermion 
VOA. Such conformal blocks will be represented by states ${w}_\Psi^{}$ satisfying 
a set of invariance conditions analogous to the conformal Ward identities. 

\subsubsection{Free fermions}
The free fermion super VOA is generated by fields $\psi_s(z)$, $\bar{\psi}_s(z)$, $s=1,\dots,N$,
The fields $\psi_s(z)$ will be arranged into a row vector $\psi(z)=(\psi_1(z),\dots,\psi_N(z))$, 
while $\bar\psi(z)$ will be our notation for the column vector with components $\bar{\psi}_s(z)$. 
The modes of $\psi(z)$ and $\bar\psi(z)$, introduced as
\begin{equation}
\psi(z)=\sum_{n\in\BZ}\psi_{n} z^{-n-1}\,,\quad
\bar\psi(z)=\sum_{n\in\BZ} \bar\psi_{n} z^{-n}\,,
\end{equation}
are row and column vectors 
with components $\psi_{s,n}$ and $\bar{\psi}_{s,n}$ satisfying the commutation relations
\begin{equation}
\{\,\psi_{s,n}\,,\,\bar\psi_{t,m}\,\}=\de_{s,t}\de_{n,-m}\,,\qquad
\{\,\psi_{s,n}\,,\,\psi_{t,m}\,\}=0\,,\qquad\{\,\bar\psi_{s,n}\,,\,\bar\psi_{t,m}\,\}=0\,.
\end{equation}
We will here consider a representation generated from a highest weight vector ${e}_0$ 
satisfying  
\begin{equation}
\psi_{s,n}\,{e}_0=0\,,\quad n\geq 0\,,\qquad
\bar\psi_{s,n}\,{e}_0=0\,,\quad n>0\,.
\end{equation}
The Fock space $\CF$ is generated from ${e}_0$ by the action of the modes $\psi_{s,n}$, $n<0$, 
and $\bar\psi_{s,m}$, $m\leq 0$.

\subsubsection{Free fermion conformal blocks from solutions to the Riemann-Hilbert problem}\label{FF-RH}

We are going to construct states ${w}_\Psi^{}$ in the 
dual $\CF'$ of the fermionic Fock space $\CF$, 
characterised by a set of Ward identities defined from a solution $\Psi(y)$ of the 
RH problem. 
Let us define the following 
infinite-dimensional spaces of multi-valued vector functions on $\tilde{C}_{0,n}$:
\begin{equation}\label{RbarRdef}
\begin{aligned}
&\mathcal{R}=\big\{\,v(y)\cdot\Psi(y);\;\,v(y)\in\mathbb{C}^N\otimes\mathbb{C}[\mathbb{P}^1\!\setminus\!\{\infty\}]\,\big\}\,,\\
&\bar{\mathcal{R}}=\big\{\,\Psi^{-1}(y)\cdot \bar{v}(y);\;\,\bar{v}(y)\in\mathbb{C}^N\otimes
\mathbb{C}[\mathbb{P}^1\!\setminus\!\{\infty\}]\,\big\}\,,
\end{aligned}
\end{equation}
where $v$ and $\bar{v}$ are row and column vectors with $N$ components, respectively, 
and $\mathbb{C}[\mathbb{P}^1\!\setminus\!\{\infty\}]$ is the space of meromorphic functions
on $\mathbb{P}^1$
having poles at $\infty$ only. The elements of the space $\CR$ represent solutions of
a generalisation of the  RH problem where the condition of regularity at infinity has been dropped.

The vectors ${w}_\Psi^{}$ we are about to define are
required to  satisfy the conditions
\begin{equation}\label{FFWard}
\psi[\bar{g}]\,{w}_\Psi^{}=0\,,\qquad
\bar{\psi}[g]\,{w}_\Psi^{}=0\,,
\end{equation} 
where $g\in\CR$, $\bar{g}\in\bar{\CR}$, and  the operators $\psi[\bar{g}]$ are constructed as 
\begin{equation}\label{psigdef}
\psi[\bar{g}]\,=\,\frac{1}{2\pi\mathrm{i}}\int_{\CC} dz\; \psi(z)\cdot \bar{g}(z)\,,\qquad
\bar{\psi}[g]\,=\,\frac{1}{2\pi\mathrm{i}}\int_{\CC} dz\; g(z)\cdot \bar{\psi}(z)\,,
\end{equation}
with $\CC$ being a circle separating $\infty$ from $z_1,\dots,z_n$.
One may recognise the identities \rf{FFWard} as analogs of the 
conformal Ward identities defining Virasoro conformal blocks where the role of 
the Virasoro algebra is taken by the free fermion VOA, and the role of the space of vector 
fields on $C$ is taken by the sets $\CR$ and $\bar{\CR}$. 
The definition of the state ${w}_\Psi^{}$ by means of the 
identities \rf{FFWard} can be considered as a fermionic 
analog of the one-point localisation discussed in Section \ref{OneptLoc} for the 
case of the Virasoro algebra.

It can easily be shown that the vector ${w}_\Psi^{}$ is defined uniquely up to 
normalisation by the identities \rf{FFWard}. Using
the notation $\langle {v},{w}\rangle_\CF$
for the pairing between a vector ${v}\in\CF$ and a vector  ${w}$ in the dual $\CF'$ of $\CF$.
this means  that
the identities \rf{FFWard} can be used to calculate the values of $\langle {v},{w}_\Psi^{}\rangle_\CF$
for ${w}_\Psi^{}\in\CF'$ satisfying \rf{FFWard} and arbitrary ${v}\in\CF$ 
in terms of $\langle {e}_0,{w}_\Psi^{}\rangle_\CF$. It is not hard to check that this is the case.
This implies that the space of conformal blocks 
for the free fermionic VOA is one-dimensional.

\subsubsection{Explicit construction of conformal blocks for the free fermion VOA}

Associated to a solution $\Psi(y)\equiv\Psi(y_0,y)$ to the Riemann-Hilbert problem as formulated above,
we have constructed a free fermion conformal block ${w}_\Psi^{}$. 
Let  $G(x,y)$ be the matrix which has the two-point function
\begin{equation}\label{FFtwopt}
\big\langle \,\bar{\psi}_s(x)\psi_t(y) \,\big\rangle_\Psi \equiv
\frac{\langle \,{e}_0\,,\,\bar{\psi}_s(x)\psi_t(y)\,{w}_\Psi^{}\,\rangle}{\langle \,
{e}_0\,,\,{w}_\Psi^{}\rangle},
\end{equation}
as its matrix element in row $s$ and column $t$. We are now going to show that $G(x,y)$
has a very simple relation to the solution $\Psi(x,y)$ of the Riemann-Hilbert problem, namely
\begin{equation}\label{GfromPsi}
G(x,y)=\frac{1}{x-y}\Psi(x,y)=\frac{\mathbb{I}}{x-y}+R(x,y)\,.
\end{equation}
In order to see this, let us expand $G(x,y)$ around $x=\infty$ and $y=\infty$, 
respectively:
\begin{equation}\label{G-exp}
\begin{aligned}
&G(x,y)=\sum_{l\geq 0} y^{-l-1} \bar{G}_l(x)\,\qquad \bar{G}_l(x)=  -x^l\,\mathbb{I}+\sum_{k>0}x^{-k}R_{kl}\,,\\
&G(x,y)=\sum_{k> 0} x^{-k} {G}_k(y)\,,\qquad {G}_k(y)= y^{k-1}\,\mathbb{I}+\sum_{l\geq 0}y^{-l-1}R_{kl}\,.
\end{aligned}
\end{equation}
Note that rows of the matrix-valued functions $G_k(y)$, $k>0$, and the columns 
of $\bar{G}_l(x)$, $l\geq 0$, defined in this way  generate 
a basis for the spaces $\CR$ and $\bar{\CR}$ introduced in \rf{RbarRdef}, respectively.
We claim that a vector ${w}_\Psi^{}$ satisfying \rf{FFWard} can be represented 
in terms of the expansion coefficients $R_{kl}$ introduced in \rf{G-exp} explicitly as
\begin{equation}
{w}_\Psi^{} 
=\exp\bigg(-\sum_{k>0}\sum_{l\geq 0} \psi_{-k}\cdot R_{kl}\cdot \bar{\psi}_{-l}\bigg) {e}_0\,.
\end{equation}
This can be verified by a straightforward computation. 
In a similar way one may furthermore check that the two-point function defined by this conformal block
is exactly the function $G(x,y)$ introduced in \rf{GfromPsi} above.

\subsubsection{Chiral partition functions as tau-functions}\label{taufromZ}

Out of the free fermion VOA one may define a representation of the Virasoro algebra
by introducing the energy-momentum tensor as
\begin{equation}\label{FFVir}
T(z)=\frac{1}{2}\lim_{w\ra 0}\sum_{s=1}^N\bigg(\pa_z\psi_s(w)\bar{\psi}_s(z)+\pa_z\bar{\psi}_s(w)\psi_s(z)+\frac{1}{(w-z)^2}\bigg)\,.
\end{equation}
Conformal blocks for the free fermion VOA represent conformal blocks for the 
Virasoro algebra defined via \rf{FFVir}.  We will use the conformal Ward identities 
to demonstrate that the isomonodromic tau-functions coincide with the chiral partition functions 
$\langle {e}_0,{w}_\Psi^{}\rangle$.

To this aim let us compare the Taylor-expansion of matrix-functions 
$\Psi(y)$ solving the Riemann-Hilbert problem to the one of the two-point function \rf{FFtwopt}.
We have, on the one hand,
\begin{equation}
\Psi(y)\,=\,{\rm id}+(y-y_0)J(y_0)+\frac{1}{2}(y-y_0)^2
\big[J^2(y_0)+\pa_yJ(y_0)\big]+\CO((y-y_0)^3)\,,
\end{equation}
where $J(y)=(\Psi(y))^{-1}\pa_y\Psi(y)=(\Psi(y))^{-1}A(y) \Psi(y)$. The 
residues of the trace part of $J^2(y)$ are the Schlesinger Hamiltonians, 
\begin{equation}
\frac{1}{2}\;\underset{y=z_r}{\rm Res}\,({\rm tr} J^2(y))=
\frac{1}{2}\;\underset{y=z_r}{\rm Res}\,({\rm tr}A^2(y))=\sum_{s\neq r}
\frac{{\rm tr}(A_rA_s)}{z_r-z_s}\,.
\end{equation}
Note, on the other hand, that the same expansion of $\Psi(y)$ around $y_0$ can be 
calculated using the OPE of the two fermionic fields in \rf{FFtwopt}. 
At second order in the expansion around $y_0$ one finds 
the energy-momentum tensor $T(y)$ in the trace part of the expansion of $G(x,y)$ around 
$x=y$. 
Using only the Ward identities one can show\footnote{In order to derive \rf{FFCWI} 
one may start by noting that it is built into the definitions that the fermion two point function 
permits an analytic continuation to the universal cover of $C_{0,n}$ with singularities of the 
form specified by \rf{asym}  only at $x=z_r$ or $y=z_r$, $r=1,\dots,n$. 
Using the fact that the energy-momentum tensor \rf{FFVir} appears in the 
trace part of the free fermion OPE one may define the expectation value $t(y)$ of the energy-momentum 
tensor, and show that it is holomorphic on $C_{0,n}$ with singular behaviour near $z_r$ of the form 
\rf{FFCWI}. This implies that the free fermion conformal block is a particular  conformal block on $C_{0,n}$ 
for  the Virasoro algebra defined via  \rf{FFVir}, represented in terms of the 
one-point localisation discussed in Section \ref{OneptLoc}. 
It follows from \rf{modFSconn} that
the residues of $t(y)$ get represented in terms the derivatives $\pa_{z_r}$. Working out the details 
of this argument is a good
exercise.}
 that 
\begin{equation}\label{FFCWI}
\big\langle\,{e}_0\,,\,T(y)\,
{w}_\Psi^{}\,\big\rangle_\CF^{}
=\sum_{r=1}^n \left(\frac{\De_r}{(y-z_r)^2}+\frac{1}{y-z_r}\pa_{z_r}\right)
\langle\,{e}_0,{w}_\Psi^{}\,\rangle_\CF^{},
\end{equation}
where $\De_r=\mathrm{tr}(D_r^2)$, with $D_r$ being the diagonal matrices introduced in \rf{asym}.

The definition of the tau-function $\tau(\mathbf{z})$ relates the Schlesinger-Hamiltonians $H_r$
to the derivatives of $\log\tau(\mathbf{z})$ with respect to $z_r$, $r=1,\dots,n$, while the
residues of the poles of the expectation value of  $T(y)$ near $y=z_r$ are identified with the 
derivatives of the chiral partition functions via the definition of the conformal blocks.
Comparison shows that we have $H_r=\pa_{z_r}\log\langle{e}_0,{w}_\Psi^{}\rangle$ implying $\tau(\mathbf{z})=
\langle{e}_0,{w}_\Psi^{}\rangle$, as
claimed.


In the one-point localisation discussed in Section \ref{OneptLoc} for the case of the Virasoro algebra
one identifies the 
derivatives with respect to $z_r$ with operators $L_{-1}^{(r)}$ acting on the vacuum representation 
which here gets replaced by the 
Fock space representation 
of the free fermion VOA. The isomonodromic deformation flows thereby get identified with 
commuting flows in the fermionic Fock space generated by commuting subalgebras of the 
Virasoro algebra. 

\subsubsection{Isomonodromic flows in infinite Grassmannians}

The space $\CR$ may be recognised as an element 
of the Grassmannian of the Hilbert space $\CH=L^2(S^1)\otimes \BC^N$, consisting of all closed 
subspaces $W$ of $\CH$ such that 
\[
\begin{aligned}
&\text{(i) the orthogonal projection $W\ra \CH_+$ is a Fredholm operator, and }\\
&\text{(ii) the orthogonal projection $W\ra \CH_-$ is a compact operator.}
\end{aligned}\hspace{3cm}
\]
A subspace $W$ can be defined by specifying a basis for $W$.  The set of all
subspaces $W$ satisfying the conditions above forms an infinite-dimensional manifold 
called the infinite Grassmannian.

Generalising the construction in Section \ref{FF-RH} slightly,
one gets a one-to-one correspondence between subspaces $W$ in the infinite Grassmannian
and states ${w}_W$ in the fermionic Fock space. Abelian subalgebras in the free fermion VOA
define commuting flows in the fermionic Fock space. 
The matrix  elements $\langle {e}_0,{w}_W\rangle_{\CF}^{}$ can be used to represent the tau-functions
of various integrable hierarchies, see e.g. \cite{AGMV,Di} for reviews.

In order to describe the isomonodromic deformation flows as flows in infinite Grassmannians  one should 
choose as subspaces $W\subset\CH$ the spaces $\CR$ considered above.

\subsubsection{Discussion}

Relations between the Riemann-Hilbert problem and the theory of free fermions
were first established in \cite{SMJ} based on an explicit construction of
fermionic twist fields. The relation between the 
vertex operator constructions of \cite{SMJ} and conformal field theory was
discussed in \cite{Mo} establishing the relation between tau-functions
and  expectation values of fermionic twist fields with the help of
the fermionic construction of the energy-momentum tensor.

The approach described above follows a somewhat different route: 
Assuming that a solution to the Riemann-Hilbert
problem exists, we have outlined a simple way to represent the iso\-mono\-dromic tau-function
as a fermionic matrix element using only the defining Ward-identities. This  explains in particular how the  
isomonodromic deformation problem fits into the general framework for integrable hierarchies
based on the  infinite Grassmannians \cite{SW}. Our approach
is also related to the work \cite{Pa}  identifying the isomonodromic tau-functions 
as determinants of certain Cauchy-Riemann operators. 

It remains to find more explicit and
computable descriptions of the state ${w}_\Psi^{}$. 
More recent versions
of the fermionic twist field construction \cite{ILT,GM,GL} fulfil this task, leading to  
explicit representations
for the tau-functions in the case  $N=2$. We will in the following describe 
the approach of \cite{ILT}, followed by a brief guide to other vertex operator
constructions.


\subsection{Trace coordinates for moduli spaces of flat connections}\label{Darboux}

As indicated above, 
we will temporarily restrict attention to the case $N=2$, and furthermore assume  that the 
connection $A(y)$ is traceless, which implies that its holonomies are elements of $G={\rm SL}(2,\BC)$.
The goal of this subsection is to introduce useful coordinates for the space of monodromy 
data in this case which will be used in the construction of a solution to the Riemann-Hilbert 
problem in Section \ref{sec_solRHP} below.
It will be fairly easy to remove the condition of vanishing trace later, while the generalisation 
of the construction described in Section \ref{sec_solRHP}
to $N>2$ represents an interesting open problem.

The Riemann-Hilbert correspondence between flat connections $\pa_y-A(y)$ and
representations $\rho:\pi_1(C_{0,n})\ra G$ discussed above relates 
the moduli space $\CM_{\rm flat}(C_{0,n})$ of flat
connections on $C_{0,n}$ to the so-called character
variety
$\CM_{\rm char}(C_{0,n})={\rm Hom}(\pi_1(C_{0,n}),{\rm SL}(2,\BC))/{\rm SL}(2,\BC)$.
Useful sets of coordinates for
$\CM_{\rm flat}(C_{0,n})$ are given by the trace functions
$L_{\ga}:=\operatorname{\rm tr}\rho(\ga)$ associated to 
simple closed curves $\ga$ on $C_{0,n}$. 
Minimal sets of trace functions that can be used to
parameterise $\CM_{\rm flat}(C_{0,n})$ can be identified using
pants decompositions. 

\begin{figure}[h]
\epsfxsize13.5cm
\centerline{\epsfbox{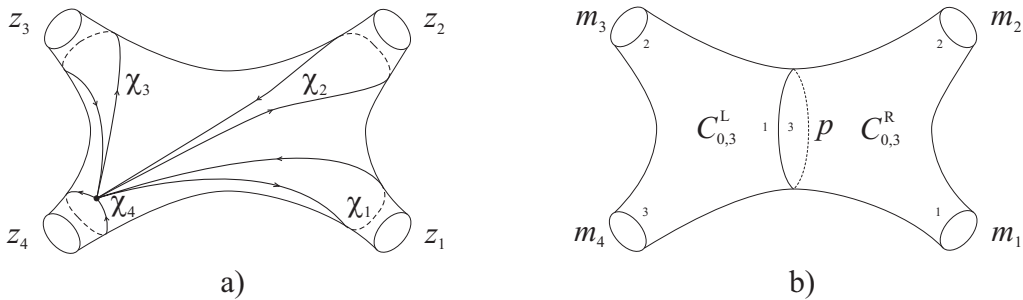}}
\caption{\it Basis of loops of $\pi_1(C_{0,4})$ and the decomposition $C_{0,4}=C_{0,3}^L\cup C_{0,3}^R$.}
\label{c04}\vspace{.3cm}
\end{figure}

To simplify the exposition, we shall restrict attention to the case $n=4$ 
in the following. By using pants decompositions one may easily generalise the 
following definitions to the cases with $n>4$.
Conjugacy classes of irreducible representations of $\pi_1(C_{0,4})$ are uniquely specified by
seven invariants
\begin{subequations}
\begin{align}\label{Mk}
&L_k=\operatorname{Tr} M_k=2\cos2\pi m_k,\qquad k=1,\ldots,4,\\
&L_s=\operatorname{Tr} M_1 M_2,\qquad L_t=\operatorname{Tr} M_1 M_3,\qquad L_u=\operatorname{Tr} M_2 M_3,
\end{align}
\end{subequations}
generating the algebra of invariant polynomial functions on $\CM_{\rm char}(C_{0,n})$. The monodromies $M_r$ are associated to the
curves $\ga_r$ depicted in Figure \ref{c04}. 
These trace functions satisfy the quartic equation
\begin{align}
 \label{JFR}
& L_1L_2L_3L_4+L_sL_tL_u+L_s^2+L_t^2+L_u^2+L_1^2+L_2^2+L_3^2+L_4^2=\\
 &\nonumber \quad=\left(L_1L_2+L_3L_4\right)L_s+\left(L_1L_3+L_2L_4\right)L_t
+\left(L_2L_3+L_1L_4\right)L_u+4.
\end{align}

The affine algebraic variety defined by (\ref{JFR}) is a concrete representation for
the character variety of $C_{0,4}$. For fixed choices of $m_1,\ldots,m_4$ in \rf{Mk}
one may use
equation \rf{JFR} to describe the character variety as a cubic surface in $\BC^3$.
This surface
admits a parameterisation in terms of coordinates $(\la,\kappa)$ of the form
\begin{align}\label{classWT}
L_s\,=\,2\cos 2\pi \lambda\,,\qquad
\begin{aligned}
& (\sin (2\pi  \lambda))^2\,L_t\,=\,
C_t^+(\lambda)\,e^{i\kappa}+C_t^0(\lambda)+C_t^-(\lambda)\,e^{-i\kappa}\,,\\
&  (\sin (2\pi  \lambda))^2\,L_u\,=\,
C_u^+(\lambda)\,e^{i\kappa}+C_u^0(\lambda)+C_u^-(\lambda)\,e^{-i\kappa}\,,
\end{aligned}
\end{align}
where
\begin{subequations}\label{Cepdef}
\begin{align}
C_u^{\pm}(\la)&=-4\prod_{s=\pm 1} \sin\pi(\la+s(m_1\mp m_2))\sin\pi(\la+s(m_3\mp m_4))\,,\\
C_u^0(\la)&={2 }\,
\big[\cos 2\pi  m_2 \cos 2\pi m_3 + \cos 2\pi  m_1 \cos 2\pi m_4\big]\\
&\quad- {2 \cos 2\pi  \la}
\big[\cos 2\pi  m_1 \cos 2\pi m_3 + \cos 2\pi  m_2 \cos 2\pi m_4\big]
\, .\notag
\end{align}
\end{subequations}
together with similar formulae for $C_t^{k}$, $k=\pm,0$.
Explicit formulae expressing the monodromy 
matrices $M_r=M_r(\la,\kappa)$, $r=1,\dots,4$, 
in terms of close relatives\footnote{The relation between our coordinates $(\la,\kappa)$ 
and the coordinates used in \cite{Ji} is given in \cite[Eqn. (6.69)]{ILT}.}
of our coordinates $(\la,\kappa)$
defined above can be found in \cite{Ji}.
These coordinates are also closely related to the coordinates used
in \cite{NRS}.

\subsection{Solving the Riemann-Hilbert problem using CFT}\label{sec_solRHP}

We will now describe how to solve the Riemann-Hilbert problem in terms of 
conformal blocks for the Virasoro-algebra $\mathsf{Vir}_c$ at $c=1$.
For the case $c=1$ we shall
replace the parameters $\al_r$ and $\be$ used above by variables $m_r$ and $p$ giving
the conformal dimensions as $\De_{m_k}=m_k^2$, for $r=1,\dots,4$, 
and $\De_{p}=p^2$. The chiral vertex operator
$V^{m}_{p_2,p_1}(z)$ maps $\CV_{p_1}$ to $\CV_{p_2}$. We shall use the 
notation 
$\mathsf{h}_{s}(y)$ for the chiral vertex operator associated to the degenerate representation 
$\CV_{\frac{1}{2}}$ which
maps $\CV_{p}$ to $\CV_{p-s/2}$, $s=\pm 1$, for all $p$. 

We will find it convenient to
assume that the vertex operators
$V^{m}_{p_2,p_1}(z)$ are normalized by 
\begin{equation}\label{normcond}
V^{m}_{p_2,p_1}(z)\,v_{p_1}\,=\,
N(p_2,m,p_1)\,z^{\De_{p_2}-\De_{p_1}-\De_m}\,
\big[\, v_{p_2}+\CO(z)\,\big]\,,
\end{equation}
with $N(p_3,p_2,p_1)$ being chosen as
\begin{align}\label{Ndef}
&N(p_3,p_2,p_1)\,=\,\\
&\;\;=\,\frac{G(1+p_3-p_2-p_1)G(1+p_1-p_3-p_2)G(1+p_2-p_1-p_3)G(1+p_3+p_2+p_1)}{G(1+2p_3)G(1-2p_2)G(1-2p_1)}\,,
\notag\end{align}
where $G(p)$ is the Barnes $G$-function that satisfies $G(p+1)=\Ga(p)G(p)$.
\begin{quote}
{\it Out of the vector 
\begin{subequations}\label{dualblocks}
\begin{equation}\label{FDdef}
{v}^{\rm\sst D}_{\epsilon}(\la,\kappa):=
\sum_{{n}\in\BZ}\,e^{i{n}\kappa}\;{v}^{}_{\epsilon}(\la+n)\,,\quad
{v}^{}_{\epsilon}(p):=
V^{m_3}_{m_4+\frac{\ep}{2},p}(z_3)\,V^{m_2}_{p,m_1}(z_2)\,V^{m_1}_{m_1,0}(z_1)\,e_{0}\,.
\end{equation}
let us define a) the matrix $\Psi(y;y_0)$ which has elements
\begin{equation}\label{Psi-constr}
\Psi_{s's}(y;y_0):=\,\frac{\pi s'(y_0-y)^{\frac12}}{\sin2\pi m_4}
\frac{\big\langle\,v_{m_{4}}\,,\,\mathsf{h}_{-s'}(y_0)\mathsf{h}_{s}(y)\,
{v}^{\rm\sst D}_{{s-s'}}(\la,\kappa)\,\big\rangle}
       {\big\langle\,v_{m_{4}}\,, \,{v}^{\rm\sst D}_0(\la,\kappa)\,\big\rangle}\,,
\end{equation}
and b) the function 
\begin{equation}\label{taufourier}
\tau(\,z\,)\,=\,
\big\langle\,v_{m_{4}}\,,\,{v}^{\rm\sst D}_0(\la,\kappa)\,\big\rangle\,.
\end{equation}
\end{subequations}
}\end{quote}
We claim that
$\Psi_{s's}(y;y_0)$ represents
the solution to the Riemann-Hilbert problem which has monodromy 
matrices $M_r(\la,\kappa)$ parameterised by the complex numbers
$(\la,\kappa)$,  
while $\tau(\mathbf{z})$ is the associated tau-function.
The proof of this statement is given in \cite{ILT}. 
This result yields in particular a proof of the relation between
the tau function for Painlev\'e VI and Virasoro conformal blocks
discovered in \cite{GIL}.
At this point we only remark that the prefactor in (\ref{Psi-constr})
ensures the normalization $\Psi(y_0;y_0)=1$.

In order to show that \rf{dualblocks} represents a solution to the 
Riemann-Hilbert problem we mainly need to compute the 
monodromies along the closed curves $\ga_r$.  It is straightforward to
check that the analytic continuation with respect to the variable $y$ can be 
represented in terms of a  composition of the fusion and braiding operations
introduced in Section \ref{fusbr}.
The main ingredient 
for carrying out this computation used in \cite{ILT} are the fusion relations
\rf{F-deg}. The value of the normalisation \rf{normcond} adopted above 
is that it turns all the Gamma-functions appearing in the relations \rf{F-deg}
into trigonometric functions.

While the details of this calculation are a bit technical, it is fairly easy to
understand the role of the Fourier-transformation in the definition \rf{FDdef}.
The relations between conformal blocks constructed from 
different gluing patterns imply in particular relations of the form 
\begin{equation}
\mathsf{h}_{-s_1}(y)\,V^{m}_{p_2-\frac{s_1}{2},p_1}(z_2)\,=\,\sum_{s_2=\pm}B_{s_1s_2}^{} 
V^{m}_{p_2,p_1-\frac{s_1}{2}}(z_2)\mathsf{h}_{s_2}(y)\,.
\end{equation}
Using such relations to compute the monodromy 
of the vector $\mathsf{h}_{s_2}(y){v}^{}_{\epsilon}(p)$
along the contour $\ga_u$
encircling $z_2$ and $z_3$
one finds  a result of the form 
\[
\mathsf{h}_{s_1}(\ga_u.y)\,{v}^{}_{\epsilon+s_1\frac{1}{2}}(p)\,=\,\sum_{s_2=\pm}
M_{s_1s_2}(p,\mathsf{T}_p)\,\mathsf{h}_{s_2}(y)\,{v}^{}_{\epsilon+s_2\frac{1}{2}}(p)\,,
\]
where $\mathsf{T}_p$ is the shift operator acting as
$\mathsf{T}_p{v}^{}_{\epsilon}(p)={v}^{}_{\epsilon}(p+1)$. 
This means that the monodromy matrices of degenerate fields
are {\it operator-valued}, acting nontrivially on the spaces of conformal blocks.
The Fourier-transformation in \rf{FDdef} {diagonalises} $\mathsf{T}_p$. It may 
furthermore be checked
that the traces of the monodromy matrices 
depend on $\la$ only via $e^{2\pi \mathrm{i}\la}$,
which is unaffected by the Fourier transformation. We conclude that the  
Fourier-transformation in \rf{FDdef} maps the operatorial realisation of the 
trace functions on spaces of conformal blocks to the classical expressions \rf{classWT}
used to define the coordinates $(\la,\kappa)$. 

By means of an argument very similar to the one described in Section \ref{taufromZ} 
above, one may then
show that the function $\tau(\mathbf{z})$ defined in \rf{taufourier}
represents the isomonodromic tau-function. A completely different method to prove this
result was presented in \cite{BS}.

\subsection{Non-abelian fermionisation}

We will now see how the free fermionic representation of tau-functions described in Section \ref{sec:FF}
is related to a simple generalisation of the bosonic  construction in Section \ref{sec_solRHP}.

To this aim let us introduce an additional  free field $\vf_0$,
\[
\vf_0(w)\vf_0(z)\,\sim\,-\frac{1}{2}\log(w-z)\,.
\]
Note furthermore that we have
\begin{equation}
\De_{-b/2}\big|_{b=i}\,=\,\frac{1}{4}\,,\qquad
\De_{-b}\big|_{b=i}\,=\,1\,.
\end{equation}
Let us construct the fields
\begin{equation}
\psi_s(z):=\,e^{i\vf_0(z)}\mathsf{h}_s'(z),\qquad
\bar{\psi}_s(z):=s\,e^{-i\vf_0(z)}\mathsf{h}_{-s}'(z),\qquad s=\pm 1,
\end{equation}
where $\mathsf{h}'_s(z)$ are related to the fields $\mathsf{h}_s(z)$ by a change of normalisation
\begin{equation}
\mathsf{h}_+'(z)=\mathsf{h}_+(z),\qquad
\mathsf{h}_-'(z)=\frac{\pi}{\sin 2\pi \mathsf{p}} \mathsf{h}_-(z)\,,
\end{equation}
with $\mathsf{p}$ satisfying $\mathsf{p} v_{p}=  p v_p$ for all $v_p\in \CV_p$.
The  fields  $\psi_s(z)$, $\bar{\psi}_s(z)$ have the free fermion OPE
\begin{subequations}
\label{FF-OPE}
\begin{align}
\psi_s(w)\psi_{s'}(z)\,& \sim\,{\rm regular}\,,\\
\psi_s(w)\bar\psi_{s'}(z)\,& \sim\,\frac{\de_{s,s'}}{w-z}\,.
\end{align}
\end{subequations}
This means that the fields $\psi_s(w)$, $\bar\psi_s(w)$ generate
a representation of the fermionic VOA.

A straightforward generalisation of the construction described in Section \ref{sec_solRHP}
will allow us to represent the solution of the  Riemann-Hilbert problem for monodromies in $\mathrm{GL}(2)$
in the form 
\begin{equation}
\Psi_{s's}(y;y_0):=\,(y_0-y)
\frac{\langle\,v_{0}\,,\,\bar{\psi}_{s'}(y_0)\psi_{s}(y)\,
{w}_\Psi^{}\,\rangle_\CF}
 {\langle\,v_0\,, \,{w}_\Psi^{}\,\rangle_\CF}\,,
\end{equation}
where $w_\Psi$ is a state in the fermionic Fock space 
which can be represented in bosonised form
as the tensor product 
of the state defined in \rf{FDdef} with a state in the Fock space generated by the modes of $\vf_0$
created by a product of exponential vertex operators.

This furnishes a more explicit description of the state  $w_\Psi$ 
we had associated  in Section \ref{sec:FF} to a solution $\Psi(y)$ of the 
Riemann-Hilbert problem. 
One may, in particular,  use the explicit formulae for the series expansions
of Virasoro conformal blocks provided by the AGT-correspondence \cite{AGT,AFLT} to obtain very detailed
information on the isomonodromic tau-functions that had previously been unknown, as illustrated
by the examples studied in \cite{GIL}.

Another approach was recently described in \cite{GM}. 
A generalisation of the constructions presented in Section \ref{sec:FF} can be used to 
construct fermionic twist fields from the solutions of the 
Riemann-Hilbert problem on the three-punctured sphere. The state $w_\Psi$  can then
be created from the vacuum by a composition of these fermionic twist fields. 
A similar construction was used in \cite{GL} to prove 
the formulae for the isomonodromic tau-functions furnished by the 
AGT-correspondence without explicit reference to conformal field theory.

\subsection{Implications for bosonic CFT} \label{Impl}

Going back to the Schlesinger equations (\ref{Schlesinger}), note that the 
three points $z_1$, $z_3$, $z_4$ can be mapped to $0$, $1$ and $\infty$ using
M\"obius transformations. The Schlesinger system then reduces to 
the Painlev\'e VI equation
\begin{align}\label{PVVI}
&-\frac12\left(z(z-1)\zeta''\right)^2=\\
&\nonumber=\operatorname{det}
\left(\begin{array}{ccc}
2m_1^2 & z\zeta'-\zeta & \zeta'+m_1^2+m_2^2+m_3^2-m_4^2 \\
z\zeta'-\zeta & 2m_2^2 & (z-1)\zeta'-\zeta \\
\zeta'+m_1^2+m_2^2+m_3^2-m_4^2 & (z-1)\zeta'-\zeta & 2m_3^2
\end{array}\right),
\end{align}
satisfied by the logarithmic derivative of the tau function
\begin{equation}
\zeta(z)=\displaystyle z (z-1)\frac{d}{dz}\ln \tau.
\end{equation}

The representation (\ref{taufourier})
of $\tau(\mathbf{z})\equiv \tau_s(z)$ as a Fourier transform of the $c=1$ Virasoro
conformal block can be written as
\begin{equation}
\label{tauc04}
\tau_s(z)=\sum_{n\in\BZ}\langle\, v_{m_4}\,,\,V_{m_4,\la_s+n}^{m_3}\left(1\right)V_{\la_s+n,m_1}^{m_2}\left(z\right)\,e_{m_1}\,\rangle\,e^{in\kappa_s}.
\end{equation}
We have 
renamed $(\la,\kappa)$ into $(\la_s,\kappa_s)$ in order to indicate the gluing 
pattern on which the definition of these coordinates was based.
Assuming without loss of generality that $-\frac12<{\rm Re}(\la)<\frac12$, 
and taking into account the normalization (\ref{Ndef}) of the chiral vertex operators,
we may deduce the asymptotic behaviour 
\begin{align}
\nonumber\tau_s(z)=&\sum_{n=0,\pm1}N(m_4,m_3,\la_s+n)
N(\la_s+n,m_2,m_1)\,e^{in\kappa_s}\,
z^{(\la_s+n)^2-m_1^2-m_2^2}\\
&+O\big(z^{\la_s^2-m_1^2-m_2^2+1}\big).
\label{Jimbo}\end{align}
This is equivalent to Jimbo's 
asymptotic formula \cite[Theorem~1.1]{Ji} expressing the asymptotic
behavior of the Painlev\'e VI tau-function in terms of monodromy data. 

It is perfectly possible, of course, to replace the construction \rf{Psi-constr} based on the 
gluing pattern $\Gamma_s$ by a similar construction using the gluing pattern
$\Gamma_u$. The definition of the tau-function $\tau_u(z)$ 
associated with the gluing pattern
$\Gamma_u$ has asymptotics of the form 
\begin{align}
\nonumber\tau_u(z)=&\sum_{n=0,\pm1}N(m_4,\la_u+n,m_1)N(\la_u+n,m_3,m_2)\,
e^{in\kappa_u}\,(1-z)^{(\la_u+n)^2-m_2^2-m_3^2}\\
&+O\big((1-z)^{\la_u^2-m_2^2-m_3^2+1}\big),
\label{Jimbo-u}\end{align}
expressed in terms of coordinates $(\la_u,\kappa_u)$ defined by 
expressions similar to \rf{classWT}, but with
trace function $L_u$ now represented simply as $L_u=2\cos(2\pi \la_u)$.
Higher order terms
in the expansion \rf{Jimbo-u} are uniquely
determined by the Painlev\'e equation \rf{PVVI}.

The pairs $(\la_s,\kappa_s)$ and $(\la_u,\kappa_u)$ are two different sets of 
coordinates for the character variety, and must therefore be related by a change of
coordinates.\footnote{Beautiful descriptions of this change of coordinates can be found
in \cite{NRS,ILTy}.} For given monodromy data $\mu$ one may use the differential equation 
\rf{PVVI} to determine the tau-function uniquely up to a multiplicative 
constant. It follows that
the tau-functions $\tau_s(z;\mu)$ and $\tau_u(z;\mu)$ associated to the same monodromy 
data $\mu$ must coincide up to a multiplicative constant,
\begin{equation}\label{taufusion}
\tau_s(z,\mu)\,=\,F(\mu)\tau_u(z,\mu)\,.
\end{equation}
An explanation for this 
observation is offered by the relation to the conformal 
blocks of the free fermion VOA described above. As the spaces of free fermion 
and free boson conformal blocks are 
one-dimensional, the fusion relations for these conformal blocks simplify to the form \rf{taufusion}.
The factor of proportionality $F(\mu)$ in \rf{taufusion} is independent of $z$, but 
depends in a highly nontrivial way
on the monodromy data $\mu$. An explicit 
formula for $F(\mu)$ has been proposed in 
\cite{ILTy} and a proof was given in \cite{ILP}.

This result has interesting consequences for the theory of conformal 
blocks of the Virasoro algebra at $c=1$.
Note that the Fourier-transformations used to 
construct $\tau_s(z;\mu)$ and $\tau_u(z;\mu)$ out of conformal blocks can be 
regarded as changes of basis in the space of conformal blocks. 
Equation \rf{taufusion} expresses the fact that the fusion transformation becomes
{\it diagonal} in the basis defined by the Fourier-transformations.
It is possible to invert the  Fourier-transformation in \rf{tauc04}, allowing the
authors of \cite{ILTy} to compute the fusion transformations of 
Virasoro conformal blocks at $c=1$ explicitly.

It would be very interesting if these remarkable 
results could be used to demonstrate the crossing 
symmetry of the correlation functions of the $c=1$ generalised minimal model
defined by the three point functions \rf{DOZZm}
analytically.


\subsection{Further relations to integrable models$^{++}$}

Apart from the relations between CFT and the isomondromic deformation problem 
described above, there 
exist further important relations to quantum integrable models. We want to indicate some of them
here.

\subsubsection{Semiclassical Limit of Virasoro conformal blocks and the Garnier system} 

It seems interesting to note that conformal field theory is related to the isomonodromic deformation
problem in the classical limit $c\ra\infty$. 

The Schlesinger system has an alternative description 
known as the Garnier system 
describing  the isomonodromic deformations of the 
second order ODE  naturally associated to the matrix differential equation $(\pa_y-A(y))s=0$. 
We refer to \cite{IKSY} for a review and further references.

As shown in \cite{T17a} one may 
describe the leading asymptotics of Virasoro conformal blocks on $C_{0,n}$ with $n-3$  
insertions of degenerate representations in terms of the generating function 
for a change of coordinates between two natural sets of 
Darboux coordinates for the Garnier system. One set of coordinates is  natural
for the Hamiltonian  formulation of the Garnier system  \cite{IKSY}, the other coordinates 
are the complex Fenchel-Nielsen
coordinates for $\CM_{\rm flat}(C_{0,n})$ introduced in Section \ref{Darboux}. 
The result of \cite{T17a} characterises the leading classical asymptotics of 
Virasoro conformal blocks completely and clarifies in which sense conformal field theory
represents a quantisation of the isomonodromic deformation problem.

\subsubsection{Yang's functions for the Hitchin systems}

The Yang's function was introduced in \cite{YY} as a useful potential for the Bethe ansatz
equations in a certain integrable system. More recently it was realised in \cite{NS} that the 
quantisation conditions in much larger classes of integrable systems can be described in terms
of potentials generalising the Yang's function.  An  large class of classically integrable systems 
is provided by the Hitchin systems introduced in \cite{Hi} associated to the  choice of a Riemann surface $C$ 
and a Lie algebra $\fg$. The Hamiltonians of the Hitchin system have been 
quantised in the work of Beilinson and Drinfeld on the geometric Langlands program
using methods from conformal field theory. A review 
can be found in \cite{Fr}.

More recently it was realised in \cite{T10} that the Yang's functions for the 
Hitchin systems associated to $\fg=\mathfrak{sl}_2$ are given by the 
$c\ra\infty$ limits of the Virasoro conformal blocks associated to the Riemann surface $C$. 
This leads to a geometric characterisation of the Yang's function as generating
function of the variety of opers within $\CM_{\rm flat}(C)$, as independently proposed in \cite{NRS}.
There exist natural 
quantisation conditions for the quantum Hitchin system 
which can be re-expressed in terms of the Yang's function
\cite{T17b}.

\subsubsection{Integrable structure of conformal field theories}

Important relations between conformal field theory and 
quantum integrable models follow from the observation originally made in
\cite{SY,EY} and later generalised in \cite{FF} that large
abelian subalgebras can be defined as the subspaces of various VOAs 
generated by the elements commuting with a suitable set of screening operators.
These screening operators can be identified with the interaction terms of perturbed CFTs
in the light-cone representation.  The  elements 
commuting with the screening charges thereby get related to the  
conserved quantities in integrable perturbed CFTs. One may
in this sense regard the infinite-dimensional abelian algebras generated by the 
commuting charges of integrable perturbed CFT as the 
remnant of the conformal symmetry surviving integrable perturbations. 

This point of view was developed much further in the series of papers \cite{BLZ}
by using techniques from the theory of quantum integrable models to construct
the so-called T- and Q-operators, 
generating functions for the local- and non-local integrals of motion, combined with profound 
studies of their analytic properties.

We see that conformal field theory has various relations to both classical and quantum integrable models both
at finite central charge $c$ and in the limit $c\ra\infty$. These relations continue to 
represent an attractive topic of research in mathematical physics.

{\bf Acknowledgements.}
The author thanks the organisers of the Les Houches summer school for the invitation, and
for setting up an inspiring event. 

Special thanks to M. Alim, A. Balasubramanian and O. Lisovyy for comments on the draft.

This work was supported by the Deutsche Forschungsgemeinschaft (DFG) through the 
collaborative Research Centre SFB 676 ``Particles, Strings and the Early Universe'', project 
A10.


\newpage

\begin{thebibliography}{9999}

\newcommand{\CMP}[3]{{ Commun. Math. Phys. }{\bf #1} (#2) #3}
\newcommand{\LMP}[3]{{ Lett. Math. Phys. }{\bf #1} (#2) #3}
\newcommand{\IMP}[3]{{ Int. J. Mod. Phys. }{\bf A#1} (#2) #3}
\newcommand{\NP}[3]{{ Nucl. Phys. }{\bf B#1} (#2) #3}
\newcommand{\PL}[3]{{ Phys. Lett. }{\bf B#1} (#2) #3}
\newcommand{\MPL}[3]{{ Mod. Phys. Lett. }{\bf A#1} (#2) #3}
\newcommand{\PRL}[3]{{ Phys. Rev. Lett. }{\bf #1} (#2) #3}
\newcommand{\AP}[3]{{ Ann. Phys. (N.Y.) }{\bf #1} (#2) #3}
\newcommand{\LMJ}[3]{{ Leningrad Math. J. }{\bf #1} (#2) #3}
\newcommand{\FAA}[3]{{ Funct. Anal. Appl. }{\bf #1} (#2) #3}
\newcommand{\TMP}[3]{{ Theor. Math. Phys. }{\bf #1} (#2) #3}
\newcommand{\PTPS}[3]{{ Progr. Theor. Phys. Suppl. }{\bf #1} (#2) #3}
\newcommand{\LMN}[3]{{ Lecture Notes in Mathematics }{\bf #1} (#2) #2}
\small  \setlength{\itemsep}{-3pt}






\bibitem[AFLT]{AFLT}
V.A. Alba, V.A. Fateev, A.V. Litvinov, G.M. Tarnopolsky,
{\it On combinatorial expansion of the conformal blocks arising from AGT conjecture},
Lett. Math. Phys. {\bf 98} (2011) 33--64.

\bibitem[AGGTV]{AGGTV}
L. F. Alday, D. Gaiotto, S. Gukov, Y. Tachikawa, H. Verlinde,
{\em Loop and surface operators in $\mathcal{N}=2$ gauge theory and
Liouville modular geometry}, J. High Energy Phys. {\bf 1001} (2010) 113.



\bibitem[AGT]{AGT}
L.~F.~Alday, D.~Gaiotto, and Y.~Tachikawa,
{\em Liouville Correlation Functions from Four-dimensional Gauge Theories},
Lett. Math. Phys. {\bf 91} (2010) 167--197.



\bibitem[AGMV]{AGMV}
L. Alvarez-Gaume, C. Gomez, G. Moore, C. Vafa,
{\it Strings in the Operator Formalism}, 
Nucl. Phys. {\bf B303} (1988) 455-521. 


\bibitem[BK1]{BK} B. Bakalov, A. Kirillov, Jr., 
{ \it On the Lego-Teichm\"uller game.}  
Transform. Groups  {\bf 5}  (2000) 207--244.

\bibitem[BK2]{BK2} B. Bakalov, A. Kirillov, Jr., 
{\it Lectures on tensor categories and modular functor},
Univ. Lecture Ser. {\bf 21}, American Mathematical Society, 
Providence, RI, 2001.


\bibitem[BLZ]{BLZ} 
V. V. Bazhanov, S. L. Lukyanov, A. B. Zamolodchikov,
{\it Integrable structure of conformal field theory},
Part 1: Comm. Math. Phys. {\bf 177} (1996) 381-398,
Part 2: Comm. Math. Phys. {\bf 190} (1997) 247-278,
Part 3: Comm. Math. Phys. {\bf 200} (1999) 297-324.

\bibitem[BPZ]{BPZ} A.A.
Belavin, A.M. Polyakov, A.B. Zamolodchikov,
{\it Infinite conformal symmetry in
two-dimensional quantum field theory}. Nucl. Phys. {\bf B241} (1984) 333--380.

\bibitem[BS]{BS} M. A. Bershtein, A. I. Shchechkin,
{\it Bilinear Equations on Painlev\'e $\tau$ Functions from CFT},
Comm. Math. Phys. {\bf 339} (2015) 1021--1061.

\bibitem[Bo]{Bo}
R. E. Borcherds, 
{\it Vertex algebras, Kac-Moody algebras, and the Monster}, 
Proc. Natl. Acad. Sci. U.S.A. {\bf 83} (1986), 3068--3071.

\bibitem[CCY]{CCY}
M. Cho, S. Collier, X. Yin,
{\it Recursive Representations of Arbitrary Virasoro Conformal Blocks},
Preprint arXiv:1703.09805. 

\bibitem[CIZ]{CIZ} A. Cappelli, C. Itzykson, J.-B. Zuber,
{\it The A-D-E classification of minimal and $A^{(1)}_1$ conformal invariant theories},
Comm. Math. Phys. {\bf 113} (1987) 1-26.

\bibitem[CKLW]{CKLW}
S. Carpi, Y. Kawahigashi, R. Longo, M. Weiner,
{\it From vertex operator algebras to conformal nets and back},
Preprint arXiv:1503.01260.

\bibitem[CKLY]{CKLY}
S. Collier, P. Kravchuk, Y.-H. Lin, X. Yin,
{\it Bootstrapping the Spectral Function: On the Uniqueness of Liouville and the Universality of BTZ},
Preprint  arXiv:1702.00423.

\bibitem[DKRV]{DKRV}
F. David, A. Kupiainen, R. Rhodes, V. Vargas,
{\it Liouville Quantum Gravity on the Riemann sphere},
Comm. Math. Phys. {\bf 342} (2016) 869--907.


\bibitem[Di]{Di}
R. Dijkgraaf,
{\it Intersection theory, integrable hierarchies and topological field theory},
NATO Sci. Ser. B {\bf 295} (1992) 95-158. 

\bibitem[DF]{DF}
V. Dotsenko,  V. Fateev. 
{\it Four point correlation functions and the operator algebra in the two-dimensional conformal invariant theories with the central charge $c<1$}, 
Nucl.Phys. {\bf B251} (1985) 691--734.

\bibitem[DGOT]{DGOT} N. Drukker, J. Gomis, T. Okuda, J. Teschner,
{\em Gauge Theory Loop Operators and Liouville Theory},
J. High Energy Phys. {\bf 1002} (2010) 057.




\bibitem[DO]{DO} H.
Dorn, H.-J. Otto, {\it Two and three-point functions in Liouville theory}. 
Nucl. Phys. B {\bf 429} (1994) 375--388.

\bibitem[Du]{Du} J. Dubedat, {\it SLE and Virasoro Representations: Localization},
Comm. Math. Phys. {\bf 336} 
(2015) 695--760.

\bibitem[EY]{EY}
T. Eguchi, S.K.  Yang, {\it Deformation of conformal field theories and soliton equations}, Phys. Lett. {\bf B224} (1989) 373--378.

\bibitem[FF]{FF} B. Feigin, E. Frenkel,	
{\it Free field resolutions in affine Toda field theories}, 
Phys. Lett. {\bf B276} (1992) 79--86. 


\bibitem[FFK1]{FFK1} G. Felder, J. Fr\"ohlich,  G. Keller,
{\it On the structure of unitary conformal field theory. I. Existence of conformal blocks},
Comm. Math. Phys. {\bf 124} (1989) 417-463.

\bibitem[FFK2]{FFK2} G. Felder, J. Fr\"ohlich,  G. Keller,
{\it On the structure of unitary conformal field theory. II. Representation-theoretic approach},
Comm. Math. Phys. {\bf 130} (1990) 1-49.

\bibitem[FB]{FB} E. Frenkel, D. Ben-Zvi,
{\it Vertex algebras and algebraic curves.} Second edition. 
Mathematical Surveys and Monographs, {\bf 88}. 
American Mathematical Society, Providence, RI, 2004.

\bibitem[Fr]{Fr}
E. Frenkel, {\it Lectures on the Langlands Program and Conformal Field Theory}, 
in: Frontiers in Number Theory, Physics, and Geometry II (P. Cartier, P. Moussa, B. Julia, and
P. Vanhove, eds.), pp. 387--533. Springer, 2007.

\bibitem[FLM]{FLM} I. Frenkel, J. Lepowsky, A. Meurman,
{\it Vertex Operator Algebras and the Monster},
Academic Press, Boston, 1989.

\bibitem[FS]{FS} D. Friedan, S. Shenker,
{\it The Analytic Geometry of Two-Dimensional Conformal Field Theory},
Nucl. Phys. {\bf B281} (1987) 509--545.

\bibitem[FuRS]{FRS}
J.Fuchs, I.Runkel, C. Schweigert,
{\it TFT construction of RCFT correlators IV: Structure constants and correlation functions}, 
Nucl. Phys. {\bf B715} (2005) 539-638.

\bibitem[GIL]{GIL}
O. Gamayun, N. Iorgov, O. Lisovyy,
{\it Conformal field theory of Painlev\'e VI},
JHEP {\bf 10} (2012) 038. 

\bibitem[GL]{GL}
P. Gavrylenko, O. Lisovyy,
{\it Fredholm determinant and Nekrasov sum representations of isomonodromic tau functions},
Preprint arXiv:1608.00958v2.

\bibitem[GM]{GM} 
P. Gavrylenko, A. Marshakov,
{\it Free fermions, W-algebras and isomonodromic deformations},
Theor. Math. Phys. {\bf 187} (2016) 649-677.


\bibitem[HJS]{HJS}
L. Hadasz, Z. Jaskolski, P. Suchanek, 
{\it Modular bootstrap in Liouville field theory}, 
Phys. Lett. {\bf B685} (2010) 79--85.

\bibitem[Hi]{Hi} N. Hitchin, {\it Stable bundles and integrable systems},
Duke Math. J. {\bf 54} (1987) 91--114.

\bibitem[HL]{HL}
Y.-Z. Huang, J. Lepowsky, 
{\it Tensor categories and the mathematics of rational and logarithmic conformal field theory}, 
J.Phys. {\bf A46} (2013) 494009. 

\bibitem[H96]{Hu} Y.-Z. Huang, 
{\it Virasoro Vertex Operator Algebras, the (Nonmeromorphic)
 Operator Product Expansion and the Tensor Product Theory},
Journal of Algebra {\bf 182} (1996) 201--234.

\bibitem[IJS]{IJS}
Y. Ikhlef, J.L. Jacobsen, H. Saleur,
{\it Three-point functions in $c\leq 1$ Liouville theory and conformal loop ensembles.}
Phys. Rev. Lett. {\bf 116} (2016) 130601.

\bibitem[ILP]{ILP}
A. Its, O. Lisovyy, A. Prokhorov,
{\it Monodromy dependence and connection formulae for isomonodromic tau functions},
Preprint arXiv:1604.03082. 

\bibitem[ILTe]{ILT} N. Iorgov, O. Lisovyy, J. Teschner,
{\it Isomonodromic Tau-Functions from Liouville Conformal Blocks},
Comm. Math. Phys. {\bf 336} (2015) 671--694.

\bibitem[ILTy]{ILTy} N. Iorgov, O. Lisovyy, Yu. Tykhyy,
{\it Painlevé VI connection problem and monodromy of $c=1$ conformal blocks},
JHEP {\bf 12} (2013) 029.

\bibitem[IKSY]{IKSY}
K. Iwasaki, H. Kimura, S. Shimomura, and M. Yoshida. 
{\it From Gauss to Painlev\'e, a Modern Theory of Special Functions}, Volume {\bf E 16}. 
Aspects of Mathematics, 1991.


\bibitem[Ji]{Ji} M. Jimbo,
{\it Monodromy problem and the boundary condition for some Painlev\'e
equations.}
Publ. Res. Inst. Math. Sci.~\textbf{18} (1982), no. 3, 1137--1161.

\bibitem[Ka]{Ka} Y. Kawahigashi ,
{\it Conformal Field Theory, Tensor Categories and Operator Algebras}, 
J. Phys. {\bf A48} (2015) 303001. 


\bibitem[KL]{KL}
D. Kazhdan, G. Lusztig
{\it Tensor structures arising from affine Lie algebras. I,II:} 
J. Amer. Math. Soc. {\bf 6} (1993) 905-947, 949-1011;
{\it IV:} J. Amer. Math. Soc. {\bf 7} (1994) 383-453.

\bibitem[KR]{KR} V.G. Kac, A.K. Raina,
{\it Bombay lectures highest weight representations of infinite-dimensional Lie algebras},
Advanced Series in Mathematical Physics {\bf 2}. World Scientific, 1987.

\bibitem[KRV]{KRV} 
A. Kupiainen, R. Rhodes, V. Vargas,
{\it Integrability of Liouville theory: proof of the DOZZ Formula},
Preprint arXiv:1707.08785.


\bibitem[Mo]{Mo}
G.W. Moore, 
{\it Geometry of the string equations}, 
Comm. Math. Phys. {\bf 133} (1990) 261--304.

\bibitem[MS]{MS} G. Moore, N. Seiberg,
{\em Classical and quantum conformal field theory},
Comm. Math. Phys. {\bf 123} (1989) 177--254.


\bibitem[NS]{NS}
N. Nekrasov,  S. Shatashvili,
{\it Quantization of Integrable Systems and Four Dimensional Gauge Theories}, 
In: Proceedings of the 16th International Congress on Mathematical Physics, 
Prague, August 2009, P. Exner, Editor, pp.265-289, World Scientific 2010, p.265-289.


\bibitem[NRS]{NRS} N. Nekrasov, A. Rosly, S. Shatashvili,
{\em Darboux coordinates, Yang-Yang functional, and gauge theory}, 
Nucl. Phys. Proc. Suppl.~\textbf{216} (2011) 69--93.


\bibitem[Pa]{Pa}
J. Palmer, 
{\it Determinants of Cauchy-Riemann operators as $\tau$-functions}, 
Acta Appl. Math. {\bf 18} (1990) 199--223.

\bibitem[PS]{PS}
A. Pressley, G. Segal, {\it Loop groups}, 
Oxford University Press, 1986.

\bibitem[RiS]{RS} S. Ribault, R. Santachiara,
{\it 	Liouville theory with a central charge less than one},
JHEP {\bf 1508} (2015) 109.

\bibitem[ReS]{ReS} A. Recknagel, V. Schomerus,
{\it Boundary Conformal Field Theory and the Worldsheet Approach to D-Branes},
Cambridge Monographs on Mathematical Physics
Cambridge University Press, 2013.


\bibitem[RW]{RW} I. Runkel, G.M.T. Watts,
{\it A Nonrational CFT with c = 1 as a limit of minimal models}.
JHEP {\bf 0109} (2001) 006.

\bibitem[SY]{SY} R. Sasaki, I. Yamanaka, 
{\it Virasoro algebra, vertex operators, quantum Sine- Gordon and solvable Quantum Field theories}, 
Adv. Stud. in Pure Math. {\bf 16} (1988) 271-296.

\bibitem[SMJ]{SMJ}
M. Sato, T. Miwa, M. Jimbo, 
{\it  Holonomic quantum fields. II -- The Riemann-Hilbert Problem}, Publ. 
RIMS, Kyoto Univ. {\bf 15} (1979) 201-278.

\bibitem[S]{S}	Volker Schomerus, 
{\it Rolling tachyons from Liouville theory}, 
JHEP {\bf 0311} (2003) 043. 

\bibitem[Se]{Se} G. Segal. 
{\it The definition of conformal field theory}, in: 
{\it Topology, geometry and quantum field theory}, vol. {\bf 308} of 
London Math. Soc. Lecture Note Ser., pages 421--577. Cambridge Univ. Press, 2004.

\bibitem[SW]{SW} 
G. Segal, G. Wilson,
{\it Loop groups and equations of KdV type}, 
Publ. Math. IHES {\bf 61} (1985) 5--65.

\bibitem[T95]{T95}
J.Teschner, {\it On the Liouville three-point function}. 
Phys.Lett., {\bf B363} (1995) 63. 

\bibitem[Ten]{Ten} J.E. Tener,	
{\it Geometric realization of algebraic conformal field theories}, 
Preprint arXiv:1611.01176.

\bibitem[T01]{T01}
J.~Teschner, {\em Liouville theory revisited}, 
Class.\ Quant.\ Grav.\  {\bf 18} (2001) R153--R222.


\bibitem[T03]{T03} J. Teschner,
{\it A lecture on the Liouville vertex operators},
Int. J. Mod. Phys. {\bf A19S2} (2004) 436--458



\bibitem[T10]{T10} J. Teschner,
{\it Quantization of the Hitchin moduli spaces, Liouville theory,
and the geometric Langlands correspondence I}.
Adv. Theor. Math. Phys.  {\bf 15}  (2011) 471--564.



\bibitem[TV]{TV13} J. Teschner, G. S. Vartanov,
{\it Supersymmetric gauge theories, quantization of 
moduli spaces of flat connections, and conformal field theory}.
Adv. Theor. Math. Phys. {\bf 19} (2015) 1--135.  
 
\bibitem[T17a]{T17a} J. Teschner,
{\it Semiclassical Limit of Virasoro conformal blocks and the isomonodromic  deformation problem},
Preprint arXiv:1707.07968.

\bibitem[T17b]{T17b} J. Teschner,
{\it Quantisation conditions of the quantum Hitchin system and the real geometric Langlands correspondence},
Preprint arXiv:1707.07873.
 
 

\bibitem[YY]{YY}
C. N. Yang, C. P. Yang, 
{\it Thermodynamics of a one-dimensional system of bosons with repulsive delta-function interaction}, 
J. Math. Phys. {\bf 10} (1969) 1115.

\bibitem[Yi]{Yi} Xi Yin, {\it Conformal Bootstrap
in Two Dimensions}, Talk at String-Math 2017, available at 
https://stringmath2017.desy.de/e45470/

\bibitem[Z85]{Za} A. Zamolodchikov, 
{\it Infinite additional symmetries in two-dimensional conformal field
theory}, Theor. Math. Phys. {\bf 65} (1985) 1205-1213.

\bibitem[ZZ]{ZZ}  A.B.Zamolodchikov, Al.B.Zamolodchikov,
{\it Structure Constants and Conformal Bootstrap in Liouville Field Theory.}
Nucl. Phys. {\bf B477} (1996) 577--605.

\bibitem[Z87]{Z87} 
Al. Zamolodchikov, 
{\it Conformal symmetry in two-dimensional space: Recursion representation of conformal block}, 
Theor. Math. Phys. {\bf 73} (1987) 1088--1093.

\bibitem[Z05]{Z05} Al.Zamolodchikov
{\it On the Three-point Function in Minimal Liouville Gravity},
Preprint arXiv:hep-th/0505063. 

\bibitem[Zh]{Zh} Y. Zhu, 
{\it Modular invariance of characters of vertex operator algebras}, J. AMS {\bf 9} (1996)
237-302.

\end{thebibliography}
\end{document}